\documentclass[iop,numberedappendix]{emulateapj-rtx4}

\usepackage{mathrsfs }
\usepackage{natbib}
\usepackage{multirow}
\usepackage{amssymb}
\usepackage{apjfonts}
\newcommand{\msun}{\ensuremath{\rm M_\odot}}

\newcommand{\lya}{\ensuremath{\rm Ly\alpha}}
\newcommand{\lyb}{\ensuremath{\rm Ly\beta}}
\newcommand{\zla}{\ensuremath{z_{\rm Ly\alpha}}}
\newcommand{\zis}{\ensuremath{z_{\rm IS}}}

\def\NHI{$N_{\rm HI}$}

\def\kms{km~s$^{-1}$}
\def\cm2{cm$^{-2}$}
\def\Msun{M$_{\odot}$}
\def\hmin{h$^{-1}$}
\def\dtran{$D_{\rm tran}$}
\def\dv{$\Delta v$}
\def\absdv{$|\Delta v|$}
\def\dtd{$D_{\rm 3D}$}
\def\dhub{$D_{\rm Hubble}$}
\def\fc{$f_{\rm c}$}
\def\Fc{$F_{\rm c}$}
\newcommand{\bd}{\ensuremath{{b_{\rm d}}}}
\def\rvir{$r_{\rm vir}$}
\def\eabs{$\eta_{\rm abs}$}
\def\ltsima{$\; \buildrel < \over \sim \;$}
\def\gtsima{$\; \buildrel > \over \sim \;$}
\def\simgt{\lower.5ex\hbox{\gtsima}}
\def\simlt{\lower.5ex\hbox{\ltsima}}
\newcommand{\secpoint}{\mbox{$''\mskip-7.6mu.\,$}}

\begin{document}

\title{The Gaseous Environment of High-$z$ Galaxies: 
Precision Measurements of Neutral Hydrogen \\ 
in the Circumgalactic Medium of $z \sim 2-3$ Galaxies in the Keck Baryonic Structure Survey\altaffilmark{1}}

\author{
 Gwen C. Rudie,\altaffilmark{2}
 Charles C. Steidel,\altaffilmark{2}
 Ryan F. Trainor,\altaffilmark{2}
 Olivera Rakic,\altaffilmark{3,4} 
 Milan Bogosavljevi\'{c},\altaffilmark{5}
 Max Pettini,\altaffilmark{6,7} 
  Naveen Reddy,\altaffilmark{8,9,10}
 Alice E. Shapley\altaffilmark{11},
 Dawn K. Erb,\altaffilmark{12} and
 David R. Law\altaffilmark{13}}



\altaffiltext{1}{Based on data obtained at the W.M. Keck Observatory, which is operated as a scientific partnership among the California Institute of Technology, the University of California,  and the National Aeronautics and Space Administration, and was made possible by the generous financial support of the W.M. Keck Foundation.}
\altaffiltext{2}{Cahill Center for Astronomy and Astrophysics, California Institute of Technology, MS 249-17, Pasadena, CA 91125, USA}
\altaffiltext{3}{Leiden Observatory, Leiden University, P.O. Box 9513, 2300 RA Leiden, The Netherlands}
\altaffiltext{4}{Max-Planck-Institut f\"{u}r Astronomie, K\"{o}nigstuhl 17, 69117 Heidelberg, Germany}
\altaffiltext{5}{Astronomical Observatory, Volgina 7, 11060 Belgrade, Serbia}
\altaffiltext{6}{Institute of Astronomy, Madingley Road, Cambridge CB3 0HA, UK}
\altaffiltext{7}{International Centre for Radio Astronomy Research, University of Western Australia, 7 Fairway, Crawley WA 6009, Australia}
\altaffiltext{8}{National Optical Astronomy Observatory, 950 N Cherry Ave, Tucson, AZ 85719}
\altaffiltext{9}{Department of Physics and Astronomy, UC Riverside, 900 University Avenue, Riverside, CA 92521}
\altaffiltext{10}{Hubble Fellow}
\altaffiltext{11}{Department of Astronomy, University of California, Los Angeles, 430 Portola Plaza, Los Angeles, CA 90024, USA}
\altaffiltext{12}{Department of Physics, University of Wisconsin Milwaukee, Milwaukee, WI, 53211}
\altaffiltext{13}{Dunlap Institute for Astronomy \& Astrophysics, University of Toronto, 50 St. George Street, Toronto M5S 3H4, Ontario, Canada}

\email{gwen@astro.caltech.edu}


\shortauthors{Rudie et~al.}


\shorttitle{Neutral Hydrogen in the $z\sim2.3$ CGM}


\begin{abstract}
We present results from the Keck Baryonic Structure Survey (KBSS), a unique spectroscopic survey of the distant universe designed to explore the details of the connection between galaxies and intergalactic baryons within the same survey volumes, focusing particularly on scales from $\sim 50$ kpc to a few Mpc. The KBSS is optimized for the redshift range $z\sim 2-3$, combining S/N$\sim 100$ Keck/HIRES spectra of 15 of the brightest QSOs in the sky at $z\simeq 2.5-2.9$ with very densely sampled galaxy redshift surveys within a few arcmin of each QSO sightline. In this paper, we present quantitative results on the distribution, column density, kinematics, and absorber line widths of neutral hydrogen (\ion{H}{1}) surrounding a subset of 886 KBSS star-forming galaxies with $2.0 \simlt z \simlt 2.8$ and with projected distances $\le 3$ physical Mpc from a QSO sightline. Using Voigt profile decompositions of the full Ly$\alpha$ forest region of all 15 QSO spectra, we compiled a catalog of $\sim6000$ individual absorbers in the redshift range of interest, with $12 \le $log(\NHI)$\le21$. These are used to measure \ion{H}{1} absorption statistics near the redshifts of foreground galaxies as a function of projected galactocentric distance from the QSO sightline and for randomly chosen locations in the intergalactic medium (IGM) within the survey volume. We find that \NHI\ and the multiplicity of velocity-associated \ion{H}{1} components increase rapidly with decreasing galactocentric impact parameter and as the systemic redshift of the galaxy is approached. The strongest \ion{H}{1} absorbers within $\simeq 100$ physical kpc of galaxies have \NHI\ $\sim 3$ orders of magnitude higher than those near random locations in the IGM. The circumgalactic zone of most significantly enhanced \ion{H}{1} absorption is found within transverse distances of $\simlt 300$ kpc and within $\pm300$ \kms\ of galaxy systemic redshifts. Taking this region as the defining bounds of the circumgalactic medium (CGM), nearly half of absorbers with log(\NHI) $>$ 15.5 are found within the CGM of galaxies meeting our photometric selection criteria, while their CGM occupy only 1.5\% of the cosmic volume. The spatial covering fraction, multiplicity of absorption components, and characteristic \NHI\ remain significantly elevated to transverse distances of $\sim$2 physical Mpc from galaxies in our sample. Absorbers with \NHI\ $>10^{14.5}$ \cm2\ are tightly correlated with the positions of galaxies, while absorbers with lower \NHI\ are correlated with galaxy positions only on $\simgt$Mpc scales. Redshift anisotropies on these larger scales indicate coherent infall toward galaxy locations, while on scales of $\sim 100$ physical kpc peculiar velocities of $\Delta v\simeq \pm260$ \kms\ with respect to the galaxies are indicated. The median Doppler widths of individual absorbers within 1-3 $r_{\rm vir}$ of galaxies are larger by $\simeq 50$\% than randomly chosen absorbers of the same \NHI, suggesting higher gas temperatures and/or increased turbulence likely caused by some combination of accretion shocks and galactic winds around galaxies with M$_{\rm halo} \simeq 10^{12}$ M$_{\odot}$ at $z \sim 2-3$. 
\end{abstract}


\keywords{cosmology: observations --- galaxies: high-redshift --- galaxies: evolution ---  galaxies: formation --- intergalactic medium --- quasars: absorption lines}





\section{Introduction}

Hydrogen, comprising three quarters of the baryonic mass of the universe, is the principal component of all luminous objects in the universe. It is the fuel source for stars and therefore for star formation. Thus, in order to understand the formation and evolution of galaxies, one must understand and be able to trace the inflow and outflow of this fuel. 

There exist very poor observational constraints on the movement of baryons in and out of galaxies. At high redshift in star-forming systems, it has been argued that the outflow rate must be similar to the star-formation rate \citep{pet00} and that the inflow rate must be similar to the combined star-formation rate and outflow rate \citep{erb08, fin08}.

Recently there has been a flurry in the theoretical literature predicting the prevalence of accretion of cold gas (log($T$) $\simlt 4.5-5.5$ K) onto high-\textit{z} galaxies via filamentary ``cold flows'' \citep{bir03,ker05, ocv08,bro09,fau11b,van11b,van11a}.  In this model, the baryons stream into galaxies along the filamentary structure of the cosmic web, accreting onto galaxies without experiencing virial shocks. A wide range of predictions has been made concerning the efficiency of the transport of this material into galaxy halos,  as well as its role in fueling ongoing star formation \citep{van11a,fau11b}. Further, there may be substantial suppression of the cold accretion rate caused by galaxy-scale mass outflows, evidence for which is commonly observed in the spectra of high-$z$ star-forming systems \citep{pet01, sha03, ade05, ccs10}. Mapping the gas distribution surrounding galaxies is critical to constraining these models \citep{fau11a, fum11}, and would be a significant step toward understanding and quantifying the exchange of baryons between the sites of galaxy formation and the nearby intergalactic medium (IGM). 

There has been a large amount of recent theoretical examination of the nature of IGM absorbers and their relation to galaxies using simulations. Ly$\alpha$ is believed to broadly trace the filamentary large-scale structure \citep{cen94, zha95, mir96,her96,rau97,the98, dav99, sch01} although there are indications \citep{bar11} that winds could blow spatially extended halos of gas which may have recently been observed both in absorption \citep{ccs10} and in Ly$\alpha$ emission \citep{ccs11}. There seems to be general agreement that galactic winds are responsible for metal absorbers in the IGM. \citet{boo10} suggest that mostly low mass (M$_{\rm DM} \simlt 10^{10} $ \Msun) galaxies must be responsible for the pollution, while \citet{wie10}  suggest only half of the metals would originate from galaxies with M$_{\rm DM} \simlt 10^{11}$ \Msun.  \citet{wie10} also studied the history of the ejection of these metals and found that half of the metals observed at redshift 2 were ejected during the time between $2 < z < 3$. Using cosmological ``zoom-in'' simulations \citet{she11} found a ``Lyman Break''-type galaxy could distribute metals to 3 virial radii by $z=3$. \citet{sim11} recently considered this problem observationally, finding that indeed 50\% of the metals observed in the IGM at $z\sim2.4$ were placed there since $z\sim4.3$, i.e. in 1.3 Gyr. \citet{opp08} and \citet{opp10} studied the fate of winds using cosmological simulations and found that while galactic winds are likely responsible for the metallic species seen in the IGM, much of the outflowing gas may be bound to galaxies and may fall back in. In their simulations the recycling timescale scaled inversely with mass because winds emanating from more massive galaxies experienced greater hydrodynamic drag due to the increased abundance of dense IGM surrounding them. Further, in these simulations the largest source of gaseous fuel for star formation after $z \sim 1$ was recycled wind material.

To date, systematic attempts to jointly study high-$z$ galaxies and their intergalactic environs have been made by \citet{ade03,ade05} \citep[see also][]{cri11}. These studies focused primarily on \ion{H}{1} and \ion{C}{4} absorption in the spectra of background QSOs whose sightlines probed regions covered by $2 < z < 4$ Lyman break galaxy (LBG) surveys. This work allowed for a first glimpse of the distribution of diffuse gas surrounding forming galaxies at high redshift, and, perhaps more tantalizing, evidence for interactions between the IGM and galaxies during the epoch when galaxies are expected to be most active. Generally, \citet{ade03} and \citet{ade05} found excess \ion{H}{1} absorption out to $\approx 5-6$\hmin comoving Mpc (cMpc) of galaxies [$\sim$2 physical Mpc (pMpc) at $\langle z\rangle = 3.3$ using the cosmology adopted in this paper]. \ion{C}{4} systems were found to correlate strongly with the positions of galaxies suggesting a causal connection. Unfortunately, these papers could not consider physical properties of the gas such as its column density or temperature because the data were not of sufficiently high quality to perform Voigt profile analysis. As such these papers focused on the transmitted flux which could be applied to a wider range of data qualities. 

In this work, we provide high-accuracy analysis of the spectral regions surrounding 886 high-$z$ star-forming galaxies as seen in absorption against the spectra of background hyper-luminous QSOs using data drawn from the Keck Baryonic Structure Survey (KBSS; Steidel et al, in prep). The KBSS was specifically designed to allow for the observation of gas absorption features surrounding high-redshift star-forming galaxies, providing unique insight into the IGM/galaxy interface at high redshift. The size and quality of the KBSS sample allow us to map the distribution and properties of gas near to individual star-forming galaxies with direct physical parameters such as the column density as opposed to proxies such as the equivalent width. This paper is the first in a series designed to study the physical properties of star-forming galaxies at high redshift using Voigt profile analysis of this data sample. The complementary analysis presented by \citet{rak11b} describes the pixel statistics of the QSO spectra from the KBSS and the correlation of \ion{H}{1} optical depth with the positions and redshifts of galaxies. 

In \S \ref{data} we discuss the galaxy and QSO data and present the Voigt profile analysis of the QSO spectra in \S \ref{spectral_analysis}. The distributions of \ion{H}{1} absorbers as a function of velocity, impact parameter, and 3D distance are presented in \S \ref{analysis}. In \S \ref{cf} we consider the geometric distribution of the gas using the covering fraction and incidence of absorbers. \S \ref{mapText} focuses on 2D ``maps''  of the absorber distribution around galaxies. In \S \ref{dop} we analyze the velocity widths of individual absorbers and their correlation with the proximity of galaxies. We discuss the results and their possible interpretation in \S \ref{discussion} with a brief summary of the paper and our conclusions in \S \ref{conclusions}.

Throughout this paper we assume a $\Lambda$-CDM cosmology with $H_{0} = 70$ \kms\  Mpc$^{-1}$, $\Omega_{\rm m} = 0.3$, and $\Omega_{\Lambda} = 0.7$. All distances are expressed in physical (proper) units unless stated otherwise. We use
the abbreviation pkpc and pMpc to indicate physical units, and ckpc or cMpc for co-moving units. 
At the mean redshift of the galaxy sample ($\langle z \rangle = 2.3$), 300 pkpc is 210\hmin pkpc (physical) or $\simeq 700$\hmin ckpc; the age of the universe is 2.9 Gyr, the look-back time is 10.9 Gyr, and 8.2 pkpc subtends one arcsecond on the sky.

\label{intro}

\section{Observations}

\label{data}

The data presented in this paper are drawn from the Keck Baryonic Structure Survey (KBSS) and include a large sample of rest-UV (2188) and rest-optical (112) spectra of UV-color selected star-forming galaxies at $\langle z \rangle~\sim~2.3$. These galaxies were photometrically selected to lie in the foreground of one of 15 hyper-luminous QSOs in the redshift range $2.5 \le z_{\rm QSO} \le 2.85$ for which we have obtained high-resolution, high signal-to-noise ratio (S/N) echelle spectra. 

  The redshift range of this survey has important significance in the history of the universe -- it coincides with the peak of both universal star formation \citep[see ][]{red08} and super-massive black hole growth \citep[see ][]{ric06}.  Spectroscopic observations of star-forming galaxies during this epoch commonly exhibit signatures of strong outflowing winds \citep{pet01, sha03, ade05, ccs10}. At the same time, the baryonic accretion rate onto galaxies is predicted to be near its peak at $z\sim2.5$ (e.g., \citealt{van11a, fau11b, van11b}.) Thus, the signatures of galaxy formation within the IGM should be at their peak as well. 

The redshift range $2.0 \simlt z \simlt 2.8$ offers a number of practical advantages as well: first, the rapidly-evolving \lya\ forest has thinned enough to allow measurements of individual \ion{H}{1} systems and to enable the detection of important metallic transitions falling in the same range (notably, \ion{O}{6} $\lambda\lambda 1031$, 1036); second, the astrophysics-rich rest-frame far-UV  becomes accessible to large ground-based telescopes equipped with state-of-the-art optical spectrographs; third, the rest-frame optical spectrum is shifted into the atmospheric transmission windows in the near infrared, allowing for observations of a suite of diagnostic emission lines (H$\alpha$, H$\beta$, [\ion{O}{2}] $\lambda\lambda$3726,3729, [\ion{O}{3}]$\lambda\lambda\lambda$4363,4959,5007, and [\ion{S}{2}] $\lambda\lambda$6717,6732) neatly packed into the J, H, and K bands. 
  
\begin{deluxetable*}{lllllcccc}
\tablecaption{KBSS Central QSOs and Foreground Galaxy Samples}  
\tablewidth{0pt}
\tablehead{
\colhead{Name} & \colhead{RA} & \colhead{Dec} & \colhead{$z_{\rm QSO}$\tablenotemark{a}} & \colhead{$\lambda_{\rm min}$\tablenotemark{b}}  &  \colhead{$z_{\rm gal}$ range} & \colhead{N$_{\rm gal}$\tablenotemark{c}} & \colhead{S/N Ly$\alpha$\tablenotemark{d}} & \colhead{S/N Ly$\beta$}\tablenotemark{d}}
\startdata
 Q0100+130 (PHL957) & 01:03:11.27 &  $+$13:16:18.2  &  2.721 & 3133 &  2.0617-- 2.6838 & ~47~  & ~77   &   50  \\
 HS0105+1619 & 01:08:06.4  &  $+$16:35:50.0  &  2.652 & 3230 &  2.1561-- 2.6153 & ~53~  &127  	&   89  \\
 Q0142$-$09 (UM673a) & 01:45:16.6  &  $-$09:45:17.0  &  2.743 & 3097 &   2.0260-- 2.7060  & ~65~  & ~71  	&   45  \\
 Q0207$-$003 (UM402) & 02:09:50.71 &  $-$00:05:06.5  &  2.872 & 3227 & 2.1532-- 2.8339 & ~46~  &   ~82  	&   55  \\
 Q0449$-$1645 & 04:52:14.3  &  $-$16:40:16.2  &  2.684 &   3151 &  2.0792-- 2.6470 & ~50~  &  ~73  	&   41  \\
 Q0821+3107 & 08:21:07.62 &  $+$31:07:51.17 &  2.616 & 3239 &    2.1650-- 2.5794 & ~37~  &   ~50  	&   33  \\
 Q1009+29 (CSO 38) & 10:11:55.60 &  $+$29:41:41.7  &  2.652\tablenotemark{e} &  3186 &   2.1132-- 2.6031	& ~36~  &   ~99  	&   58  \\
 SBS1217+499 & 12:19:30.85 &  $+$49:40:51.2  &  2.704  &   3098 & 	 2.0273-- 2.6669     & ~43~  &  ~68  	&   38  \\
 HS1442+2931 & 14:44:53.67 &  $+$29:19:05.6  &  2.660 & 3152 &    2.0798-- 2.6237 & ~46~  &  ~99  	&   47  \\
 HS1549+1919 & 15:51:52.5  &  $+$19:11:04.3  &  2.843 &    3165 & 	 2.0926-- 2.8048 & ~54~ &  173  	&   74  \\
HS1603+3820 & 16:04:55.38 &  $+$38:12:01.8  &  2.551\tablenotemark{f} & 3181 &  2.1087-- 2.5066  & ~37~  &  108  	&   58  \\
 Q1623+268 (KP77) & 16:25:48.83 &  $+$26:46:58.8  &  2.5353 &   3126 &  2.0544-- 2.4999 & 133$\tablenotemark{g}$   &    ~48  	&   28  \\
 HS1700+64 & 17:01:00.6  &  $+$64:12:09.4  &  2.751 &   3138 & 2.0668-- 2.7138  & 110$\tablenotemark{g}$   &     ~98  	&   42  \\
 Q2206$-$199 & 22:08:52.1  &  $-$19:43:59.7  &  2.573 &  3084 &  2.0133-- 2.5373  & ~45~  & 	   ~88  	&   46  \\
 Q2343+125 & 23:46:28.30 &  $+$12:48:57.8  &  2.5730 &   3160 & 2.0884-- 2.5373 & ~84$\tablenotemark{g} $  &                     ~71  	&   45      
 \enddata
  \tablenotetext{a}{The redshift of the QSO}
    \tablenotetext{b}{The minimum wavelength covered in the HIRES QSO spectrum}
    \tablenotetext{c}{The number of galaxies in our LRIS survey with \dtran\ $<$ 3 pMpc and spectroscopic redshifts in the correct range given in the previous column.}
     \tablenotetext{d}{The average signal to noise ratio per pixel of the QSO spectrum in the wavelength range pertaining to CGM Ly$\alpha$ and Ly$\beta$ absorption.}
      \tablenotetext{e}{The redshift of this QSO was revised after the fitting of the HIRES spectrum was completed. The redshift assumed for the $\Delta v = -3000$ \kms\ is 2.6395}
       \tablenotetext{f}{The redshift of this QSO was revised after the fitting of the HIRES spectrum was completed. The redshift assumed for the $\Delta v = -3000$ \kms\ is 2.5420}
      \tablenotetext{g}{The photometry and spectroscopy in fields Q1623, HS1700, and Q2343 cover larger areas than a single LRIS footprint. As such, they represent the fields in which we sample the galaxy-IGM connection at \dtran\ $>$ 2 pMpc.}
     \label{field}
\end{deluxetable*}

\begin{figure}
\includegraphics[width=.95\columnwidth]{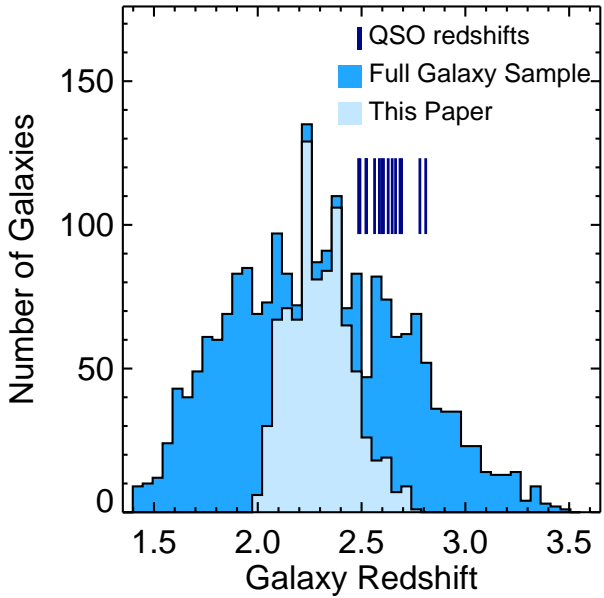}
\caption{The redshift distribution of the 2188 galaxies in the full KBSS sample (dark shaded histogram) and of the subset of 886 galaxies used for the analysis in this paper (light shaded histogram), which are those with redshifts high enough that the corresponding wavelength of the \lyb\ transition is observed in the relevant QSO spectrum, and with $z < z_{\rm QSO} - 3000$ \kms. The dark vertical line segments mark the redshifts of the 15 KBSS QSOs. }
\label{redshift_hist}
\end{figure}

\begin{figure}
\center
\includegraphics[width=.95\columnwidth]{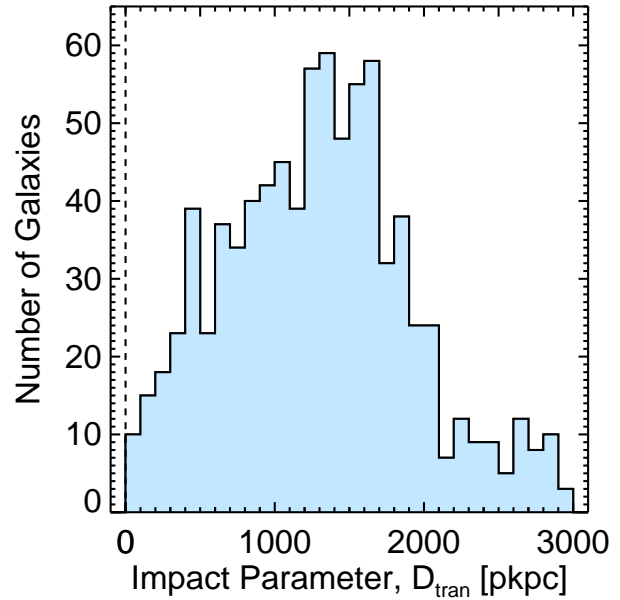}
\caption{The number of galaxies as a function of physical impact parameter \dtran\ for the sample which has appropriate redshifts to be used in this work (light shaded histogram in Figure~\ref{redshift_hist}; see Table~\ref{field}). The decline in the number of galaxies at \dtran\ $\approx$ 2 pMpc is due to the typical survey geometry of the KBSS fields as described in \S \ref{largescale}.}
\label{impact}
\end{figure}

\subsection{The Galaxy Sample}

\label{gal_sample}

The KBSS galaxy sample, a subset of which is used in this paper (Figure~\ref{redshift_hist}), is described in detail by Steidel et al (2012); here we present a brief summary. Galaxies were selected for spectroscopy using their rest-frame UV colors (i.e., LBGs) according to the criteria outlined by \citet{ccs04} and \citet{ade04} for $z \sim 2$ and by \citet{ccs03} for $z \sim 3$. In combination, these criteria have been devised to efficiently select star-forming galaxies over the redshift range $1.5 \simlt z \simlt 3.5$. In total, the KBSS galaxy sample includes $R \simeq 1000$ ($\sim$5-10\AA) rest-UV spectra of 2188 star-forming galaxies obtained at the W.M. Keck Observatory using the Keck 1 10m telescope and the blue arm of the Low Resolution Imaging Spectrometer (LRIS-B; \citealt{oke95, ccs04}.) 

Most of the observations were conducted using a 400~lines~mm$^{-1}$ grism (blazed at 3400 \AA\ in first order) in combination with a dichroic beamsplitter sending all light shortward of $\sim 6800$ \AA\ into the blue channel, where the wavelength coverage for a typical slit location was $\simeq 3100-6000$~\AA\ with a resolving power $R \simeq 800$ using 1\secpoint2 wide slits \citep{ccs10}. Some of the spectra were obtained using the d560 beamsplitter (beam divided near 5600 \AA) together with a 600 lines mm$^{-1}$ grism (4000 \AA\ blaze), typically covering $\simeq 3400-5600$~\AA\ with a resolving power $R \simeq 1300$.  Observations obtained after July 2007 made use of the Cassegrain Atmospheric Dispersion Corrector, thus minimizing the effects of differential atmospheric refraction over the spectral range of interest.  Wavelength calibration was accomplished using Hg, Cd, Zn, and Ne lamps, with zero point corrections based on night sky emission lines on each individual exposure.\footnote{An extensive discussion of the LRIS/KBSS wavelength calibration errors (including slit illumination and atmospheric dispersion) is given in \citet{ccs10}.}

The exposure times allocated to individual galaxies ranged from 1.5-7.5 hours depending on the number of separate masks containing the same target; typically galaxies were observed on either 1 or 2 masks, each mask receiving a total integration of 1.5 hours. Further details on the selection, observing strategy, and data reduction for KBSS galaxies are presented elsewhere \citep[][Steidel et al. 2012]{ccs10}.

The rest-UV spectra of LBGs are dominated by the continuum emission of O and B stars, over which are superposed numerous resonance absorption lines of metallic ions and \ion{H}{1}. The \ion{H}{1} \lya\ line at 1215.67 \AA\ may be seen in emission or absorption (and often in both). The absorption features arise in cool interstellar gas in the foreground of the OB stars; they are most commonly observed to be blue-shifted by 100 -- 800 \kms\ with respect to the systemic velocity of the stars, as measured from either rest-frame optical nebular emission lines or stellar photospheric lines in stacked spectra \citep[see][]{pet01, sha03, ade05, ccs10} or from the redshift-space symmetry of Ly$\alpha$ absorption in the nearby IGM \citep{rak11}. Common lines observed include: \ion{O}{6} $\lambda\lambda$1031,1036, \ion{Si}{2} $\lambda 1260$, $\lambda 1526$, \ion{Si}{2}+\ion{O}{1} $\lambda 1303$ (blend),\ion{Si}{3} $\lambda$1206, \ion{Si}{4} $\lambda\lambda$1393,1402, \ion{N}{5} $\lambda\lambda$1238,1242, \ion{C}{2} $\lambda$1334, \ion{C}{3} $\lambda$977, and \ion{C}{4} $\lambda\lambda$1548,1550. The profile of the Ly$\alpha$ emission or absorption line is modulated by the optical depth of the material closest to the systemic velocity of the stars, which has been shown to correlate most significantly with the baryonic mass \citep{ccs10} and the physical size (Law et al., in prep) of the galaxy. 

The sample used in this study includes 886 galaxies within the redshift range where at minimum Ly$\alpha$ and Ly$\beta$ are observed in the HIRES spectrum and with redshifts placing them at least 3000 \kms\ blue-ward of the redshift of the QSO. The latter criterion was selected to avoid proximate systems that originate within material ejected from the QSO itself and/or the region affected by its ionizing radiation field.  

A typical galaxy in the spectroscopic survey has a bolometric luminosity of $\sim 2.5\times10^{11}$ L$_\odot$ \citep{red08,red11}, a star-formation rate (SFR) of $\sim 30$ \Msun\ yr$^{-1}$ \citep{erb06c}, a stellar age of $\sim 0.7$ Gyr\ \citep{erb06b}, and a gas-phase metallicity of $\simeq 0.5$ $Z_\odot$ \citep{erb06a}. The galaxies inhabit dark matter halos of average mass $\sim 10^{12}$ \Msun\ \citep[][Trainor \& Steidel, in prep; Rakic et al, in prep]{ade05b,con08} and average dynamical masses of $\sim 7 \times 10^{10}$ \Msun\ \citep{erb06b} and generally exhibit dispersion-dominated kinematics \citep{law09}. The luminous parts of the galaxies are dominated by baryons, typically half stars and half cold gas \citep{sha05a,erb06b}, with half-light radii of $\sim 2$ pkpc \citep{law11}. The spectroscopic sample includes objects with apparent magnitudes ${\cal R} \le 25.5$, where ${\cal R}$ is equivalent to m$_{\rm AB}$(6830 \AA). At the mean redshift of the sample ($\langle z \rangle = 2.30$) the faint limit corresponds to a galaxy of 0.25L$_{\rm UV}^*$ \citep{red09}. 
 
 The redshift distribution of the galaxy and QSO sample is presented in Figure \ref{redshift_hist}, and the distribution of physical impact parameters, \dtran, between the galaxies and the QSO lines of sight is shown in Figure~\ref{impact}. 

At present, 112 galaxies in the full KBSS sample have been observed spectroscopically in the near-IR using 
using NIRSPEC \citep{mcl98} on the Keck II telescope. 87 of these galaxies lie in our chosen redshift interval. 
The NIRSPEC target selection, data, and reductions are discussed in \citet{erb06c,erb06b}. The NIR redshifts, generally based on the H$\alpha$ emission line, are estimated to be accurate to $\sim$ 60 \kms\ or $\sigma_z \simeq 0.0007$ at $z \sim 2.3$. 

\subsection{Measured and Calibrated Redshifts}

Because the most prominent features in the UV spectra of star forming galaxies are not at rest with respect to a galaxy's systemic redshift, $z_{\rm gal}$, corrections must be applied to avoid substantial systematic redshift errors. The velocity peak and centroid of the \lya\ emission line, when present, tend to be redshifted with respect to $z_{\rm gal}$ by several hundred \kms, while the strong UV absorption features ($z_{\rm IS}$) tend to be similarly blue-shifted with respect to $z_{\rm gal}$ \citep{sha03, ade03, ccs10}. These observations are generally interpreted as strong evidence for the presence of galaxy-scale outflows.  

Here we adopt estimates of galaxy systemic redshifts ($z_{\rm gal}$) computed in the manner proposed by \citet{ade05} and later updated by \citet{ccs10} and \citet{rak11}. \citet{ade05} and \citet{ccs10} analyzed the subset of the UV sample for which both rest-UV and rest-optical spectra had been obtained. They measured the average offset between redshifts defined by H$\alpha$ emission versus \zla\ and $z_{\rm IS}$ to estimate average corrections. The H$\alpha$ line traces the ionized gas in star-forming regions and is therefore a reasonable proxy for the systemic velocity of the stars, which are more difficult to measure due to the weakness of the UV photospheric absorption lines. \citet{rak11} used the QSO and galaxy data set presented here and calibrated velocity offsets appropriate for various classes of LBGs by insisting that the average IGM \lya\ absorption profiles should be symmetric with respect to galaxy redshifts. In both cases, the offsets represent those of the ensemble while in reality there is some scatter between individual objects even if their spectral morphology is similar. However, as we will demonstrate, the adopted $z_{\rm gal}$ must be generally quite accurate in order to produce the trends described below. 

 The formulae used for estimating $z_{\rm gal}$ from \zla\ and \zis\ measurements are reproduced below. For galaxies with H$\alpha$-based redshifts (87/886), we set $z_{\rm gal} = z_{\rm H\alpha}$. For galaxies which have measured \zis\, with or without the presence of \lya\ emission (691/886), 
 
 \begin{equation}
 z_{\rm gal, IS} \equiv \zis + \frac{\Delta v_{\rm IS}}{c}(1+z_{\rm IS})~,
 \label{zabs}
 \end{equation}
 where
  \begin{equation}
\Delta v_{\rm IS}  = 160~ \textrm{km s}^{-1}
  \end{equation}
is the velocity shift needed to transform the observed redshift into its systemic value, \zis\ is the measured redshift from the centroids of interstellar absorption lines, and $z_{\rm gal,IS}$ corresponds to the estimated
systemic redshift of the galaxy.
  
 For galaxies which have redshifts measured \textit{only} from Ly$\alpha$ in emission (90/886), we compute the redshift as
 
  \begin{equation}
 z_{\rm gal,\lya} \equiv \zla\ + \frac{\Delta v_{\lya}}{c}(1+\zla),
 \label{zlya}
 \end{equation}
 where
  \begin{equation}
\Delta v_{\lya} = -300~ \textrm{km s}^{-1}
  \end{equation}
 is the velocity shift needed to transform the observed redshift into its systemic value, \zla\ is the measured redshift from Ly$\alpha$, and $z_{\rm gal,\lya}$ is adopted systemic redshift.

For galaxies with measurements of both \zis\ and \zla, we verify that $\zis < z_{\rm gal,IS} < \zla$. If the corrected absorption redshift is not bracketed by the two measured redshifts (18/886 galaxies)\footnote{In these cases, \zis\  and \zla\ have very similar values and therefore taking their average generally provides a systemic redshift with less potential error than in those galaxies for which an average shift (as in equations \ref{zabs} and \ref{zlya}) is required.}, then we use the average of \zis\ and \zla :

\begin{equation}
 z_{\rm gal} \equiv \frac{\zla + \zis}{2}.
 \end{equation}

The residual redshift errors have a significant impact on our ability to interpret the kinematic information in the data; thus, their amplitude will be important to consider in the examination of the line-of-sight distribution of \ion{H}{1}. \citet{ccs10} found this method generally corrects the redshifts to within $ \sim 125 ~$\kms\ of the systemic velocity.

 \label{redshift}
 
 \subsection{QSO Observations}

The 15 hyper-luminous ($m_V \simeq 15.5-17$) QSOs in the center of the KBSS fields (Table~\ref{field}) were observed with the High Resolution Echelle Spectrometer \citep[HIRES;][]{vog94} on the Keck I telescope. All available archival data for these 15 QSOs have been incorporated, including data taken with UVES \citep{dek00} on the VLT for Q2206-199 and Q2343+125. We obtained additional HIRES observations in order to reach a uniformly high S/N ratio over the spectral range of primary interest, $3100-6000$ \AA. The final HIRES spectra have $R\simeq 45,000$ (FWHM$\simeq 7$ \kms), S/N $\sim 50-200$ per pixel, covering at least the wavelength range 3100 -- 6000\AA\ with no spectral gaps. The significant improvement in the UV/blue sensitivity of HIRES resulting from a detector upgrade in 2004 enabled us to observe Ly$\beta$ $\lambda 1025.7$ down to at least $z= 2.2$ in all 15 KBSS sightlines, and to significantly lower redshifts in many (see Table 1.) The additional constraints provided by Ly$\beta$ (and in many cases, additional Lyman series transitions) allow for much more accurate measurements of \ion{H}{1} for \NHI$=10^{14}-10^{17} $ \cm2\ (see \S\ref{spectral_analysis}); this is particularly important since these column densities are typical of \ion{H}{1} gas in the CGM at these redshifts (see \S\ref{cf}). 

The QSO spectra were reduced using T. Barlow's MAKEE package which is specifically tailored to the reduction of HIRES data. The output from MAKEE is a wavelength-calibrated\footnote{The wavelength calibration of the HIRES spectra introduce negligible error into our analysis. The HIRES spectra have calibration errors less that 0.5 \kms.} extracted spectrum of each echelle order, corrected for the echelle blaze function and transformed to vacuum, heliocentric wavelengths. The spectra were continuum-normalized in each spectral order using low-order spline interpolation, after which the normalized 2-D spectrograms were optimally combined into a single one-dimensional, continuum-normalized spectrum resampled at 2.8 \kms\ per pixel.

\begin{figure}
\center
\includegraphics[width=0.5\textwidth]{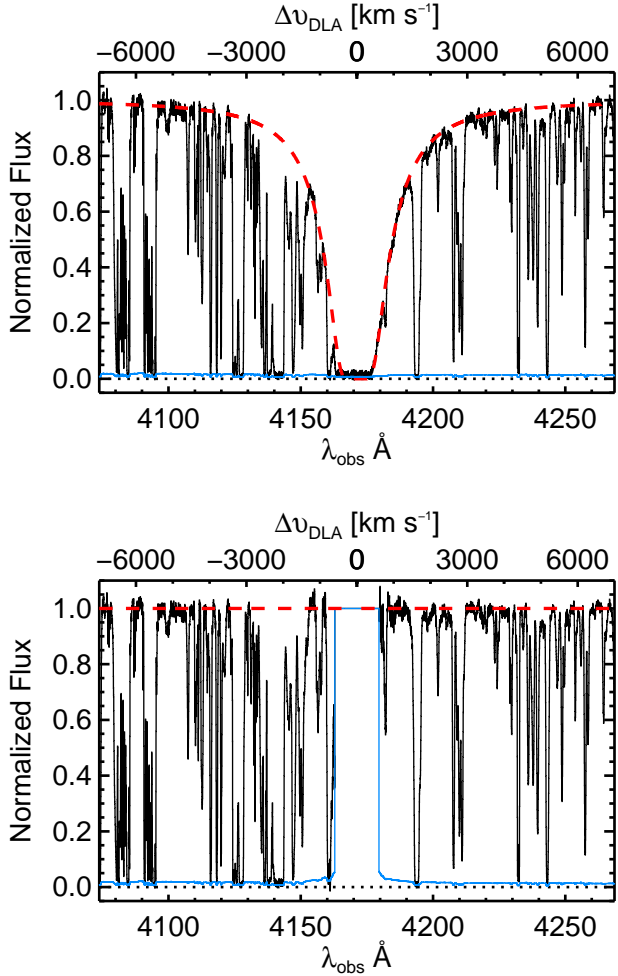}
\caption{A demonstration of our treatment of the continuum surrounding damped-Ly$\alpha$ systems. \textit{Top}: In black, the continuum-normalized HIRES spectrum of Q2343+125 showing $\pm 7000$ \kms\ surrounding the DLA. The (red) dashed line corresponds to the Voigt profile of the DLA centered at $z=2.4312$ with log(\NHI) $= 20.4$. Shown in the light (blue) curve is the error spectrum. \textit{Bottom}: The HIRES spectrum of the same QSO with the DLA profile divided out. The new error spectrum accounting for the DLA profile division is shown by the light (blue) curve. The new continuum (with the DLA divided out) is shown in the dashed (red) curve.}
\label{DLA}
\end{figure}

\label{DLA_text}

\begin{figure*}
\center
\includegraphics[width=\textwidth]{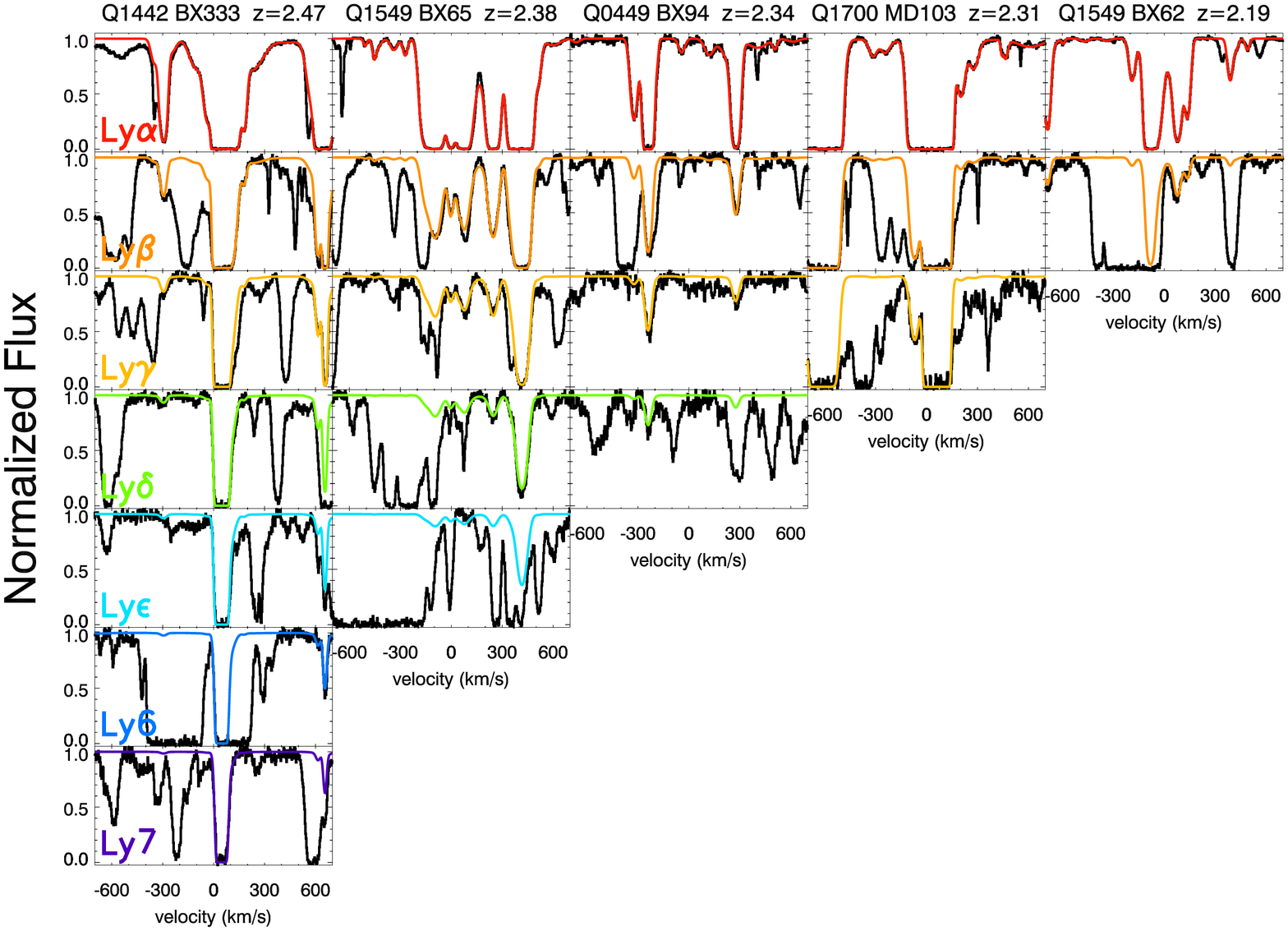}
\caption{Example fits to the QSO data surrounding the redshifts of galaxies in our sample. Displayed in black are the continuum-normalized HIRES spectra showing the Lyman series transitions within $\pm$700 \kms\ of the systemic redshift of 5 galaxies with redshifts as indicated. Over-plotted in color are Voigt profile decompositions for \ion{H}{1} absorption systems within $\pm$700 \kms\ of the galaxy redshift. Successive rows illustrate the fit to Ly$\alpha$, Ly$\beta$, etc. Absorption in the HIRES spectra that does not appear in the colored fit corresponds to absorption from metallic species or from \ion{H}{1} at a redshift far from that of the galaxy. Note that for all galaxies in our sample, the QSO spectra cover the Ly$\alpha$ and Ly$\beta$ transitions near the galaxy redshifts, and for the higher-redshift galaxies, many more transitions in the Lyman series can be measured.}
\label{hires_stack}
\end{figure*}

Three of the KBSS QSO spectra contain a damped Lyman $\alpha$ (DLA) system (\NHI $> 10^{20.3}$ cm$^{-2}$), and three contain a sub-DLA ($10^{19} \simlt$ \NHI $< 10^{20.3}$ cm$^{-2}$). Special care must be taken in the continuum fit to the regions surrounding these systems, as the damping wings of the absorption lines extend for thousands of \kms\ from line center. In these regions, we carefully fit a Voigt profile to the core of the absorber and adjusted the original continuum fit so that the Voigt profile produced a good fit to the damping wings, as shown in the top panel of Figure~\ref{DLA}. The final Voigt profile fit is divided into the true spectrum, resulting in a new spectrum where the wings of the DLA have been removed, as illustrated in the bottom panel of Figure \ref{DLA}. The re-normalized spectrum can then be used to fit additional absorption systems superposed on the damping wings. 

The redshifts of the QSOs are measured from rest-frame optical emission lines using lower-dispersion NIR spectra. The details of this procedure and the expected errors in the QSO redshifts are reported upon in Trainor \& Steidel (2011, submitted). The precise QSO redshifts do not affect our analysis.

\section{Analysis of QSO Absorption Spectra}

\label{spectral_analysis}
The process of accurately measuring \ion{H}{1} in the \lya\ forest of QSO spectra is complicated by the saturation of moderately strong absorbers and the blending of \ion{H}{1} features with other \ion{H}{1} or with lines of metallic species that happen to fall in the forest region.    

Our analysis includes a full Voigt-profile decomposition of the Ly$\alpha$ forest from the lowest redshift for which Ly$\beta$ is available in each spectrum up to $3000$~ \kms\ blue-ward of the QSO redshift; Table 1 shows the relevant redshift range used for each QSO in the sample. The cut-off at the high redshift end is to avoid \ion{H}{1} systems which could be ejected from and/or ionized by the QSO itself; the low-redshift cut is necessary due to the high frequency of systems with \NHI $\simgt 10^{14.5}$ \cm2\ that will be saturated in Ly$\alpha$\footnote{Note that the exact saturation point for the \lya\ line and for any other line depends on the line width (i.e. the Doppler parameter) and thus the internal properties of the absorbing cloud.}. For saturated systems, the fit to Ly$\alpha$ is degenerate between an increase in column density or an increase in the width of the line, \bd. The best way to resolve this degeneracy is to measure higher-order Lyman lines where decreasing oscillator strengths allow accurate \NHI\ determination.

Simultaneous fits were made to as many Lyman lines as were both (1) available in the observed spectral range and (2) needed to measure an unsaturated and uncontaminated profile in the highest-order line. The exact number of higher-order Lyman lines used therefore depended upon both the redshift of the absorber and the degree of contamination in the spectral region containing the higher-order Lyman series absorption features. Higher-order lines, whether saturated or unsaturated, were used whenever doing so provided additional constraints on the overall fit. 

Example Voigt profile fits to the \ion{H}{1} absorption in regions surrounding the systemic redshift of five galaxies from our sample are shown in Figure \ref{hires_stack}. Note that for galaxies with redshifts significantly larger than $z\sim2.2$, many higher-order Lyman transitions can be measured.

To facilitate the fitting of the Lyman $\alpha$ forest, we developed a semi-automatic line-fitting code. Briefly described, the code works with $\sim1500$ \kms\ sections of spectrum at a time\footnote{Note, the choice of 1500 \kms\ was made largely out of convenience. This velocity window is the largest window which can be easily displayed to check the goodness of fit and for which the number of components allows for a reasonable VPFIT run time.}, fitting to Ly$\alpha$ and as many higher-order lines as are accessible within the HIRES spectrum at the redshift of the \ion{H}{1} systems being fit. The algorithm first searches for systems by cross correlating a template hydrogen absorption spectrum (i.e., a single non-saturated \ion{H}{1} absorption component) with the HIRES spectrum. Peaks in the cross correlation are taken as initial estimates of the centroids of absorption lines. We fit Gaussians to these lines to estimate column densities and Doppler parameters. Residual absorption features (i.e., those inconsistent with \lya) in the Ly$\alpha$ portion of the spectrum are assumed to be metal lines. Residual absorption in the higher-order Lyman series sections of the spectra are assumed to be lower-redshift \ion{H}{1} systems. The fits begin with \lya\ at the high-redshift end of the range so that their higher-order Lyman absorption can be flagged as a known contaminant for fitting lower-redshift \ion{H}{1} absorbers.

Once estimates of the locations, column densities, and Doppler parameters of all the absorption lines are complete, they are input into the $\chi^{2}$ minimization code VPFIT\footnote{http://www.ast.cam.ac.uk/~rfc/vpfit.html; \copyright ~2007 R.F. Carswell, J.K. Webb} written by  R.F. Carswell and J.K. Webb. VPFIT simultaneously fits all transitions of \ion{H}{1} as well as the specified contaminating lines. The results are checked by eye, alterations made where the fit is inappropriate, and the process is repeated iteratively until a good fit (reduced $\chi^2 \approx 1$) is achieved. At this point, the multiple sections of spectrum are spliced together until a full fit to the forest is achieved. 

It should be noted that the Voigt profile fit to a spectrum does not represent a unique solution. In this work, we fit each set of absorbers with the minimum number of components, adding additional components only when they significantly improve the $\chi^{2}$. Median errors in \NHI\ reported by VPFIT are 0.07 dex for absorbers with 13 < log(\NHI) < 14 and 0.03 dex for absorbers with 14 < log(\NHI) < 16; however in many cases the systematic errors will exceed these values. The largest source of error in our fits to low column-density systems is uncertainty in the continuum level;  for high column-density absorbers,  it is the possibility of unrecognized sub-component structure. 

The complete decomposition of the Ly$\alpha$ forest in these 15 lines of sight includes Voigt profile fits to $>5900$ distinct \ion{H}{1} absorption systems with \NHI $> 10^{12.0}$ cm$^{-2}$, making it the largest absorber catalog ever compiled at these redshifts. It increases by an order of magnitude the number of intermediate \NHI\ absorbers measured with the additional constraint of higher-order Lyman lines.  

\section{Circumgalactic \ion{H}{1}}

\label{analysis}

\begin{figure*}
\center
\centerline{\includegraphics[height=2.5in]{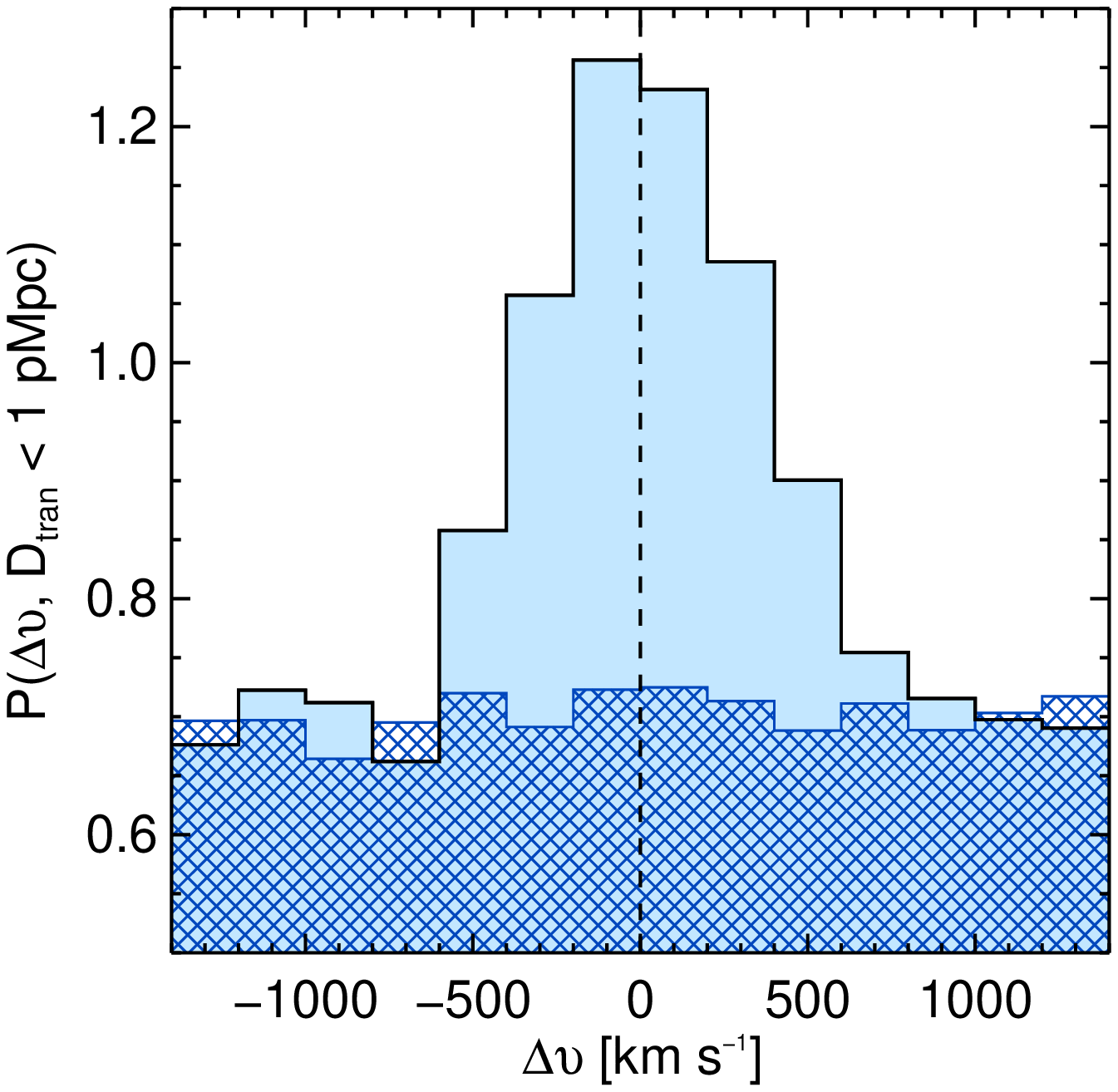}\includegraphics[height=2.5in]{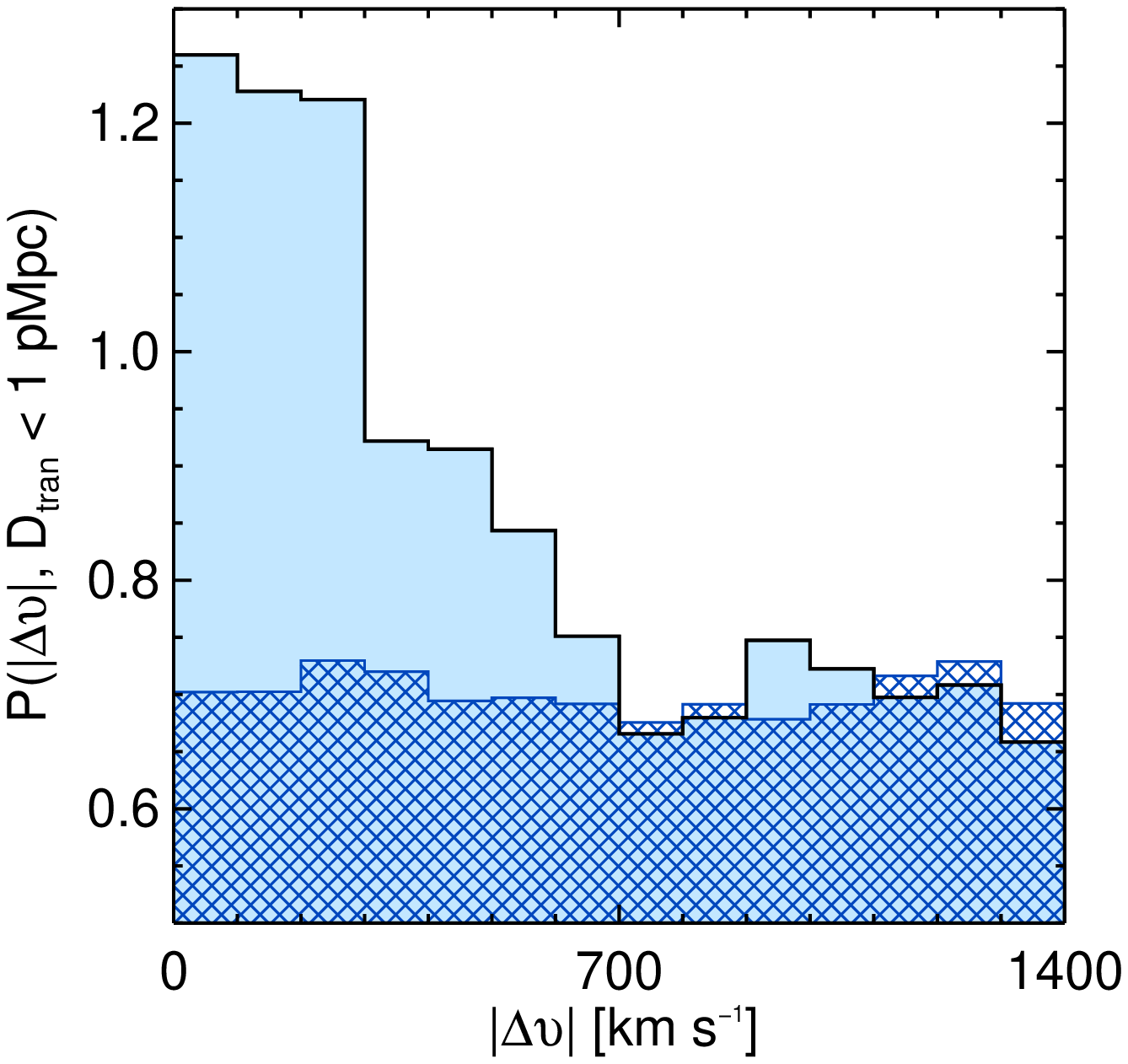}}
\caption{The velocity-space distribution of \ion{H}{1} absorption systems with respect to the systemic redshift of galaxies, normalized by the number of galaxies in the sample. Absorbers with log(\NHI) $> 13$ and within 1 pMpc of the sightline to a QSO are included. The solid histogram represents the distribution of \ion{H}{1} around galaxies, whereas the hatched histogram represents the average absorber density near randomly-chosen redshifts drawn from our galaxy redshift distribution.}
\label{vel_hist_unweight}
\end{figure*}

\begin{figure*}
\centerline{\includegraphics[height=2.5in]{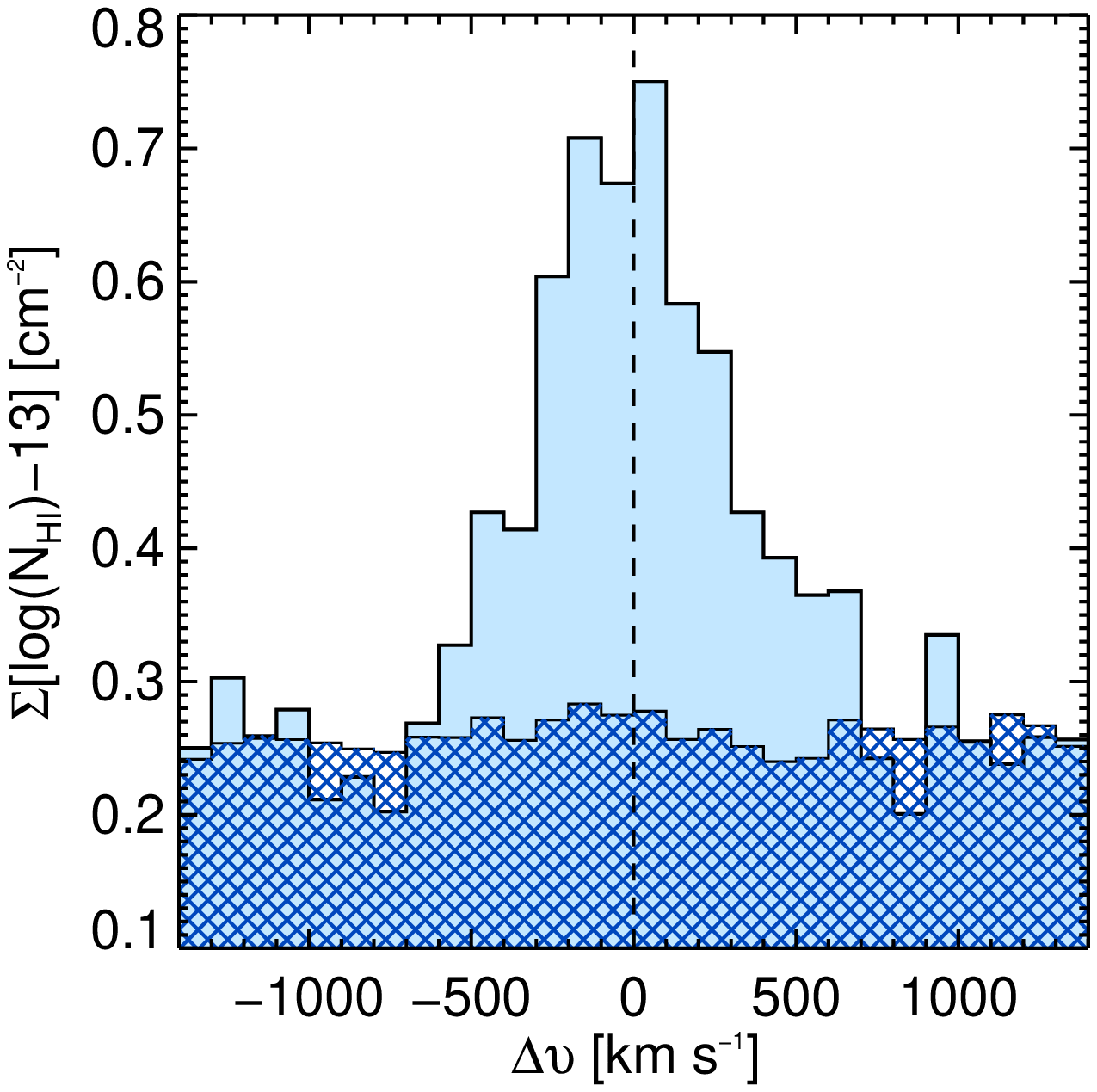}\includegraphics[height=2.5in]{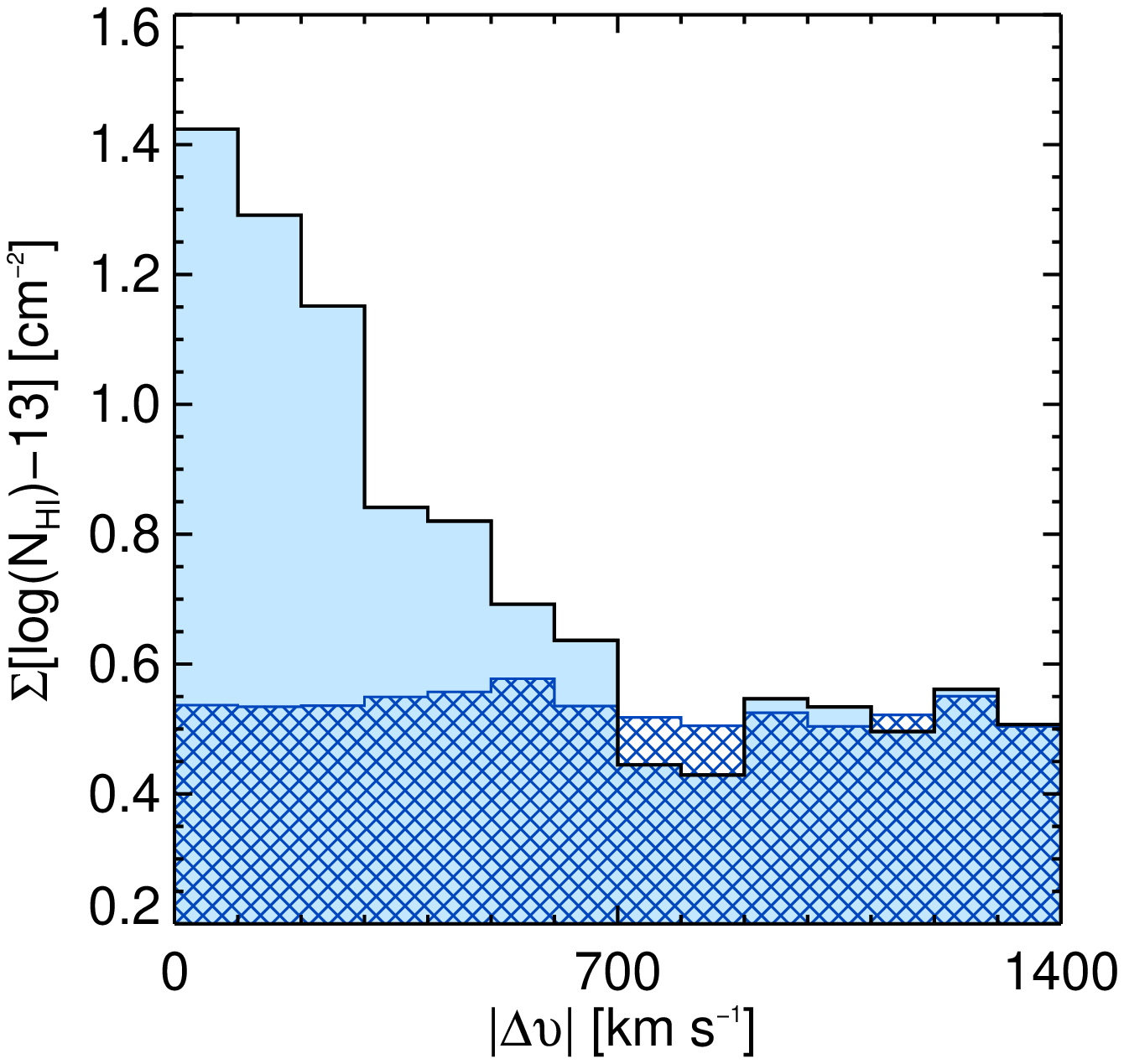}}
\caption{As in Figure \ref{vel_hist_unweight}, where the histogram is \NHI-weighted as described in the text.}
\label{vel_hist}
\end{figure*}

In the following sections, we discuss the statistics of individual \ion{H}{1} absorption systems with respect to the redshifts and transverse positions of galaxies. We do not uniquely ``assign'' each absorber to a specific galaxy or vice versa. Instead we rely on comparisons between the absorption measured close to galaxies with that typical of ``random'' locations\footnote{A ``random'' location, as described later in this section, is in effect a random redshift from our galaxy redshift catalog and an independent random position on the sky from our galaxy position catalog.} in the IGM. This allows us to quantify the significance of any apparent correlation with galaxies. 

In principle, it would be preferable to compare absorbers found close to galaxies with those found in IGM locations known to be far from galaxies; however, because the galaxy sample is spectroscopically incomplete compared to our photometrically selected targets,\footnote{Our photometric sample is also incomplete at the faintest apparent magnitudes \citep[see][]{red08}, and no attempt is made to include galaxies with ${\cal R} > 25.5$.} we do not measure the redshifts of all galaxies in our survey volume. As a consequence, we do not know which locations in 3D space \textit{do not} have a nearby galaxy. Thus, we can only compare locations near to galaxies with random locations in the IGM (irrespective of the positions of galaxies). 

In order to reproduce the absorber distributions for a typical place in the IGM, we compiled a catalog of 15,000 random locations (in both redshift space and on the plane of the sky). The redshifts are drawn from the actual galaxy redshift list and therefore reproduce the typical IGM absorption associated with the redshift ranges covered by our galaxy sample. With each redshift, we also associate a randomly drawn QSO and an impact parameter to the QSO sightline from the real galaxy impact parameter list. We can then study the distribution of absorption systems around these 15,000 random locations and thereby understand how the presence of a galaxy alters the distributions.


Below we consider the distribution of \ion{H}{1} surrounding galaxies: first along the line of sight, then on the plane of the sky, and finally as a function of 3D distance. These measurements are used to determine the relevant velocity and transverse scales of circumgalactic \ion{H}{1}.

\subsection{Velocity-Space Distribution of \ion{H}{1} Near Galaxies}

The properties of \ion{H}{1} gas near galaxies in our sample can be quantified in several ways. First, we consider the line of sight velocity distribution of absorbers relative to the redshifts of galaxies. Shown in Figure \ref{vel_hist_unweight} is the velocity distribution of all absorbers with \NHI $> 10^{13}$ \cm2\ within 1400 \kms\ in redshift and 1 pMpc in projected distance from a galaxy. We define the velocity offset, $\Delta v$, of an absorber, 
\begin{equation}
\Delta v \equiv \frac{\left(z_{\rm abs} -z_{\rm gal }\right) c}{1+z_{\rm gal}}
\end{equation}
where $z_{\rm abs}$ is the absorption system redshift and $z_{\rm gal}$ is the adopted systemic redshift of the galaxy from \S \ref{redshift}. With this definition, absorbers blue-shifted with respect to galaxies have negative \dv.

\begin{figure}
\center
\includegraphics[width=0.36\textwidth]{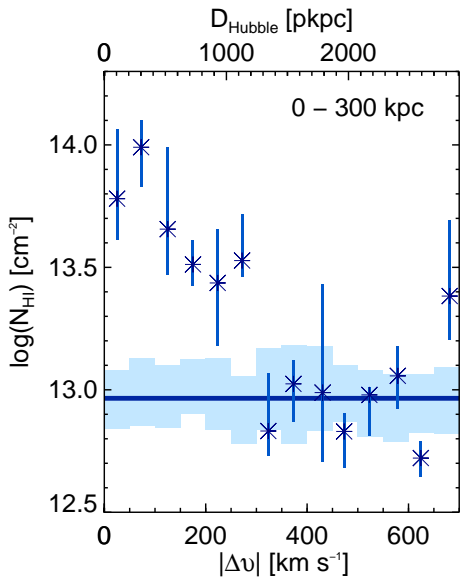}
\caption{The median value of \NHI\ as a function of the velocity offset of absorbers with respect to those galaxies with impact parameters \dtran $<$ 300 pkpc. Asterisks represent the median value of \NHI, dark vertical bars are the 1-$\sigma$ dispersion in the median determined via the bootstrap method. The dark horizontal line is the median value of \NHI\ in the random distribution. The light shaded boxes are the bootstrapped symmetric 1-$\sigma$ dispersion in the median values of the samples drawn from the random distribution. The top x-axis shows the conversion between velocity offset and Hubble distance. See Equation \ref{eq_hubble} and \S \ref{text3D} for further discussion. }
\label{inner_dist_bin300}
\end{figure}

\begin{figure}
\center
\includegraphics[width=0.45\textwidth]{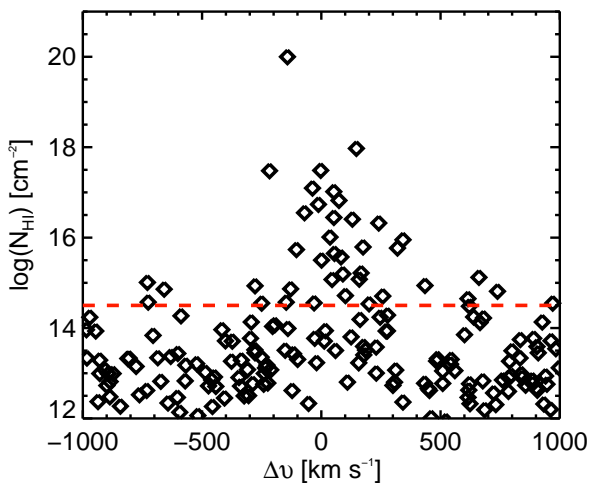}
\caption{\NHI\ as a function of \dv ~for absorbers with \dtran $< 100$ pkpc. The dashed (red) line marks the position of  log(\NHI) $=14.5$. The significance of this column density threshold is discussed in \S \ref{cf}. }
\label{N_vs_V_100}
\end{figure}

\begin{figure*}
\center
{\includegraphics[width=0.5\textwidth]{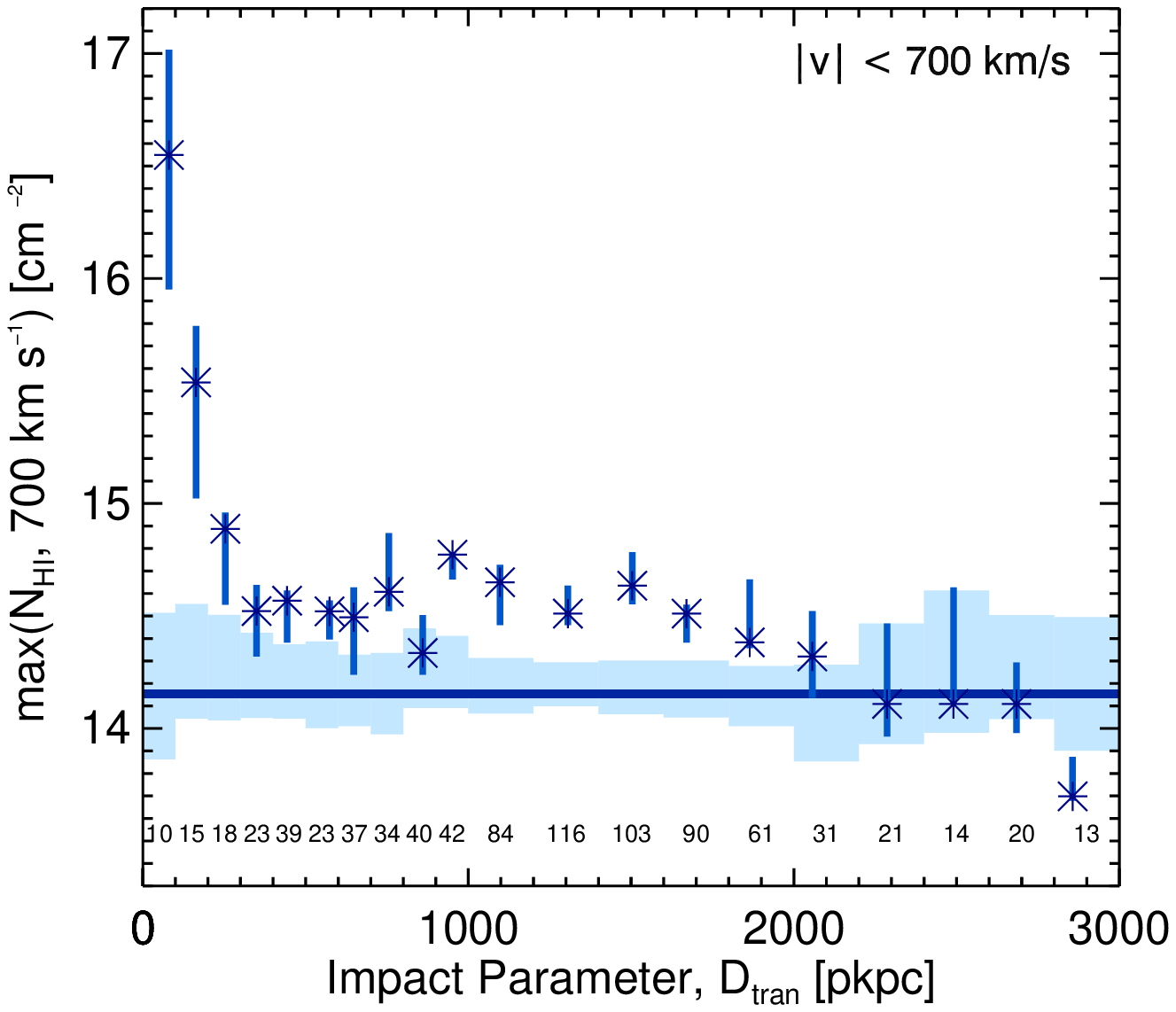}\includegraphics[width=0.5\textwidth]{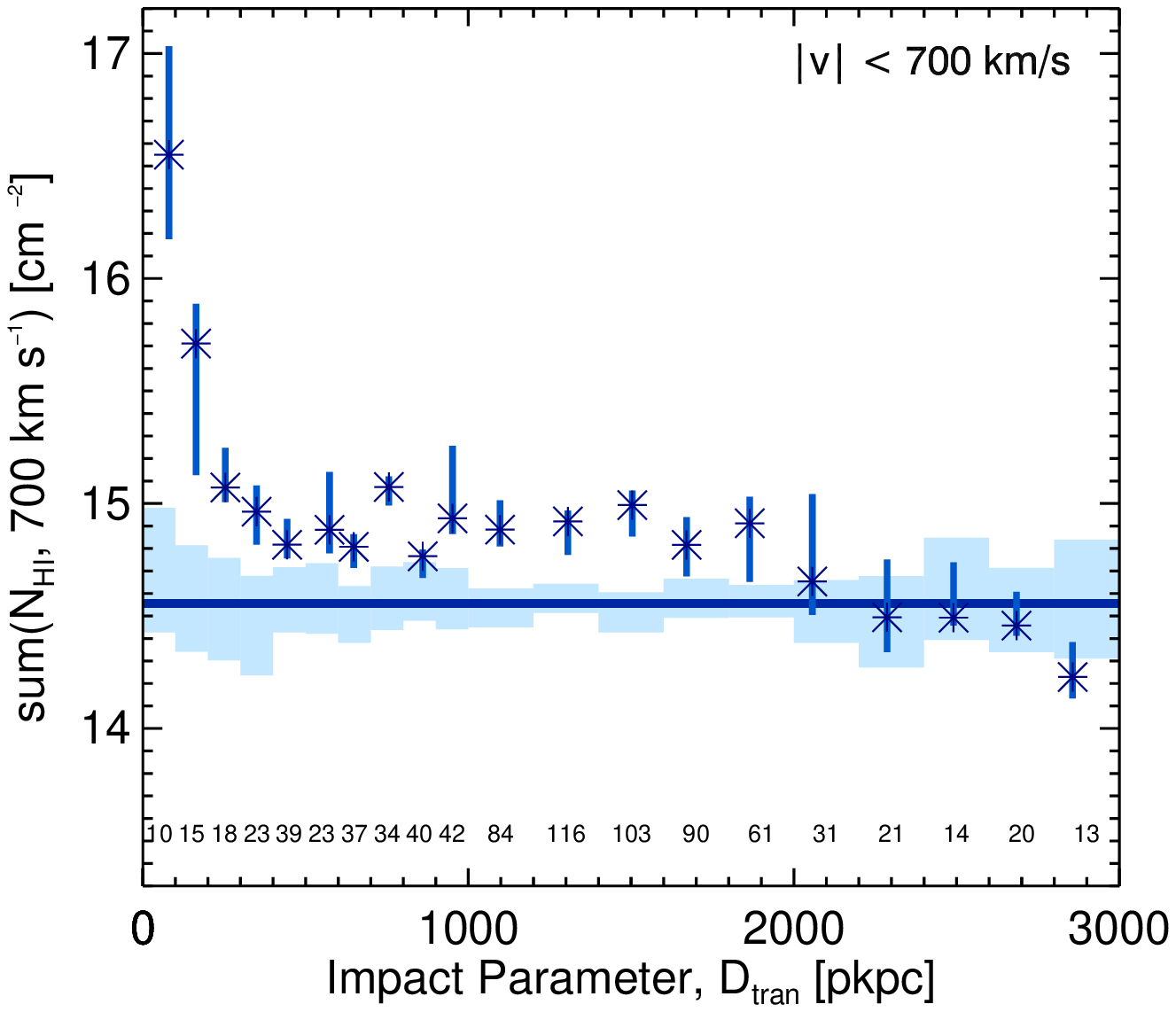}}
\caption{The log column densities of the strongest \NHI\ absorbers as a function of transverse distance.  \textit{On the left} we consider the max(\NHI) statistic, log(\NHI) of the single strongest absorber per galaxy with \absdv $<$ 700 \kms.   \textit{On the right} is the sum(\NHI) statistic, the log of the sum of the \NHI\ of all the absorbers within  \absdv $<$ 700 \kms. Asterisks represent the median value of the considered statistic in a given bin of \dtran. Dark vertical bars are their dispersions. The horizontal position of the asterisks represent the median \dtran\ of the galaxies in that bin. The number of galaxies in each bin is indicated at the bottom of the plot. The dark horizontal line is the median value drawn from the random distribution. The light shaded boxes are the bootstrapped symmetric 1-$\sigma$ dispersion in the median values of the samples drawn from the random distribution. The bin size is 100 pkpc for absorbers with \dtran $<$ 1 pMpc and 200 pkpc for those with \dtran $>$ 1 pMpc. We increase in binning to reduce the shot noise in the bins at \dtran\ $>$ 2 pMpc which have fewer galaxies due to the limited field-of-view of LRIS.}
\label{whisker}
\end{figure*}

We define $P($\dv,\dtran) as the number of absorbers per galaxy at a given \dv\ and within the specified range of impact parameters \dtran. The solid histogram in Figure~\ref{vel_hist_unweight} represents the distribution of \ion{H}{1} around galaxies, whereas the hatched histogram shows the average number of absorbers expected relative to randomly chosen redshifts, as described above. The number of absorbers is clearly higher near galaxies, with an excess peaking near $\Delta v = 0$ (the galaxy systemic redshift) and confined to $\Delta v \simeq \pm 700$ \kms.

Figure \ref{vel_hist} shows a similar pattern; in this case, each absorption system is weighted in proportion to log(\NHI) such that high-column density systems contribute more significantly to the histogram. Each absorber with log(\NHI)$>13$ contributes [log(\NHI) $-13$] to the histogram, which is normalized by the number of galaxies considered. 
Taken together, Figures~\ref{vel_hist_unweight} and \ref{vel_hist} indicate that there is an increase near galaxies of both the number and the column density of \ion{H}{1} absorbers. 
The narrow peak of the \dv\ distribution has an apparent half-width of $\sim$300 \kms, while the full excess extends to $\simeq \pm 700$ \kms\ (most clearly shown in the right-hand panel of Figure \ref{vel_hist}). 

As has been argued by \citet{sha03, ade03, ccs10, rak11}, these velocity distributions are also useful for checking our redshift calibration; their symmetry about \dv $=0$ \kms\ is a sensitive probe of systematic errors in our galaxy systemic redshift calibration, while the width of the distribution provides an upper limit on the random errors.\footnote{The width is an upper limit because it includes both redshift errors {\it and} any peculiar velocity of absorbers relative to galaxies.} A Gaussian fit to Figure \ref{vel_hist_unweight} yields a mean $\Delta v = 18 \pm 26$  with a standard deviation of $308 \pm 30$ \kms.

An alternative method of quantifying the column density dependence in Figure \ref{vel_hist} is to examine the distribution of \NHI\ as a function of \dv. Figure \ref{inner_dist_bin300} shows the median \NHI\ as a function of \dv\ for all absorbers within \dtran$\le 300$ pkpc of a galaxy\footnote{We choose 300 pkpc as our distance cut because the majority of the excess absorption is found within that zone, as described in \S \ref{tran}. }. The asterisks indicate the median value of \NHI\ for each bin in velocity space. The dark vertical lines represent the dispersion in the median computed through the bootstrap method\footnote{To bootstrap the dispersion in the median, we draw (with replacement) sets of data with the same size as the real distribution. For each set the median is computed; 100 such data sets are evaluated, and then we take the 1-$\sigma$ symmetric bounds on the distribution of the medians, i.e. the 16th and 84th percentiles.}. The value of the median column density across all velocity bins for the random distribution is shown as the dark horizontal bar; the light shaded contours are the 1-$\sigma$ bootstrapped dispersions in the median of the random sample, where we consider samples of the same size as those from the real distribution. Note that there is an enhancement by a factor of $\simeq 3-10$ in the median \NHI\ out to \absdv\ $\simeq$ 300 \kms\ relative to galaxy redshifts. However, for \absdv\ $>$ 300 \kms, the measurements are consistent with random places in the IGM.

Shown in Figure \ref{N_vs_V_100} are the individual measurement of \NHI\ as a function of \dv\ for all absorbers within 1000 \kms\ of a galaxy within \dtran$\le 100$ pkpc of a QSO sightline. Notably, the higher-\NHI\ absorbers cluster strongly near the galaxy systemic redshift with a full width of $\sim \pm 300 - 400$ \kms.

In summary, Figures \ref{vel_hist_unweight} $-$ \ref{N_vs_V_100} clearly illustrate an enhancement in both the column density and number of absorption systems near the systemic velocities of galaxies. The most significant enhancement of column density is seen within $\pm$ 300 \kms\ (Figure \ref{inner_dist_bin300}), but there is a higher number of absorbers out to at least $\pm$ 700 \kms\ (Figures  \ref{vel_hist_unweight} and \ref{vel_hist}). 
\label{vel}

\subsection{Transverse Distribution of Absorbers}
\label{transverse}

\label{tran}

In addition to the strong velocity alignment of absorption systems with the systemic redshift of nearby galaxies, there is also a significant increase in the column densities, \NHI, of individual absorbers with decreasing projected (or transverse) distance between the galaxy and the line of sight, \dtran. Here, we suppose that, as an ensemble, these galaxies show similar circumgalactic absorption signatures. Therefore, because galaxies fall at various discrete impact parameters from the QSO line of sight, we can combine the information from each galaxy to make a sparsely sampled map of the absorption as a function of \dtran\ relative to the ensemble galaxy.

We introduce two related statistics designed to trace the change in column density as a function of \dtran. Recalling  that $\pm 700$ \kms\ encompassed the bulk of the ``excess'' absorption (\S \ref{vel}), for each galaxy we define max(\NHI,700\kms) to be the value of log(\NHI) for the strongest absorber with \absdv $<$700~\kms\ of the galaxy systemic redshift. The left panel of Figure \ref{whisker} shows the median value of max(\NHI,700\kms) as a function of impact parameter. A second statistic is the logarithm of the {\it total} \NHI, sum(\NHI), of all absorbers with \absdv$<$700 \kms. The statistics of the median value of sum(\NHI,700\kms) versus \dtran\ is shown in the right-hand panel of Figure \ref{whisker}. Generally, the values of these two statistics are quite similar because the \NHI\ in most velocity windows is dominated by the single highest-\NHI\ absorber.  We consider the sum because it is most easily compared to results of numerical simulations, as it does not require the fitting of Voigt profiles to simulated data and can instead be compared to a simulation ``collapsed'' along the line of sight.

Figure \ref{whisker} clearly demonstrates that both  max(\NHI) and sum(\NHI) increase rapidly as one approaches a galaxy. In the bin corresponding to the smallest impact parameters, \dtran $< 100$ pkpc, the median value of max(\NHI) is more than two orders of magnitude higher than that of a random location. Moving outwards, the median value decreases with increasing \dtran\ to 300 pkpc, at which point the statistic ``plateaus'' and remains significantly higher than the random sample out to 2 pMpc. The plateau value in the galaxy-centric sample is max(\NHI)$\simeq 10^{14.5}$ \cm2, while that of the random distribution is max(\NHI)$\simeq 10^{14.1}$ \cm2. As we will discuss in \S \ref{largescale}, max(\NHI) begins to decline for \dtran$>2$ pMpc. 

\begin{figure}
\center
\includegraphics[width=0.5\textwidth]{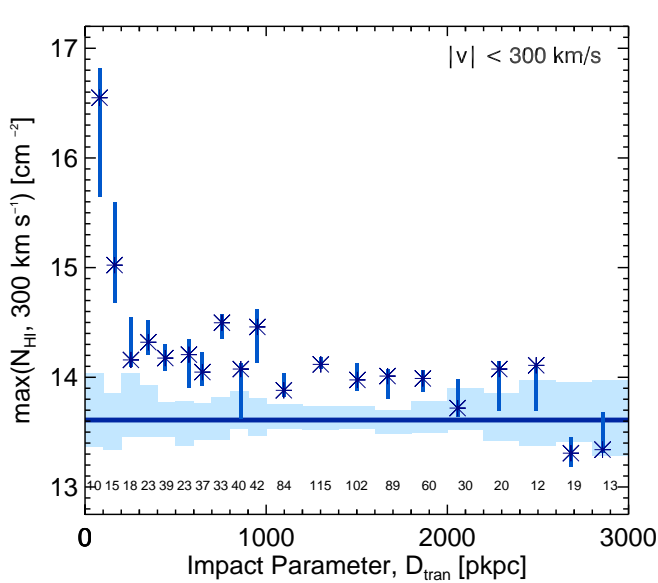}
\caption{Same as Figure \ref{whisker} but for the maximum column density absorber within $\pm$300 \kms. Changing the velocity interval considered with the max(\NHI) statistic has little effect on the observed trends.}
\label{whisker300}
\end{figure}

\begin{figure}
\center
\includegraphics[width=0.45\textwidth]{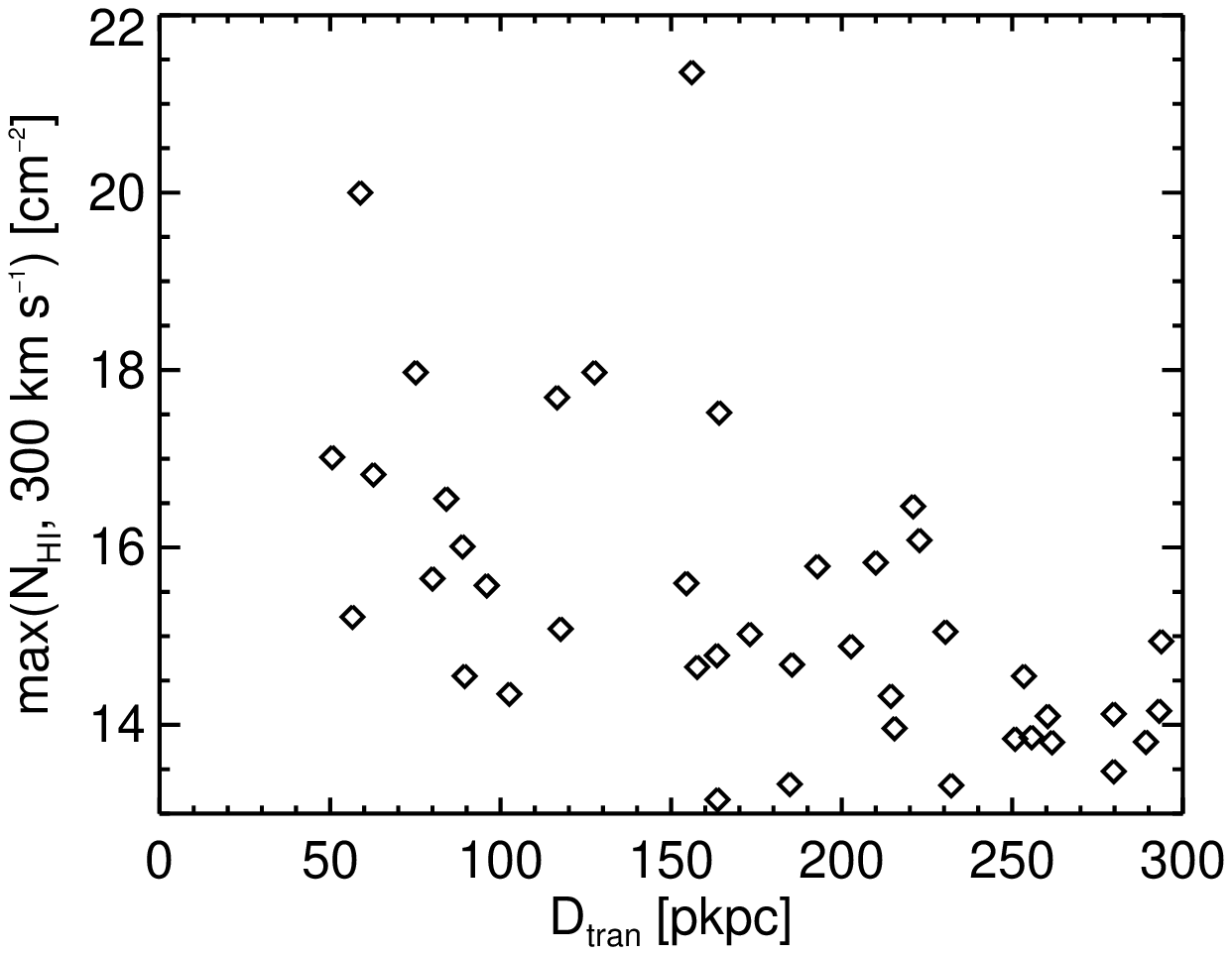}
\caption{The individual measured values of max(\NHI,300\kms) for galaxies with \dtran$<300$ pkpc.}
\label{maxN_scatter}
\end{figure}

\begin{figure}
\center
\includegraphics[width=0.45\textwidth]{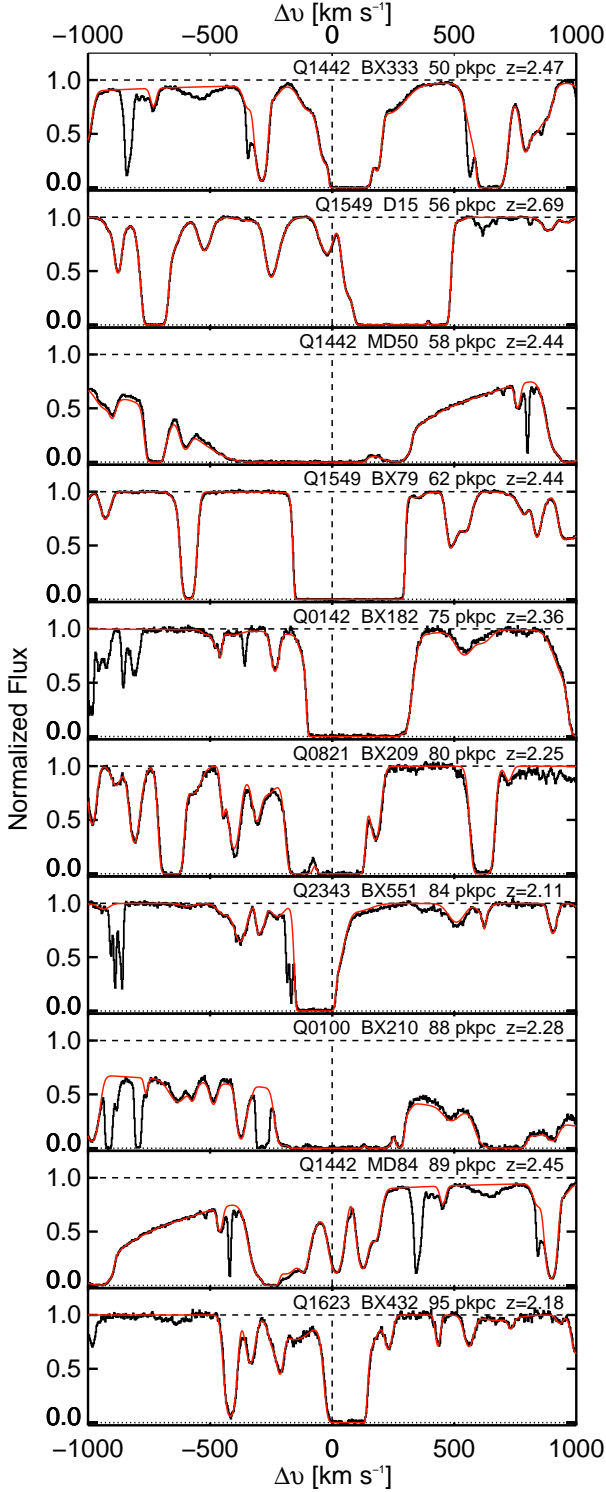}
\caption{Ly$\alpha$ absorption within $\pm$1000 \kms\ of the systemic redshift of the 10 galaxies within \dtran $<$100 pkpc of the line of sight to the QSO. The HIRES data are in black, while the red shows our Voigt profile decomposition of the \ion{H}{1} absorption near the redshift of the galaxy. The continuum and zero level of the spectrum are shown in dashed and dotted lines respectively. The systemic redshift of each galaxy is marked by the vertical dashed line at 0 \kms. Note that the continuum is depressed in some of the spectral regions surrounding Q1442-BX333, Q1442-MD50, Q0100-BX210, and Q1442-MD84 by a DLA or sub-DLA near the galaxy redshift as described in \S \ref{DLA_text}. }
\label{Lya_inner100}
\end{figure}



\begin{figure*}
\center
\includegraphics[width=0.9\textwidth]{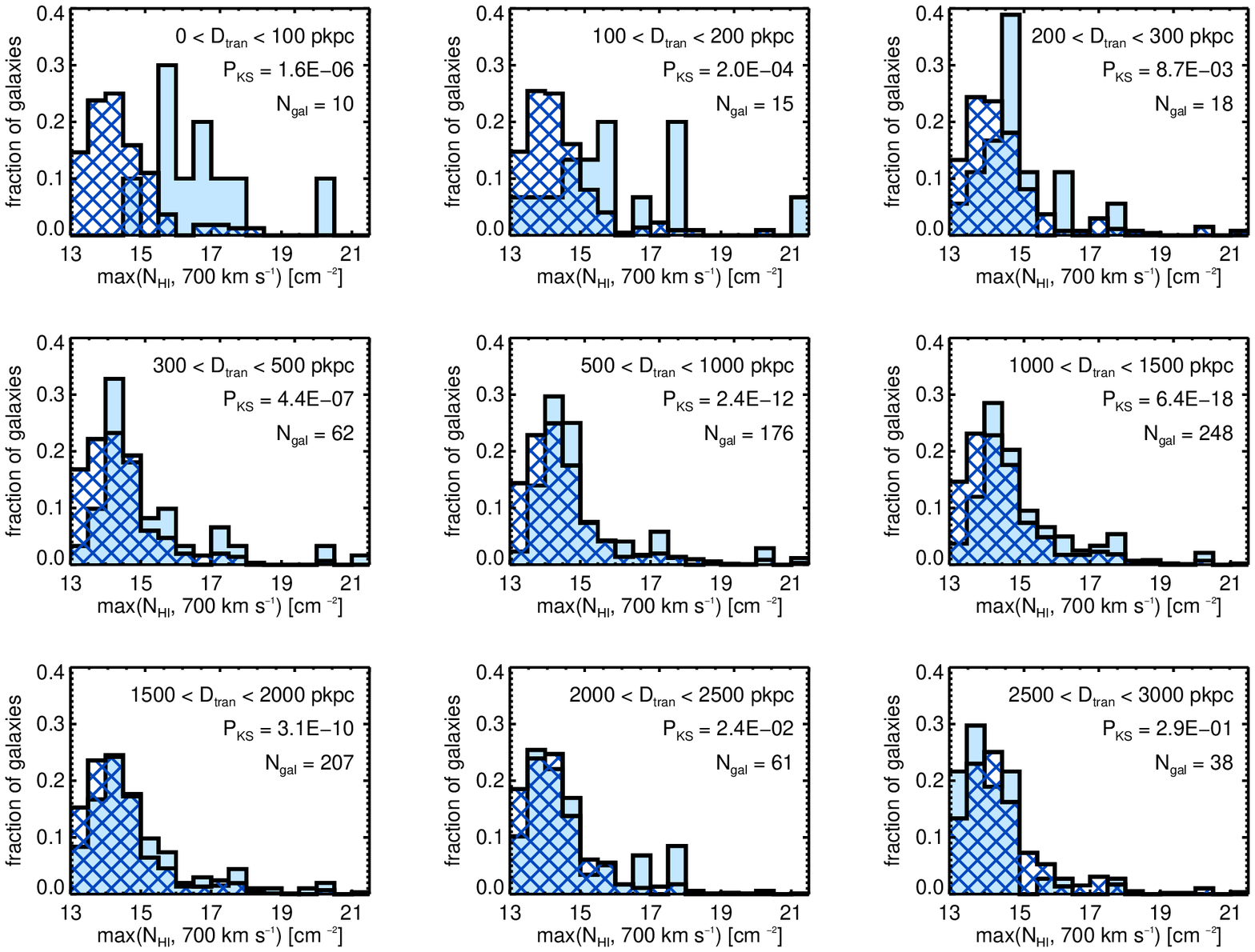}
\caption{The max(\NHI,700\kms) statistic as a function of impact parameter (different panels). The hatched histograms are the values for the random sample, the solid histograms are the values for the real sample.  These histograms quantify the variation from galaxy to galaxy of the max(\NHI) statistic at fixed impact parameter.  P$_{\rm KS}$ is the probability that the two max(\NHI) sets were drawn from the \textit{same} distribution. Notably, the last panel with the highest value of \dtran\ has the least significant departure from the random distribution. }
\label{maxN_hist}
\end{figure*}

A relevant question concerns the dependence of these statistics on the size of the velocity window considered. In Figure~\ref{whisker} we considered the maximum column density absorber within $\pm 700$ \kms\ of the systemic velocity of each galaxy. This corresponds to the full width of the velocity distribution shown in Figure \ref{vel_hist}. However, it is clear from Figures \ref{vel_hist} and \ref{inner_dist_bin300} that the majority of the excess \textit{strength} of absorption falls within $\pm 300$ \kms, especially for those systems with small impact parameters. Figure~\ref{whisker300} shows max(\NHI,300\kms); the trends are similar, though in the more restricted velocity window the peak on small scales is higher relative to random IGM locations--the median value in the first bin is 3 dex higher than the random redshift sample, and the extended floor of absorption is increased to at least .5 dex above the median of the random-redshift distribution. The more significant excess over random of the 300 \kms\ version is primarily due to the exclusion of unrelated absorbers at large velocity separation; however, we note that at \dtran$>1200$ pkpc, the differential Hubble velocity associated with this distance along the line of sight is $>300$ \kms , meaning that for large \dtran\ it may be more appropriate to adopt max(\NHI,700\kms) as the relevant statistic. Regarding the sum statistic for the smaller velocity window (not shown), the value of the plateau and that of the random sample is $\sim0.3$ dex higher for sum(\NHI,300 \kms) than for max(\NHI,300 \kms), similar to the variation in the statistics shown in Figure \ref{whisker}.

As we will argue in \S \ref{cf} and \S \ref{def}, the velocity and spatial scales of 300 \kms\ and 300 kpc capture the most significant excess in both the column density and the number of absorbers near galaxies. In Figure \ref{maxN_scatter} we provide the individual measurements  max(\NHI, 300 \kms) for all galaxies in the sample with \dtran$<300$ pkpc. Note the large intrinsic scatter in max(\NHI), even at fixed impact parameter. For the 10 galaxies with the smallest impact parameters (\dtran$< 100$ pkpc), the relevant portions of the QSO spectra within $\pm$1000 \kms\ of Ly$\alpha$ at $z_{\rm gal}$ are reproduced in Figure \ref{Lya_inner100}.

\subsubsection{The large-scale distribution of \ion{H}{1}}

\label{largescale}

We now consider the larger-scale distribution \ion{H}{1} around galaxies. Unfortunately, the sampling of galaxies with \dtran $>2$ pMpc is comparatively poor in our sample due to the survey geometry of most of the KBSS fields (typically $5.5 \times 7.5$ arcmin on the sky). This scale is imposed by the footprint of LRIS; with each field roughly centered on the bright QSO, the maximum observed impact parameter would be $\simeq 2.15$ pMpc at $\langle z \rangle = 2.3$. However, three out of 15 survey fields (see Table \ref{field}) were imaged with other instruments covering larger angular sizes and thus provide information on larger transverse scales. In Figures \ref{whisker} and \ref{whisker300} we use wider bin sizes for absorbers with \dtran $>$ 1 pMpc in order to consider the large-scale distribution. This reduces the shot noise in the bins with \dtran $>$ 2 pMpc. Figures \ref{whisker} and \ref{whisker300} demonstrate that max(\NHI) remains higher than the global median value in the IGM (dark horizontal line) out to $\sim$2 pMpc, and then begins to decline. For larger \dtran, the data suggest column densities at or below that of random places in the IGM. 

Again, considering the degree of scatter in max(\NHI) at fixed impact parameter, Figure \ref{maxN_hist} shows max(\NHI,700\kms) for various bins in \dtran, as indicated. The top row of panels correspond to \dtran$< 300$ pkpc, while the bottom two rows of panels consider larger impact parameters. Each panel shows the Kolmogorov-Smirnov probability that the two histograms are drawn from the same parent distribution. Notably, only the $2 < $\dtran$ < 3$~pMpc bins have a distribution of max(\NHI) consistent with the random sample.

Thus, we have shown that the column density of \ion{H}{1} peaks sharply at the position of galaxies in the transverse direction, that the width of the peak is $\simeq 300$ pkpc, and that there remains a significant excess of \ion{H}{1} gas to \dtran$\simeq 2$ pMpc. In \S \ref{discussion} we discuss the implications of these results.

\subsection{3D Distribution of \NHI}
\label{text3D}

The 3D distance, \dtd, is computed using the quadrature sum of the physical impact parameter (\dtran) and the line-of-sight distance calculated assuming the velocity differences \dv\ are due entirely to the Hubble flow, 
 
\begin{equation}
D_{\rm Hubble}(\Delta v,z) \equiv \frac{\Delta v}{H(z)}.
\label{eq_hubble}
\end{equation}
The 3D distance is therefore
\begin{equation}
D_{\rm 3D}(\Delta v,z,D_{\rm tran}) \equiv \sqrt{(D_{{\rm {tran}}})^2 + \left(\frac{\Delta v}{H(z)}\right)^2} ~~,
\end{equation}
where $H(z)$ is given by
\begin{equation}
H(z) = H_0 \sqrt{\Omega_{\rm m}(1+z)^3 + \Omega_\Lambda}
\end{equation}
such that $H(z=2.3)$ in our cosmology is
\begin{equation}
H(z=2.3) = 240~ {\rm km~ s}^{-1}~ {\rm Mpc}^{-1}.
\end{equation}
In this formalism, each absorber has a unique \dtd\ with respect to a galaxy in the same field.

\begin{figure}
\center
\includegraphics[width=0.45\textwidth]{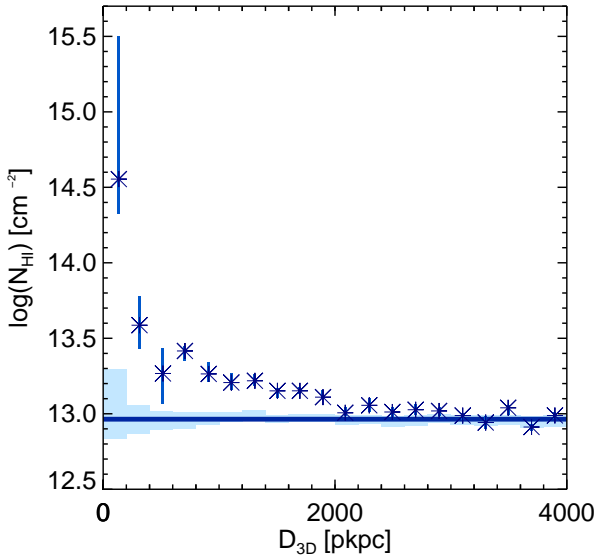}
 \caption{The median column density of all absorption systems within $\pm$1400 \kms\ of a galaxy as a function of the 3D distance between the absorber and the galaxy.  The symbols have the same meaning as those in Figure \ref{whisker}. Note the steeply rising column densities at small \dtd\ and that the median value remains above that of a random location (horizontal bar) out to $\sim$3 pMpc.}
\label{3Dwhisker}
\end{figure}

The 3D distance, due to its strong dependence on \absdv, requires that absorbers have \textit{both} small \dtran\ \textit{and} very small values of \absdv\ in order to populate bins at small values of \dtd. As a result, \dtd\ has the effect of isolating those absorbers \textit{most likely} to be associated with the galaxy without imposing a velocity cut. 

Figure \ref{3Dwhisker} shows that the median \NHI\ stays above that of an average place in the universe out to \dtd\ $\simeq$~3 pMpc.\footnote{Note, this is consistent with the quadratic sum of the extent of the velocity and transverse distance excesses seen in previous sections.} The decline of \NHI\ as a function of \dtd\ is quite smooth, but again strongly peaked at the position of galaxies.  The pixel analysis of the KBSS sample recently completed by \citet{rak11b} studies in detail the 3D distribution of \ion{H}{1} optical depths. There is excellent agreement between the optical depths measured in \citet{rak11b} and the \NHI\ trends shown in Figure \ref{3Dwhisker}.  The smoothness of the decline is caused by the shifting of absorbers with small \dtran\ and modest \absdv\ into bins at larger \dtd. Thus, the signal appearing in the inner 300 pkpc in Figure \ref{whisker} is distributed across a larger number of bins in \dtd\ as a result of the velocity distribution of the absorbers.

The nature of the velocity width of the excess absorption will be discussed at length in \S \ref{mapText}; however, it should be noted that all measurements of \absdv\ rely on the accuracy of the galaxy redshifts and in addition are affected by whatever peculiar velocities are present, whether due to random motion, inflows, or outflows. 

\subsection{Connection to Galaxy-Galaxy Pair Results}

\citet{ccs10} presented a sample of 512 close angular pairs of galaxies with different redshifts (drawn primarily from the same KBSS catalogs used in the present paper), using the spectrum of the background galaxy to probe gas associated with that in the foreground. They were able to measure the strength of absorption from \ion{H}{1} and several metallic species over a range in impact parameter \dtran$=8-125$ pkpc. The principal advantage of this method is that it allows for probes at very small angular separation ($ \theta < 5''$)--obtaining statistical results for galaxies at such small separations from QSOs is difficult due both to the ``glare'' of the QSO and the relative rarity of QSO-galaxy pairs with very small separations (the smallest QSO sightline--galaxy separation is \dtran $= 50$ pkpc or $\theta \approx 6''$) . As such, the galaxy pair results are highly complementary to the QSO sightline study described in this paper. 

Using background {\it galaxies} instead of QSOs results in two important differences between these studies: first, background galaxies have a projected ``beam size'' of $\sim$ 1 kpc at the location of the foreground galaxy, whereas background QSOs have a projected beam of order $\sim 1$ pc.  As a result, the absorption seen against background galaxies measures a combination of the covering fraction of gas on kpc scales and the column density of absorbers. Further, background galaxies are faint and therefore only low-dispersion, low-S/N spectra can be obtained. \citet{ccs10} thus used stacks of background galaxy spectra shifted into the rest frame of the foreground galaxies to quantify the average absorption profile surrounding these galaxies. The lower-resolution spectra do not separate into individual components as would be found in individual QSO spectra. As such, the galaxy pair method allows for the measurement of equivalent widths ($W_0$) only. As discussed by \citet{ccs10} interpretation of $W_0$ is complicated by its sensitivity to the covering fraction and velocity extent of the absorbing material, and its relative {\it insensitivity} to ionic column density. 

\citet{ccs10} measured $W_0$ of \ion{H}{1}, \ion{C}{4}, \ion{C}{2}, \ion{Si}{4}, and \ion{Si}{2} in bins of impact parameter (\dtran). The large $W_0$, particularly at \dtran $\simlt 40$ pkpc, seemed to require high line-of-sight velocity spread in the gas-- likely higher than could be easily explained by gravitationally-induced motions, but easily accounted for by the high velocities observed ``down the barrel'' in the spectra of LBGs, which typically reach  $\Delta v \sim 800$ \kms\ for galaxies with properties similar to those in our sample.  


\citet{ccs10} found that a simple geometric and kinematic model of outflows from LBGs could account simultaneously for the behavior of $W_0$ as a function of \dtran\ as well as the shape of the blue-shifted line profiles of strong interstellar absorption features observed along direct sightlines to the same galaxies. The presence of detected \ion{C}{2} absorption to \dtran\ $\sim 90$ pkpc was cited as evidence for a non-negligible covering fraction of \ion{H}{1} gas having \NHI $> 10^{17}$ \cm2 at such large galactocentric distances.  \citet{ccs10} remarked that 90 kpc is very close to the expected virial radius \rvir\ for LBGs similar to those in the current sample. Our max(\NHI) statistic agrees well with this inference (Figure \ref{whisker}); in fact, we find  a covering fraction $\simeq 30$\% for absorbers with \NHI $>10^{17.2}$ \cm2\ for $r \simlt r_{\rm vir}$-- see \S~\ref{sim}.

In addition, \citet{ccs10} used the same HIRES spectra as this work to show that the $W_0 (\lya)$ measured from the relevant portions of the high resolution spectra, averaged in the same way as the background galaxy spectra, are consistent with an extrapolation to \dtran\ $\simeq 250$ pkpc of the trend seen in the galaxy-galaxy pairs results; a similar conclusion was reached by \citet{rak11b}, also using equivalent width analysis of the QSO spectra, where a smooth trend was noted out to \dtran $> 1$ pMpc.  


\subsection{Comparison to previous studies at $z>2$}

\citet{ade03,ade05} conducted the first systematic studies of high-$z$ galaxies and their surrounding IGM using sightline surveys of the $2 < z < 4$ IGM paired with large LBG surveys designed to probe galaxies in the same volume. \citet{ade03} analyzed the transmitted flux in the Ly$\alpha$ forest of background QSOs, evaluated near the redshifts of survey galaxies. At the time, this was most easily accomplished using $z \sim 3$ LBGs and $z \simeq 3.5$ background QSOs. These authors did not attempt Voigt profile decompositions (as in the present work) but focused on transmitted flux because it made for easier comparisons to theory and because the spectra covered the \lya\ transition only, making measurements of column density or optical depth difficult due to limited dynamic range for any \NHI$\simgt 10^{14.5}$\cm2. Based on 8 QSO sightlines covering $3 \simlt z \simlt 3.6$, \citet{ade03} found that excess \ion{H}{1} absorption (i.e., lower transmitted flux than the average IGM at the same redshift) was present within $\simlt 5$\hmin comoving Mpc (cMpc) of galaxies [1.7 physical Mpc (pMpc) at $\langle z\rangle = 3.3$ using the cosmology adopted in the present work], and an intriguing but not highly significant lack of \ion{H}{1} very close to these galaxies ($<0.5$\hmin cMpc or $<170$ pkpc). \ion{C}{4} absorption was observed to be correlated with galaxy positions to out to 2.4\hmin cMpc (or 800 physical kpc). The cross-correlation of \ion{C}{4} systems with LBGs was found to be similar to the LBG auto-correlation, suggesting that metal enriched IGM and galaxies shared the same volumes of space. The strongest \ion{C}{4} absorbers were so strongly correlated with LBG positions that the authors concluded that they must be causally connected to one another.

\begin{figure*}
\center
\includegraphics[width=\textwidth]{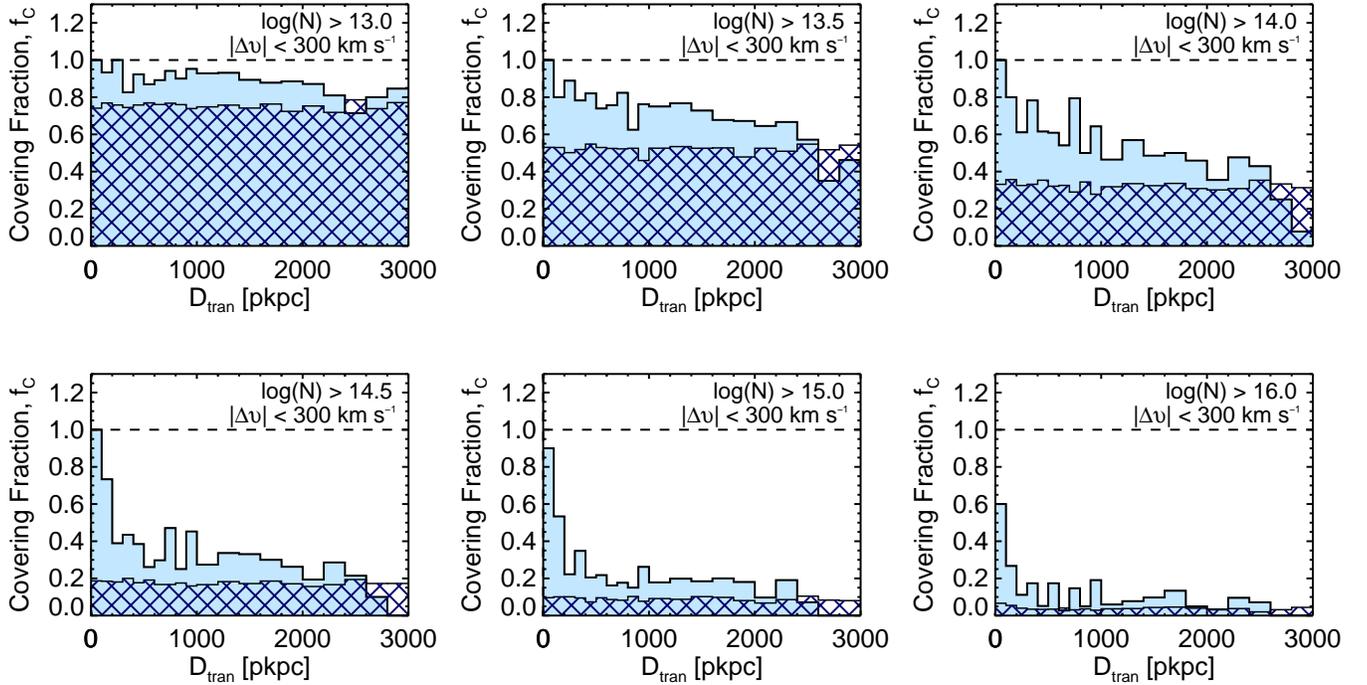}
\caption{The covering fraction, \fc, of absorbers for various \NHI\ thresholds (different panels) as a function of \dtran. The solid histogram represents the fraction of galaxies with an absorber of a given \NHI\ or greater, within $\pm$300 \kms.  The hatched histogram represents \fc\ at random locations in the IGM; the horizontal dashed line marks \fc\ = 1 (100\% covering). The bin size is 100 pkpc for absorbers with \dtran\ $<$ 1 pMpc and 200 pkpc for those with \dtran\ $>$ 1 pMpc. }
\label{covFrac}
\end{figure*}

\citet{ade05} extended similar studies to $1.8 \leq z \leq 3.3$, using a larger number of QSO sightlines and a wider range of QSO spectra for the analysis. Based once again on the transmitted flux statistics, the large-scale excess \NHI\ near galaxies was consistent with the results of \citet{ade03}, but the ``turnover'' of \ion{H}{1} absorption on the smallest scales was not confirmed using the larger sample at somewhat lower redshift. Generally, the transmitted flux was decreasing with galactocentric distance, though it was still the case that $\sim 30$\% of the galaxies with the smallest impact parameters showed a lack of strong absorption, interpreted as evidence that the gas is clumpy. The results for \ion{C}{4} were extended, and it was shown that the correlation with galaxies grows increasingly strong as $N_{\rm CIV}$ increases; the correlation length of $N_{\rm CIV} \simgt 10^{12.5}$ \cm2\ absorption systems was similar to that of the autocorrelation length of the galaxies. The authors attempted to correlate the observed IGM properties with galaxy properties, but no significant correlations were found given the relatively small sample of galaxies at small impact parameters.

There is no overlap in the data sample used in this paper with that of \citet{ade03} which focused on higher-redshift galaxies and QSOs. Three of the fields (Q1623, HS1700, and Q2343) included in our analysis were also included in \citet{ade05} which considered similar redshifts to this work; however, we have increased the S/N of the QSO spectra and also added many galaxy redshifts to our catalogs for these fields since the earlier analysis.  The most surprising result of the \citet{ade03} sample, especially in light of the work presented here, was the reported \textit{deficit} of absorption found  within 0.5 $h^{-1}$ cMpc or 170 pkpc of galaxies.  \citet{ade05}, with a larger data set, found that 1/3 of galaxies had relatively-little \ion{H}{1} absorption (consistent with the pixel statistics of our sample presented in \citealt{rak11b}), but that the majority of galaxies were associated with significant absorption. In our sample, only 1/21 galaxies with \dtran $<$ 170 pkpc has a sum(\NHI, 300 \kms) value less than the median of the random sample. As such we concur with the argument of \citet{ade05} suggesting that the reported deficit was due to the small sample size presented. Further, we note that a measure of \NHI\ shows that the majority of galaxies do have excess \ion{H}{1} absorption in their surroundings.

On larger physical scales, \citet{ade03} and \citet{cri11} considered the large-scale distribution of \ion{H}{1} absorbers at $z\approx3$ while \citet{ade05} studied that at $z\approx2.5$. All found evidence for increased absorption to \dtd\ $\approx$ 5-6 $h^{-1}$ cMpc or $\sim$ 2 pMpc. This scale is roughly consistent with our measurements of the 3D distribution of absorption presented in Figure \ref{3Dwhisker}, as well as with the trends in optical depth vs. \dtd\ presented in \citet{rak11b}.

\section{The Covering Fraction and Incidence of \ion{H}{1}}

\label{cf}

\begin{figure*}
\center
\includegraphics[width=\textwidth]{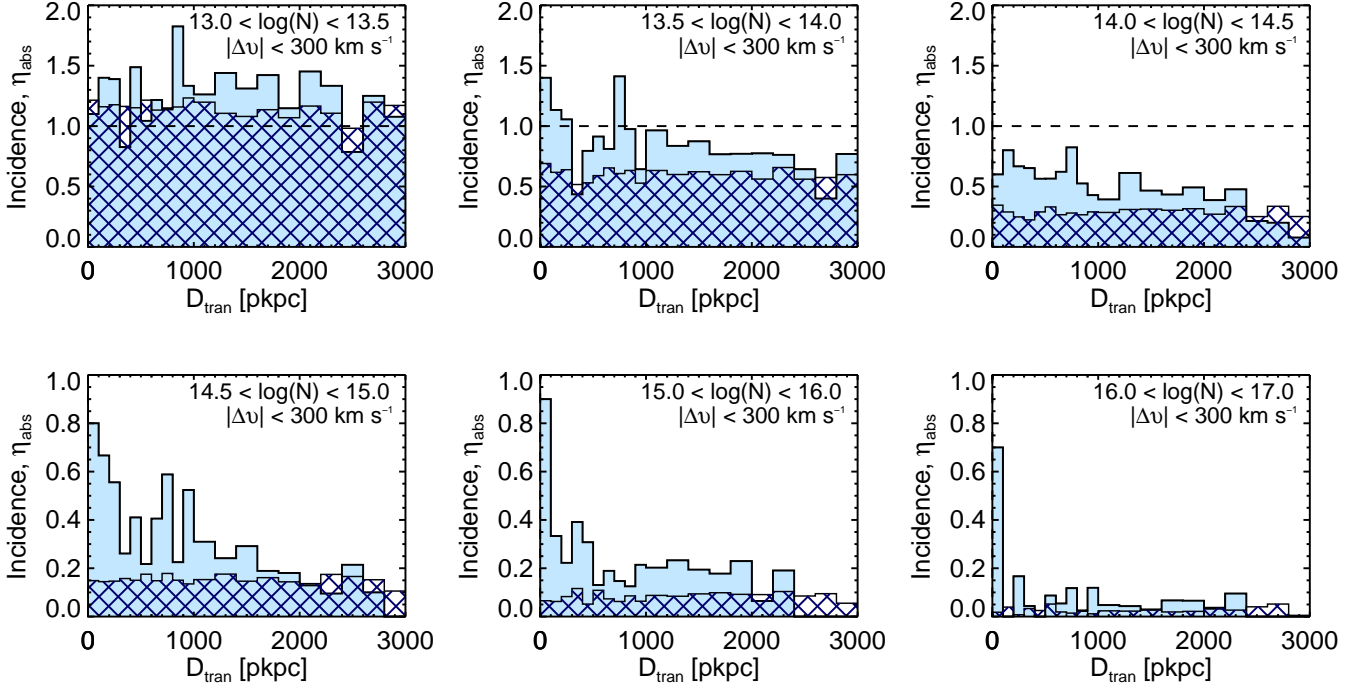}
\caption{The incidence of \ion{H}{1} absorbers, \eabs, as a function of impact parameter, \dtran. The solid histogram represents the mean number of absorbers per galaxy within $\pm$300 \kms\ at the given distance, whereas the hatched histogram represents the average incidence of absorbers near randomly-chosen redshifts. The different panels show various ranges of \NHI. The bin size is 100 pkpc for absorbers with \dtran\ $<$ 1 pMpc and 200 pkpc for those with \dtran\ $>$ 1 pMpc. The incidence of absorbers exceeds the random distribution for \NHI\ $> 10^{13.5}$ \cm2 to \dtran\ $>$ 2 pMpc. Note that only absorbers with log(\NHI) $>14.5$ show strong association with the positions of galaxies. Table \ref{excessProbNHI} gives the data values for \fc\ and P$_{\rm E}$ determined from these distributions.}
\label{vpack}
\end{figure*}

\begin{deluxetable*}{ccccccc}
\tablecaption{Absorber Incidence (\eabs) and Excess Probability (P$_{\rm E}$) for \absdv$< 300$ \kms \tablenotemark{a}}
\tablewidth{0pt}
\tablehead{
\colhead{log(\NHI) range\tablenotemark{b}} & \colhead{\eabs} &  \colhead{P$_{\rm E}$} & 
 \colhead{\eabs} & \colhead{P$_{\rm E}$} & \colhead{\eabs} & \colhead{P$_{\rm E}$} \\
 \colhead{[\cm2]} & \multicolumn{2}{c}{\dtran $ < 0.1$ pMpc} & \multicolumn{2}{c}{\dtran $ < 0.3$ pMpc} & \multicolumn{2}{c}{\dtran $ < 2$ pMpc} }
\startdata
13.0 $--$13.5 &  1.1$\pm$ 0.3 & -0.1$\pm$ 0.3 &  1.3$\pm$ 0.2 &  0.1$\pm$ 0.1 &  1.33$\pm$ 0.04 &  0.17$\pm$ 0.04 \\
13.5 $--$14.0 &  1.4$\pm$ 0.4 &  1.0$\pm$ 0.5 &  1.2$\pm$ 0.2 &  0.8$\pm$ 0.3 &  0.87$\pm$ 0.03 &  0.44$\pm$ 0.06 \\
14.0 $--$14.5 &  0.6$\pm$ 0.2 &  0.7$\pm$ 0.7 &  0.7$\pm$ 0.1 &  1.4$\pm$ 0.4 &  0.53$\pm$ 0.03 &  0.80$\pm$ 0.09 \\
14.5 $--$15.0 &  0.8$\pm$ 0.3 &  4.4$\pm$ 1.9 &  0.7$\pm$ 0.1 &  3.4$\pm$ 0.8 &  0.32$\pm$ 0.02 &  1.03$\pm$ 0.13 \\
15.0 $--$16.0 &  0.9$\pm$ 0.3 & 12.7$\pm$ 4.6 &  0.4$\pm$ 0.1 &  4.9$\pm$ 1.4 &  0.22$\pm$ 0.02 &  1.59$\pm$ 0.20 \\
16.0 $--$17.0 &  0.7$\pm$ 0.3 & 38.2$\pm$14.8 &  0.2$\pm$ 0.1 &  9.7$\pm$ 3.4 &  0.06$\pm$ 0.01 &  1.78$\pm$ 0.40 
\enddata
\tablenotetext{a}{The values in this table refer to Figure \ref{vpack}}
  \tablenotetext{b}{log(\NHI) range: The range of \NHI\ considered; And the incidence, \eabs, and excess probability, P$_{\rm E}$ considered over three ranges of \dtran. The quoted uncertainties in \eabs\ and P$_{\rm E}$ are calculated assuming Poisson statistics.}
         \label{excessProbNHI}
\end{deluxetable*}

As discussed in \S \ref{vel}, the scale of the strongest correlation between \NHI\ and the positions of galaxies is found within 300 \kms\ of $z_{\rm gal}$. Adopting this value as the characteristic velocity scale for circumgalactic gas, we can examine other useful measures of the gas distribution around galaxies. The covering fraction (\fc) and the incidence of absorbers (\eabs) are two ways of quantifying the geometry of the distribution as a function of both impact parameter and \NHI.

First, we define the covering fraction, \fc(\dtran, N$_0$), as the fraction of galaxies in a bin of impact parameter, \dtran, that have an absorber within \absdv $< 300$ \kms\ with \NHI~$>$ N$_0$. This is equivalent to the geometric fraction of the area of an annulus centered on the galaxy that is covered by gas with \NHI$>$N$_0$ and \absdv $<$ 300 \kms. This quantity measures the \textit{variation} within the sample of the decline of \NHI\ as a function of \dtran.

A related quantity is the incidence of absorbers, \eabs, defined to be the number of absorbers \textit{per galaxy} within a given range of \NHI\ and \dtran, and with \absdv $< 300$ \kms. Because this quantity can be greater than unity, it has larger dynamic range and so allows for a more-complete picture of the \textit{average multiplicity} of absorbers at locations close to galaxies. Also, because we consider \eabs\ in differential bins of \NHI, it can be used to measure the degree to which absorbers of a given \NHI\ associate with galaxies.\footnote{To illustrate the difference between \fc\ and \eabs, suppose that all absorbers have the same \NHI. Then for a sample of two galaxies, one with 3 absorbers and the other with none, \fc=0.5 and \eabs = 1.5.}

Figure \ref{covFrac} illustrates the dependence of \fc\ on \dtran\ for various thresholds of \NHI\ (different panels). Within 100 pkpc, \fc $> 0.5$ even for \NHI $> 10^{16}$ \cm2. Also, \textit{every galaxy} has an absorber with log(\NHI) $> 14.5$ within 100 pkpc and \absdv $< 300$ \kms. The distribution within 2.5 pMpc of lower-\NHI\ absorbers is  relatively uniform, with \fc $>$ 0.5. However, at larger threshold \NHI\ (bottom three panels) we see that \fc\ is high ($\simgt 0.5$) only within $\sim$200 pkpc. We will return to our measured values of \fc\ in \S \ref{sim} and \S \ref{lowZ} where we compare our measurements to those made at low-$z$ and also to results from numerical simulations.

Figure \ref{vpack} shows the average incidence of absorbers, \eabs, as a function of distance, \dtran, and ranges of \NHI\ (different panels).  Notably, in all panels (i.e., at all columns densities) the average \eabs\ is higher than random for \dtran $<$ 2 pMpc.  For absorbers with \NHI $> 10^{14.5}$ cm$^{-2}$ (three bottom panels), there is a clear peak at small values of \dtran. No similar peak is present in the distributions of absorbers with \NHI $< 10^{14.5}$ cm$^{-2}$ (top panels). Clearly absorbers with \NHI $>$ 10$^{14.5}$ cm$^{-2}$ are more tightly correlated with the positions of galaxies than absorbers of lower column densities. 

These distributions can also be used to determine the excess probability (over that of a random place in the IGM) of intersecting an absorber of a given column density, within a range of \dtran\ and within some \absdv. The excess probability, P$_{\rm E}$, is defined through comparison to the random distribution:
\begin{equation}
{\rm P_{E}} = \frac{\eta_{\rm abs} - \eta_{\rm abs,random}}{\eta_{\rm abs,random}}
\end{equation}
Table \ref{excessProbNHI} summarizes the measured values of \eabs\ and P$_{\rm E}$ for absorbers binned in \NHI\ for the velocity window \absdv $< 300$ \kms\ and three distance cuts: \dtran $< 100$ pkpc, \dtran $< 300$ pkpc, and \dtran $< 2$ pMpc.

\begin{figure}
\center
\includegraphics[width=0.45\textwidth]{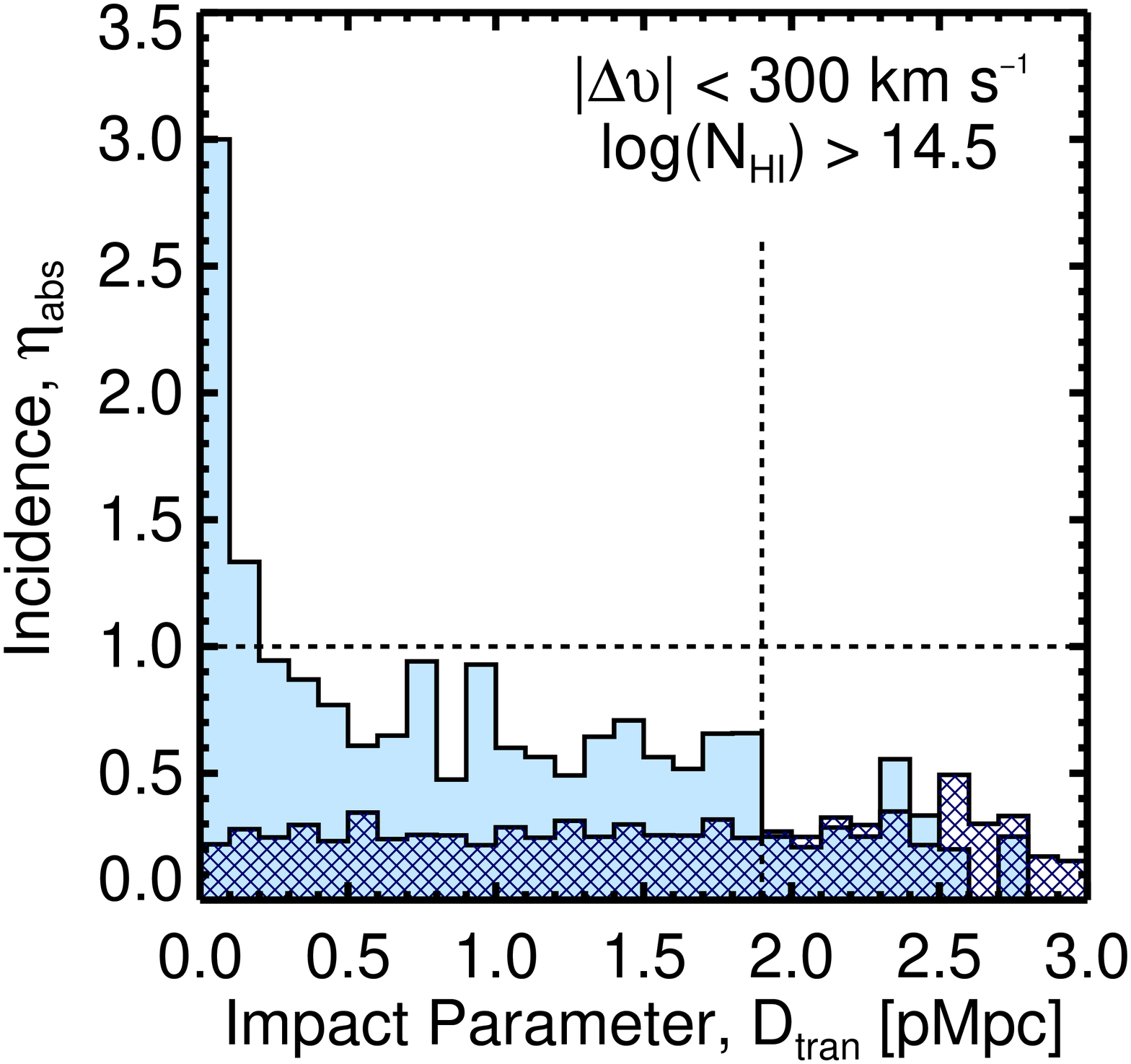}
\caption{The incidence of absorbers, \eabs, with \NHI\ $>$ 10$^{14.5}$ cm$^{-2}$ as a function of impact parameter. The solid histogram represents the distribution of these high-\NHI\ absorbers for \absdv\ $<$ 300 \kms. The hatched histogram represents the average incidence of the same absorbers near randomly-chosen redshifts. The vertical dotted line indicates the distance at which the incidence of the real distribution becomes comparable with that of the random distribution, \dtran\ $\approx$ 2 pMpc. The dotted horizontal line marks \eabs\ $=1$. }
\label{cf145}
\end{figure}


\subsection{Absorbers with \NHI\ $>$ 10$^{14.5}$ cm$^{-2}$ }

\label{clustered}

We have shown above that absorbers with \NHI\  $>$ 10$^{14.5}$ \cm2 appear to trace the positions of galaxies in our sample with high fidelity. The incidence of absorbers with \NHI\ $>$ 10$^{14.5}$ cm$^{-2}$ and \absdv\ $<$ 300 \kms, as shown in Figure \ref{cf145}, nicely encapsulates the ``shape'' of the CGM.  Similar to Figure \ref{whisker} for \dtran\ $<$ 300 pkpc, one sees rising values of \eabs\ as the galactocentric distance is reduced.  From 300 pkpc $<$ \dtran\ $<$ 2 pMpc, \eabs\ reaches a plateau value with \eabs~$\gtrsim$ 0.5 [2 pMpc = 1.4 h$^{-1}$ pMpc $\approx$ 4.6 h$^{-1}$ cMpc (at $z = 2.3$)]. For \dtran\ $>$ 2 pMpc, \eabs\ drops to values consistent with the average IGM.
  
\begin{figure*}
\center
\includegraphics[width=0.8\textwidth]{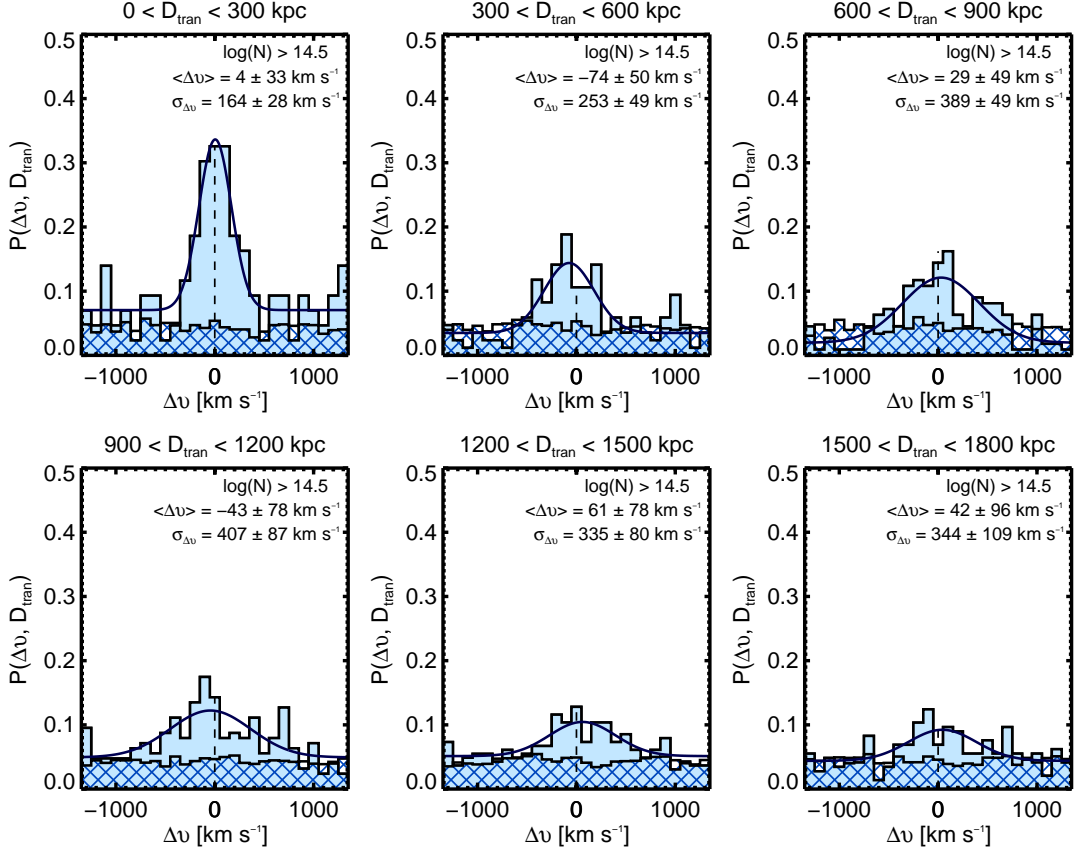}
\caption{The velocity of absorbers with \NHI $> 10^{14.5}$ cm$^{-2}$ in bins of \dtran, as indicated. Note that \NHI\ $> 10^{14.5}$ cm$^{-2}$ absorbers are $>$ 4 times more likely to be found within $\pm$300 \kms\ and 300 kpc of a galaxy than at a random place in the IGM. See Table \ref{excessProb} for \eabs\ and the excess probability for all of the panels.}
\label{exprob_noword}
\end{figure*}

Recalling the quantity P(\dv,\dtran), defined in \S \ref{vel} as the probability, per galaxy, of intersecting an absorber at a given \dv\ and within the specified range of \dtran, here we consider the velocity distribution of absorbers with \NHI $>$10$^{14.5}$ cm$^{-2}$ in bins of \dtran\ as shown in Figure \ref{exprob_noword}.

Fitting a Gaussian to the excess absorbers near galaxies compared to random gives a deviation ($\sigma_{\langle\Delta v\rangle}$) of 187 \kms\ again suggesting that the velocity scale of the excess is \dv $\approx \pm$300 \kms. 

Integrating the distribution in the first panel of Figure \ref{exprob_noword} corresponding to \dtran ~$<$ 300 kpc results in an incidence averaged over $\pm$350 \kms\ ($\sim2\sigma$) and 300 pkpc, \eabs~$= 1.6$. Similarly, comparing the value of \eabs\ measured in the real and random distribution one infers that strong absorbers (\NHI $> 10^{14.5}$) are $>$ 4 times (P$_{\rm E} = 4.1$) more likely to be found within 300 kpc and $\pm 350$ \kms\ of a galaxy in our sample, than at a random place. The parameters of the Gaussian fits in Figure \ref{exprob_noword}, as well as the inferred \eabs\ and P$_{\rm E}$, are listed in Table \ref{excessProb}.

In this section, we have shown that the covering fraction, \fc, of absorbers with log(\NHI) $<$ 14.5 is roughly uniform and $>$ 0.5 out to \dtran\ $\approx$ 2 pMpc, but that those with higher \NHI\ only have \fc\ $>$ 0.5 within \dtran\ $<$ 200 kpc (Figure \ref{covFrac}). Consideration of the incidence, \eabs, indicates that absorbers with  log(\NHI) $>$ 14.5 are more directly related to galaxies (Figure \ref{vpack}). Those absorbers nicely encapsulate the ``shape'' of the CGM (Figure \ref{cf145}) and are $>$ 4 times more likely to be found within \absdv\ $<$ 350 \kms\ and \dtran\ $<$ 300 pkpc of a galaxy in our sample than at a random place in the IGM (Figure \ref{exprob_noword}).

\begin{deluxetable}{cccccc}
\tablecaption{Incidence (\eabs) and Excess Probability for \NHI $> 10^{14.5}$ \cm2 \tablenotemark{a} }
\tablewidth{0pt}
\tablehead{
\colhead{\dtran\tablenotemark{b}} & \colhead{$\langle$\dv$\rangle$} &  \colhead{$\sigma_{\langle\Delta v\rangle}$} & \colhead{\dv\ window} & \colhead{\eabs} & \colhead{P$_{\rm E}$} \\
 \colhead{[pkpc]} & \colhead{[\kms]} & \colhead{[\kms]} & \colhead{[\kms]} & \colhead{ } & \colhead{ } }
\startdata
     ~~~0--300~  &      ~~4$\pm$33  &     164$\pm$~28  &      -350 to  350  &  1.70$\pm$0.02 &  4.1 \\	  
    ~300--600~  &      -74$\pm$50  &     253$\pm$~49  &      -550 to  450  &  1.07$\pm$0.01 &  1.19\\  
    ~600--900~  &      ~29$\pm$49  &     389$\pm$~49  &      -750 to  850  &  1.31$\pm$0.01 &  0.80 \\  
    ~900--1200  &      -43$\pm$78  &     407$\pm$~87  &      -850 to  750  &  1.60$\pm$0.01 &  1.14\\  
    1200--1500  &      ~61$\pm$78  &     335$\pm$~80  &      -650 to  750  &  1.20$\pm$0.01 &  0.76 \\  
    1500--1800  &      ~42$\pm$96  &     344$\pm$109  &      -650 to  750  &  1.06$\pm$0.01 &  0.63  
  \enddata
\tablenotetext{a}{The values in this table refer to Figure \ref{exprob_noword}.}
  \tablenotetext{b}{\dtran:The \dtran\ window considered; $\langle$\dv$\rangle$: the Gaussian velocity centroid and its associated error; $\sigma_{\langle\Delta v\rangle}$: the standard deviation of the Gaussian fit and its associated error; \dv\ window: the velocity window used to compute \eabs ~and P$_{\rm E}$ (comparable to $\pm 2 \sigma$); \eabs: the incidence, P$_{\rm E}$: the excess probability. The quoted uncertainties in \eabs\ are calculated assuming Poisson statistics. The error in P$_{\rm E}$ is dominated by the variation in individual realizations of the random sample: typically the variation in P$_{\rm E}$ is $\sim0.05$ except for the first bin (0 $<$ \dtran $<$ 300 pkpc) where the variation is $\sim0.2$.}
       \label{excessProb}
\end{deluxetable}

\subsection{Covering Fractions: Comparison with Simulations}

Motivated by the desire to predict the observational signatures of cold accretion streams, two recent theoretical papers have considered the covering fraction of absorbers of various \NHI\ surrounding galaxies.  \citet{fau11a} consider the covering fraction of Lyman Limit (LLS, 17.2 $<$ log(\NHI) $<$ 20.3) and Damped Lyman Alpha (DLA,  log(\NHI) $>$ 20.3) absorbers originating within cold streams near two simulated star-forming galaxies at $z=2$ using cosmological zoom-in simulations that do not include galactic winds. Since these authors explicitly \textit{did not} consider galactic winds, their covering fractions are approximately\footnote{These values are only a lower-limit in the case that outflowing winds do not affect the gas distribution within the filamentary cold streams. \citet{van11a} found that the wind prescription in simulations has little effect on the amount of gas that accretes onto halos but does affect the amount which is delivered into the ISM of the galaxy itself. \citet{fau11b} argue that energetic winds can undergo hydrodynamical interactions with cold-streams removing some of the inflowing gas.} a lower limit on the expected values.  They considered a galaxy slightly less massive than those in our sample (M$_{ \rm DM} =  10^{11.5}$ \Msun), as well as one comparable to those in this work  (M$_{ \rm DM} =  10^{12}$ \Msun).


For comparison, we consider the fraction of sightlines at a given \dtran\ for which sum(\NHI,700~\kms) falls within the specified column density range. This is most akin to the results of \citet{fau11a} as they measure the column density of absorbers after projecting their simulated cube onto a 2D plane. Note that here we use differential bins in \NHI (17.2 $<$ log(\NHI) $<$ 20.3), and since we are considering sum(\NHI) rather than max(\NHI), we use \Fc\ to denote the associated covering fraction (to differentiate it from the \fc\ measured earlier in this section).

\begin{deluxetable}{lccc}
\tablecaption{\Fc: Comparison with \citet{fau11a}}
\tablewidth{0pt}
\tablehead{
\colhead{Sample}&\colhead{log(\NHI) [\cm2]} & \colhead{\Fc ($<$1 \rvir)} & \colhead{\Fc ($<$2 \rvir)}} 
\startdata
\multirow{2}{*}{M$_{\rm DM}=  10^{12}$ \Msun}& $17.2 - 20.3$ & 11\% & \nodata \\
&$ > 20.3$ & 4\% & \nodata \\
\\
\multirow{2}{*}{M$_{\rm DM}=  10^{11.5}$ \Msun}& $17.2 - 20.3$ &12\% & 4\%\\
&$ > 20.3$ & 3\% & 1\% \\
\\
\multirow{2}{*}{This work\tablenotemark{a}} &$17.2 - 20.3$ & 30$\pm$14\% & 24$\pm$9\%\\
&$ > 20.3$ & 0$^{+10}_{-~0}$\%\tablenotemark{b} & 4$\pm$4\%
 \enddata
  \tablenotetext{a}{The \Fc\ tabulated here from ``this work''  are the fraction of galaxies with a sum(\NHI) statistic in the column density range. This is subtly different from the \fc\ presented earlier. Note that while we use a velocity cut of \absdv $<$ 700 \kms\ here, the results would be the same if we used a $\pm$300 \kms\ window. Had we considered \fc\ as opposed to \Fc, we would have calculated 20\% for the LLS covering within both 1 and 2 \rvir. The uncertainties quoted for "this work" are 1-$\sigma$ errors calculated assuming \Fc\ follows a binomial distribution.}
  \tablenotetext{b}{We calculate the 1-$\sigma$ upper limit on this non-detection as \\$1 - (1-0.68)^{1/(1+n)}$ where $n=10$ is the number of systems considered.}
       \label{fau_table}
\end{deluxetable}

Our measurements are compared with the \citet{fau11a} simulation results in Table \ref{fau_table}.  Particularly in the case of LLS gas, we find $\sim 3$ times higher \Fc\ within \rvir\ and $\sim 6$ times higher within 2\rvir\ compared with the simulations. However, if the simulations results are treated as lower limits, then clearly they are consistent with the observations. 

\citet{fum11} also considered the covering fraction of \ion{H}{1} surrounding 6 LBG-type galaxies at $z=2-3$ using cosmological zoom-in simulations with galactic winds included (though they are relatively weak with their particular implementation). They consider absorbers within \rvir~and 2\rvir\ for various thresholds in log(\NHI) : $>15.5$, $>17.2 $, $>19.0$, and $ > 20.3$ \cm2. A comparison with the observations is given in Table \ref{fum_table}.  The models of \citet{fum11} clearly under-predict the presence of $15.5 < $log(\NHI) $<17.2$ gas surrounding galaxies. Similar to \citet{fau11a}, they also seem to under predict the radial extent of gas with \NHI $\simlt 10^{19}$ \cm2.

\begin{deluxetable}{lccc}
\tablecaption{Covering Fraction: Comparison with \citet{fum11}}
\tablewidth{0pt}
\tablehead{
\colhead{Sample}&\colhead{log(\NHI)  [\cm2]} & \colhead{\Fc\ ($<$1 \rvir)} & \colhead{\Fc\ ($<$ 2 \rvir) }} 
\startdata

\multirow{4}{*}{Fumagalli et al.}& $>15.5$ &38\% &22\% \\
&$>17.2 $& 16\%&7\% \\
&$>19.0$& 6\%&3\% \\
&$ > 20.3$ & 3\%&1\% \\

\\
\multirow{4}{*}{This work\tablenotemark{a}} &$>15.5 $ & 90$\pm$9\% & 68$\pm$9\% \\
&$>17.2$& 30$\pm$14\% &28$\pm$9\% \\
&$>19.0$& 10$\pm$9\% & 8$\pm$5\% \\
&$ > 20.3$ & 0$^{+10}_{-~0}$\%& 4$\pm$4\% 
 \enddata
  \tablenotetext{a}{Values in ``this work'' are computed as described in Table \ref{fau_table}.  Were we to consider \absdv\ $< 300$ \kms\ rather than \absdv\ $< 700$ \kms, only the \Fc\ for log(\NHI) $>$ 15.5 within 2 \rvir\ would change. The value for 300 \kms\ would be 60$\pm$10\%.}

       \label{fum_table}
\end{deluxetable}

\label{sim}

\subsection{Evolution of the CGM from $z \sim 2.3$ to $z \simlt 1$}

\label{lowZ}

\begin{deluxetable}{lccccc}
\tablecaption{\ion{H}{1} Covering Fraction: Comparison with Low-$z$ Studies\tablenotemark{a}}
\tablewidth{0pt}
\tablehead{
\colhead{Sample}&\colhead{$W_0(Ly\alpha)$} &\colhead{log(\NHI)} & \colhead{\dtran} & \colhead{\absdv} & \colhead{\fc}\\
\colhead{ }&\colhead{m\AA} & \colhead{[\cm2]} & \colhead{pkpc} & \colhead{\kms} & \colhead{}} 
\startdata

$\begin{array}{l l} \textrm {Chen et al.}\\ 0.1 < z < 0.9 \\ \end{array}$ & 350 & $\simgt14$ & <330 & 500 & 72\% \\
\\ \hline \\
$\begin{array}{l l} \textrm {This work} \\ z \sim 2.3\\ \end{array}$ &\nodata & >14 & <330 & 500 & 83$\pm$5\% \\
\\ \hline\hline
\\
\multirow{4}{*}{$\begin{array}{l l} \textrm {Prochaska et al.}\\ 0.005 < z < 0.4\\ \end{array}$ } & 50 &$\simgt$13 & <300 &400 & 96\% \\
& 300 & $\simgt$14 & <300 &400 & 70\% \\
& 50 & $\simgt$13 & <1000 &400 & 70\% \\
& 300 & $\simgt$14 & <1000 & 400 & 38\% \\
\\  \hline \\
\multirow{4}{*}{$\begin{array}{l l} \textrm {This work} \\ z\sim2.3 \\ \end{array} $} & \nodata &>13 & <300 &400 & 100$^{+0}_{-3}$\%\tablenotemark{b} \\
&\nodata & >14 & <300 &400 & 81$\pm$6\% \\
& \nodata & >13 & <1000 &400 & 95$\pm$1\% \\
&\nodata & >14 & <1000 & 400 & 70$\pm$3\% \\
\\ \hline\hline
\\
\multirow{11}{*}{$\begin{array}{l l} \textrm {Wakker \& Savage} \\ z < 0.017 \\ \end{array} $} &50  & $\simgt$13 & <200	&   400	&	      75\%\\
&50  & $\simgt$13 &< 400	  & 400	&	      81\%\\
&50  & $\simgt$13 & <1000	 &  400 	&	      49\%\\
&50  & $\simgt$13 & <2000	  & 400  	 &	      48\%\\
&50  & $\simgt$13 & <3000	  & 400 	&	      39\%\\
&\\
&300   & $\simgt$14 & <200     &     400       &      46\%\\
&300  & $\simgt$14 & <400        &  400          &      44\%\\
&300  & $\simgt$14 & <1000       &  400         &         24\%\\
&300  & $\simgt$14 & <2000       &  400           &     22\%\\
&300  & $\simgt$14 & <3000        & 400            &      20\%\\

\\ \hline \\
\multirow{11}{*}{$\begin{array}{l l} \textrm {This work} \\ z\sim2.3 \\ \end{array} $}&\nodata & $>13$ & $< 200$ &      400 & 100$^{+0}_{- 4}$\%\tablenotemark{c} \\
&\nodata & $>13$ & $< 400$ &      400 &  98$\pm$2\% \\
&\nodata & $>13$ & $<1000$ &      400 &  95$\pm$1\% \\
&\nodata & $>13$ & $<2000$ &      400 &  94$\pm$1\% \\
&\nodata & $>13$ & $<3000$ &      400 &  94$\pm$1\% \\
&\\ 
&\nodata & $>14$ & $< 200$ &      400 &  92$\pm$5\% \\
&\nodata & $>14$ & $< 400$ &      400 &  82$\pm$5\% \\
&\nodata & $>14$ & $<1000$ &      400 &  71$\pm$3\% \\
&\nodata & $>14$ & $<2000$ &      400 &  64$\pm$2\% \\
&\nodata & $>14$ & $<3000$ &      400 &  61$\pm$2\% \\

\\ \hline \hline

 \enddata
  \tablenotetext{a}{The distance binning is cumulative, so \fc\ is the covering fraction of gas with \dtran\ $<$ the distance listed in the table and \absdv\ $<$ the velocity listed in the table. The low-$z$ samples generally do not include Voigt profile fits, but rather measure equivalent widths, $W_0$. The approximate value of \NHI\ associated with each $W_0$ limit is listed.The uncertainties quoted for "this work" are 1-$\sigma$ errors calculated assuming \fc\ follows a binomial distribution.}
   \tablenotetext{b}{We calculate the 1-$\sigma$ lower limit on this 100\% detection as \\$(1-0.68)^{1/(1+n)}$ where $n=43$ is the number of systems considered.}
\tablenotetext{c}{We calculate the 1-$\sigma$ lower limit on this 100\% detection as \\$(1-0.68)^{1/(1+n)}$ where $n=25$ is the number of systems considered.}
       \label{low_z_table}
\end{deluxetable}

There have been numerous studies of the low-$z$ galaxy-IGM connection. In the interest of brevity, below we compare our measured \fc\ with a small number of studies with comparable size and statistical power. 

\begin{figure*}
\center
\includegraphics[width=0.7\textwidth]{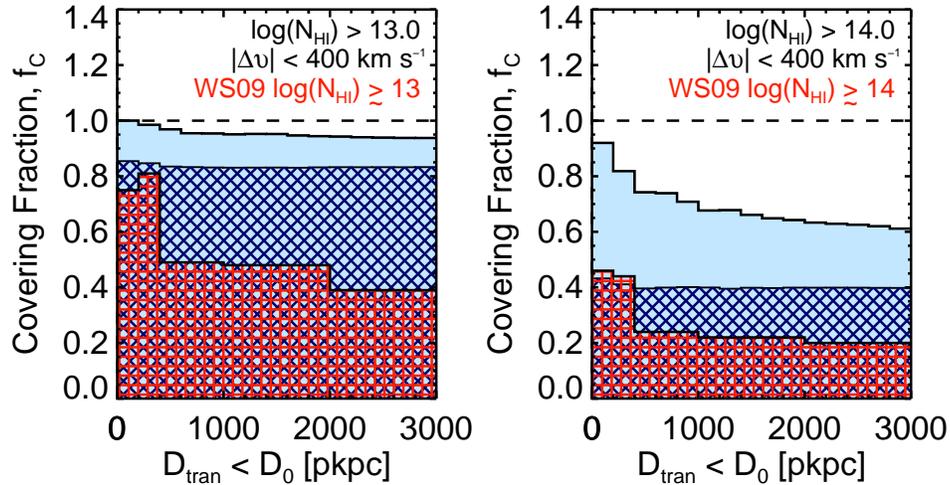}
\caption{The covering fraction of high- and low-redshift absorbers in cumulative bins of \dtran\ (i.e., at each value of \dtran, D$_0$, we consider all absorbers with smaller \dtran, as opposed to those absorbers with D$_{0} <$ \dtran\ $<$ D$_{1}$.) In the (light-blue) solid histogram we plot our high-$z$ data for \fc\ close to galaxies. The (dark-blue) cross-hatched histogram shows the distribution at random places in the high-$z$ IGM. The (red) vertically-hatched histogram shows the distribution of \fc\ of low-$z$ absorbers taken from \citet{wak09}. Here we consider the same cut in \NHI\ at both epochs ($W_0 =$ 50m\AA\ is equivalent to \NHI\ $= 10^{13}$ \cm2 and similarly $W_0 =$ 300m\AA\ is equivalent to \NHI\ $= 10^{14}$ \cm2).}
\label{wakker_same}
\end{figure*}

As discussed by \citet{dav99} and \citet{sch01},  due to the expansion of the universe, the collapse of structure, and the evolution in the intensity of the metagalactic ionizing background, the \NHI\ associated with a fixed overdensity declines with redshift (e.g., an absorber at $z=0$ would have \NHI\ $\sim$ 20 times lower than would be measured at the same overdensity at $z \approx 2.3$). As a consequence, absorbers with the same \NHI\ at different redshifts trace different structures, and have very different incidence rates. For example, there are no regions of the \lya\ forest at $z\sim 2.3$ that have zero \ion{H}{1} absorbers in a velocity window of $\pm700$ \kms\ -- every galaxy in our sample can be associated with (generally) a large number of absorbers. This is not the case at low-$z$, where absorption-line spectra are sparsely populated and generally galaxies would be associated with very few (possibly no) \ion{H}{1} absorbers. Similarly, the continued growth of structure and the rapid decline in the star-formation rate density toward $z \sim 0$ means that a typical low-$z$ galaxy is expected to be in a very different place on its evolutionary sequence compared to a ``typical'' galaxy at $z >2$. Large differences are also expected for gas fractions and baryonic accretion rates at low-$z$ compared to high. Thus, one might reasonably expect circumgalactic gas to differ as well. 

With these caveats in mind, we compare the covering fraction, \fc, of the $\langle z \rangle = 2.3$ sample with results at low-$z$.

\begin{figure*}
\center
\includegraphics[width=0.7\textwidth]{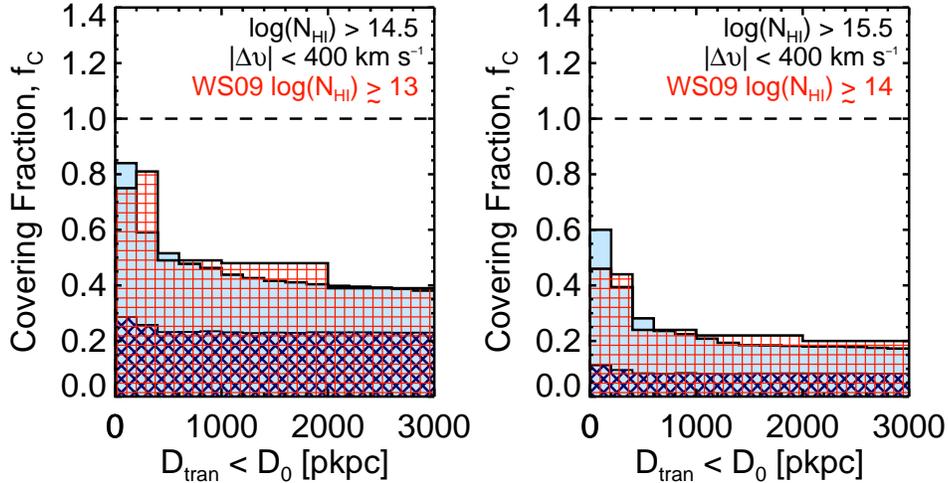}
\caption{Same as Figure \ref{wakker_same}, but here we consider different \NHI\ cuts for the high-$z$ sample. We have found the \NHI\ lower limit for the high-$z$ sample which best reproduces the trend of the low-$z$ \fc\ as a function of \dtran. We can see that the trends of high-$z$ absorbers of 1.5 dex higher \NHI\ are well matched to the distributions of low-$z$ absorbers.}
\label{wakker_diff}
\end{figure*}

\subsubsection{The CGM at $0.1 < z < 0.9$}

\citet{che01} considered  47 galaxies with $ \langle z \rangle = 0.36$ and \dtran$ < 330$ pkpc \footnote{The \dtran\ ranges quoted here have been adjusted from the cosmology considered in \citet{che01} to the cosmology used in this paper at the mean redshift of the sample, $\langle z \rangle = 0.36$.}.  Their galaxy sample covered a wide range of properties, with 68\% having L$_{\rm B} > 0.25$ L$_{\rm B}^*$. The QSO spectra used were generally not of sufficient quality to allow for Voigt profile decompositions and thus the authors measured the equivalent width ($W_0$) of Ly$\alpha$ absorption. \citet{che01} considered absorbers to be associated with galaxies if \absdv $\simlt$ 500 \kms\ \citep{mor06}.

For $W_0 > 350$ m\AA\ (\NHI $\simgt 10^{14}$ \cm2) and \dtran$ < 330$ pkpc, 34/47  galaxies in their sample had detectable Ly$\alpha$ absorption, corresponding to \fc$ = 0.72$. In the $\langle z \rangle = 2.3$ sample there are 48 galaxies with \dtran $<$ 330 pkpc. Using the same velocity window \absdv $<$ 500 \kms, we find 40/48 (83\%) have Ly$\alpha$ absorption with \NHI $\ge 10^{14}$ \cm2, consistent with the measurements from \citet{che01}. For clarity, the results from the low-$z$ studies as well as our measurements are reproduced in Table \ref{low_z_table}.

\subsubsection{The CGM at $0.005 < z < 0.4$}

\citet{pro11}  studied the association of $z < 0.2$ galaxies in 14 fields surrounding background QSOs. Similar to the \citet{che01} study, they generally relied on conversions of published $W_0$ measurements into \NHI\ using an assumed \bd. In total, their galaxy survey included 37 galaxies with \dtran $<$ 300 pkpc and 1200 galaxies in all with $\langle z \rangle = 0.18$.  They use a velocity window of \absdv $<$ 400 \kms. For the 26 galaxies with L $>$ 0.1 L$^*$ within 300 pkpc, 25 had accompanying Ly$\alpha$ absorption with \NHI $> 10^{13}$ \cm2, or \fc$=$96\%.  In our sample, there are 43 galaxies with \dtran $<$ 300 pkpc, all of which have a \NHI $\ge 10^{13}$ \cm2 absorber within \absdv $<$ 400 \kms\ (\fc=100\%).

These values appear consistent; however, the covering fraction of  \NHI $\ge 10^{13}$ \cm2 gas is a very poor measure of the extent of the CGM at high redshift. For example, if we consider the same velocity window (\absdv $<$ 400 \kms) and the same \NHI\ threshold (\NHI $> 10^{13}$ \cm2), we find \fc $=$ 95\% within 1 pMpc and  \fc $=$ 94\% within 2 pMpc; i.e., there is no appreciable change in the \fc\ of gas of this \NHI\ with increasing distance. \citet{pro11} find \fc=70\% for \dtran\ $<$ 1 pMpc in their sample.

Measurements with a higher \NHI\ threshold have larger dynamic range; for \NHI $>10^{14}$ \cm2, \citet{pro11} measure \fc$=$70\% for \dtran $<$ 300 pkpc, and \fc$=$38\% for 1 pMpc. For our $\langle z \rangle \sim 2.3$ sample, we measure \fc$=81$\% for \dtran$<$300 pkpc and \fc$=$70\% for \dtran$<$1 pMpc.

\subsubsection{The CGM in the Local Universe: $cz < 6000$ \kms }

\citet{wak09} (hereafter, WS09) studied the gaseous distribution around local galaxies ($cz < 6000$ \kms) along 76 lines of sight. The galaxy catalog considered included $\sim20,000$ local galaxies. The absorber catalog included 115 intergalactic Ly$\alpha$ absorbers (i.e., not from the Galaxy) over the same redshift range. They calculate \fc\ in cumulative distance bins (e.g. \dtran $<$ 200 pkpc). 

For comparison to the WS09 sample, we use a velocity window \absdv $<$ 400 \kms\ with respect to galaxy redshifts in our high redshift sample. From WS09 (Table 10) we use the values of \fc ~measured for the galaxy sample with L $>$ 0.25 L$^*$. \footnote{Our spectroscopic sample includes galaxies to $\mathscr{R}=25.5$. This is equivalent to  L $>$ 0.25 L$^*$ at $z \approx 2.3$.} They consider two classes of absorbers, those with $W_0 >$ 50m\AA\ (\NHI $> 10^{13}$ \cm2) and $W_0 >$ 300 m\AA\ (\NHI $\simgt 10^{14}$ \cm2). We make a direct comparison to our sample \textit{with the same \NHI\ threshold} in Figure \ref{wakker_same} and Table \ref{low_z_table}. Our measurements are shown in the solid histogram, the random high-$z$ sample is plotted in the (dark-blue) cross-hatched histogram, and the WS09 sample is represented by the (red) vertically-hatched histogram. One can clearly see that at all \dtran\ and for both thresholds in \NHI, the $z\sim0$ sample shows significantly less covering than the high-$z$ sample. 

As mentioned above, absorbers of fixed \NHI\ are theoretically expected to trace higher-overdensity gas at low redshift than at high redshift. By considering various \NHI\ thresholds in our own data, we found that \fc\ of the low-$z$ $W_0 >$ 50m\AA (\NHI $> 10^{13}$ \cm2) sample was most comparable to high-$z$  absorbers with \NHI $> 10^ {14.5}$ \cm2; similarly, the WS09 $W_0 >$ 300m\AA (\NHI $\simgt 10^{14}$ \cm2) can be matched to our high-$z$ absorbers with \NHI $> 10^ {15.5}$ \cm2. Note that we chose the \NHI\ threshold which best reproduced \fc\ measured on $\sim$Mpc scales. 

These two comparisons, shown in Figure \ref{wakker_diff}, exhibit remarkably similar patterns in the covering fraction of \NHI\ as a function of \dtran. It therefore appears that low-$z$ absorbers with log(\NHI) = N$_0$ \textit{on average} trace the same physical regions around galaxies as gas with log(\NHI) $ = $ N$_0 + 1.5$ at $\langle z \rangle = 2.3$. While we do not directly measure universal overdensity, these results lend observational support to the predictions by \citet{dav99} and \citet{sch01} that gas at the same universal overdensity at low- and high-$z$ would have lower \NHI\ at low-$z$. These measurements also suggest that the physical size of the CGM around a typical galaxy is 300 pkpc at both $z \sim 0$ and $\langle z \rangle = 2.3$. 

\section{Mapping the Circumgalactic Medium}

\label{mapText}

\begin{figure*}
\center
\includegraphics[width=\textwidth]{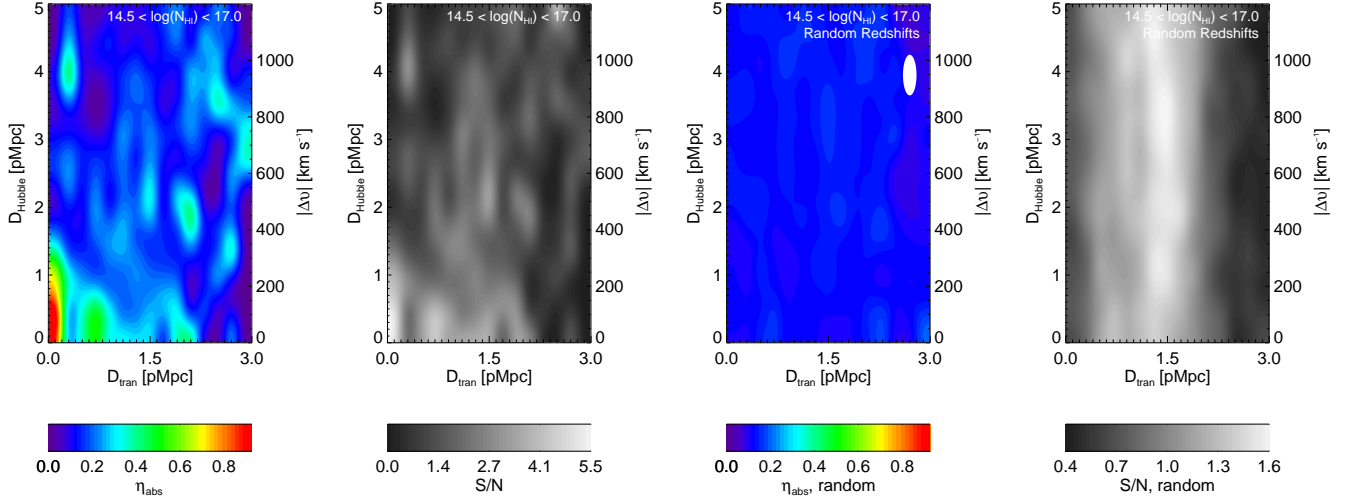}
\caption{The incidence of $14.5 < $ log(\NHI) $< 17$ absorbers. \textit{Far Left}: Map of the \eabs\ with respect to the positions of galaxies (at the origin).  \textit{Second from Right}: The distribution at random locations in the IGM. The greyscale panels show the S/N ratios of the maps calculated as discussed in \S \ref{mapText}. The ``bin'' size for this map is 200 x 200 pkpc and the smoothing scale is 200 kpc in \dtran\ and 140 \kms\ (600 pkpc) along the line of sight. The FWHM of the gaussian smoothing beam is represented by the ellipse in the top right-hand corner of the 3rd panel from the left. Note that there is a strong peak in the incidence of high column density absorbers at close galacto-centric distances. The color bars in the \eabs\ maps have been re-normalized by $\chi_v$ = 6.3 in order to match the values of  \eabs ~summed over \absdv $<$ 300 \kms. The vertical structure in the S/N maps is caused by the variation in the number of galaxies in each bin of \dtran. }
\label{asymMap14.5_17_200}
\end{figure*}

\begin{figure*}
\center
\includegraphics[width=\textwidth]{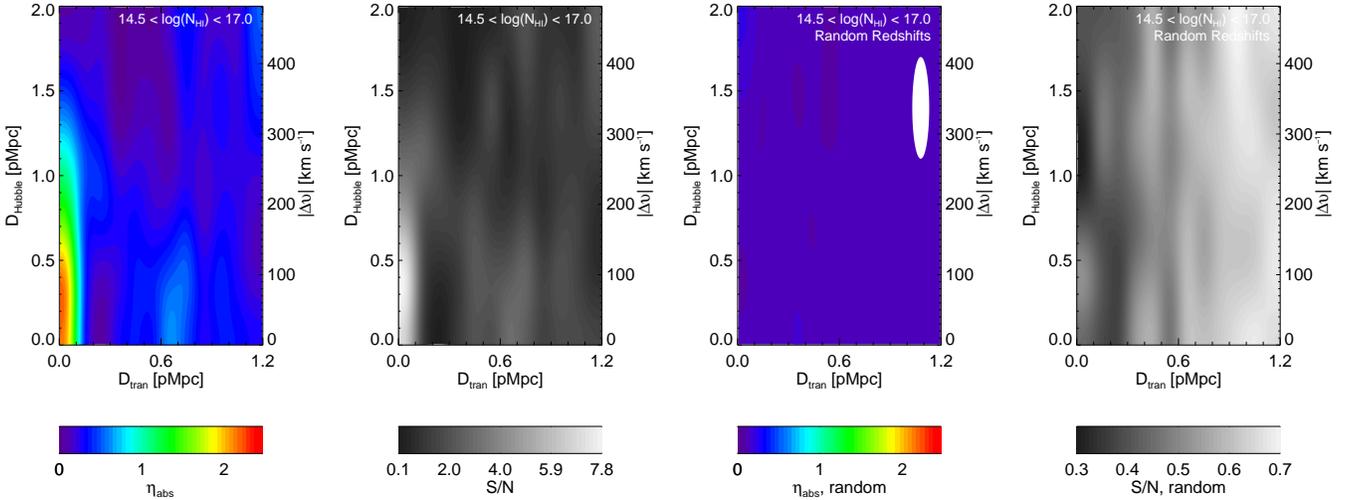}
\caption{Same as Figure \ref{asymMap14.5_17_200} but zoomed in to show the small scale distribution. The ``bin'' size for this map is 100 x 100 pkpc. The smoothing scale is 100 kpc in impact parameter and 140 \kms\ (600 pkpc) along the line of sight. The color bars for the incidence maps have been re-normalized $\chi_v$ = 12.6 in order to match the \eabs ~summed over \absdv $<$ 300 \kms.}
\label{asymMap14.5_17_zoom}
\end{figure*}

\begin{figure*}
\center
\includegraphics[width=\textwidth]{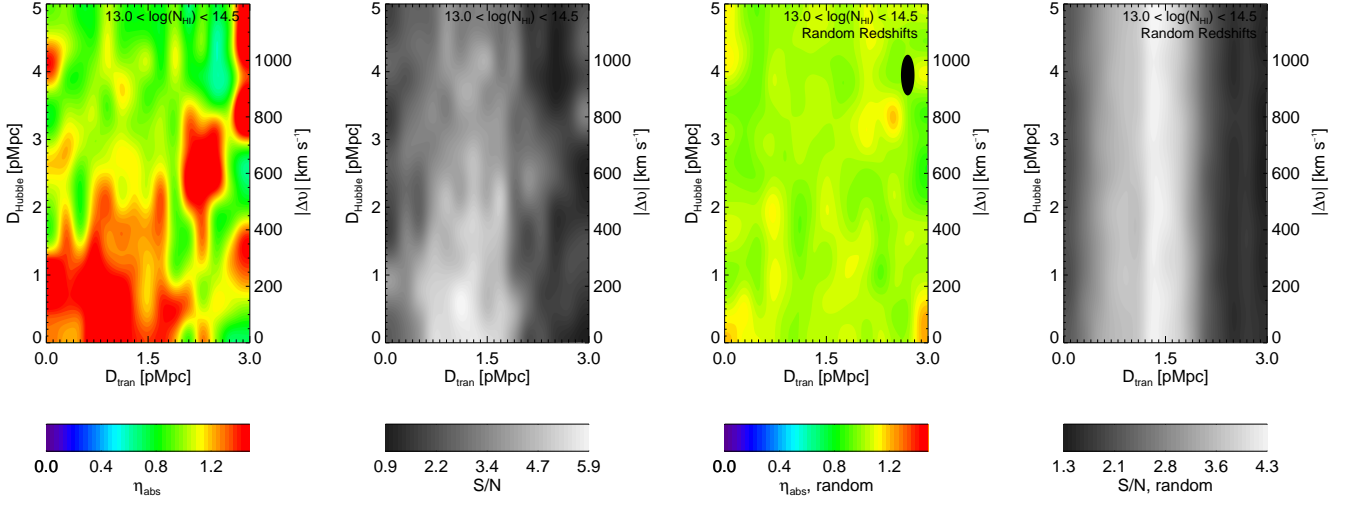}
\caption{Same as Figure \ref{asymMap14.5_17_200} but for low column density absorbers with $13 < $ log(\NHI) $< 14.5$. Note, that the incidence of these absorption systems is higher in the large-scale regions surrounding galaxies, similar to that of high column density systems. However, unlike the high column density absorbers, there is no strong peak in the incidence rate of low-\NHI ~systems at small galacto-centric distances. The color bars for the incidence maps have been re-normalized $\chi_v$ = 6.3 in order to match the \eabs ~summed over \absdv $<$ 300 \kms.}
\label{asymMap13_14.5_200}
\end{figure*}

\begin{figure*}
\center
\includegraphics[width=\textwidth]{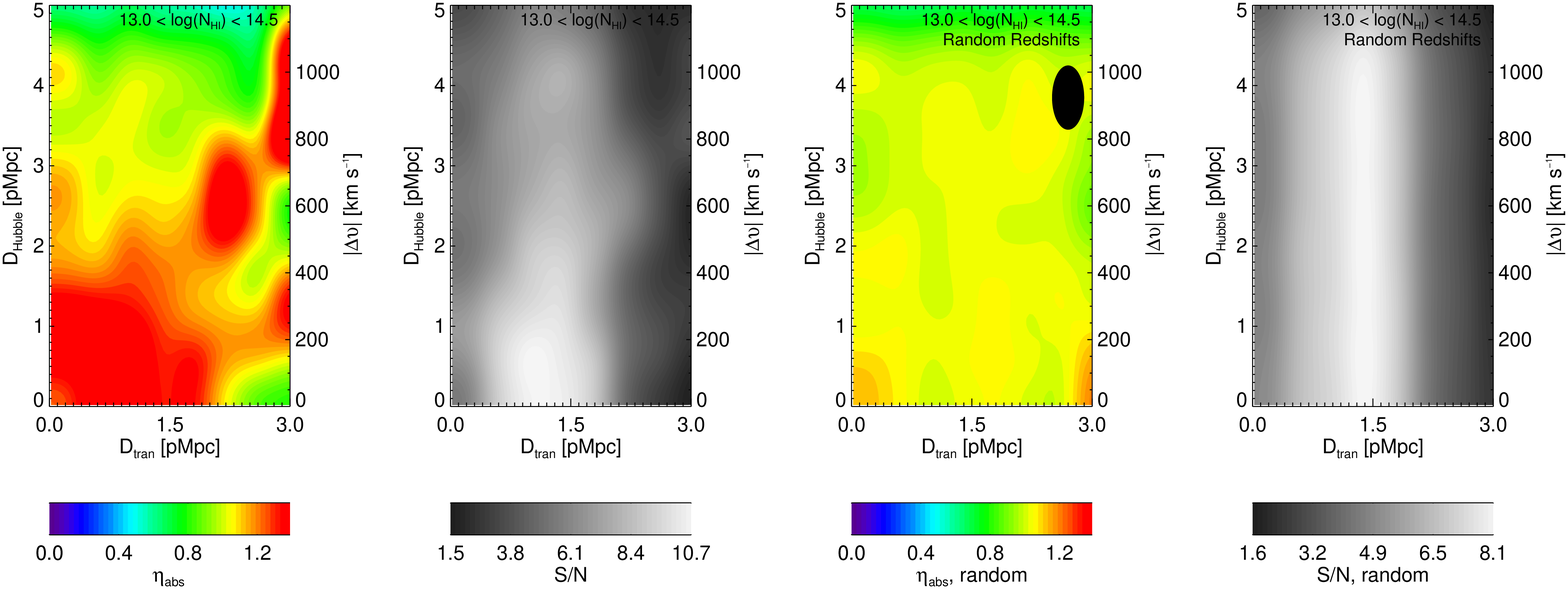}
\caption{Same as Figure \ref{asymMap13_14.5_200} but with 400 x 400 pkpc binning. Note that the line-of-sight distribution is compressed compared to the transverse distribution. The color bars for the incidence maps have been re-normalized $\chi_v$ = 3.15 in order to match the \eabs ~summed over \absdv $<$ 300 \kms.}
\label{asymMap13_14.5_400}
\end{figure*}

Another way to visualize the distribution of neutral hydrogen surrounding star-forming galaxies is to ``map'' the distribution along the transverse direction and the line of sight at the same time. As in \S \ref{text3D}, we use \absdv\ to compute a line-of-sight distance assuming it is entirely due to the Hubble flow. In Figures~\ref{asymMap14.5_17_200} - \ref{asymMap13_14.5_400}, the vertical axes represent the \dhub\ velocity scale line-of-sight distribution of gas. The horizontal axes correspond to the physical impact parameter between the galaxy and the QSO line of sight, \dtran. Colors encode the incidence, \eabs, of the gas in a given pixel in the map and the black and white shading represents the signal-to-noise ratio (S/N) with which \eabs\ is detected. 

The S/N is determined using Poisson statistics. For a given bin in \dtran\ and \dhub, we count the total number of absorption lines found in the random distribution, $n_{\rm abs,ran}$, and the number of random locations considered at that impact parameter, $n_{\rm gal,ran}$. We also count the number of real galaxies at that impact parameter, $n_{\rm gal}$. The ``noise'' level of the map is then taken to be:
\begin{equation}
\sigma_{\rm sig} = \frac{\sqrt{n_{\rm gal}\left(n_{\rm abs,ran} / n_{\rm gal,ran}\right)}}{n_{\rm gal}},
\end{equation}
where the quantity inside the square root is the number of absorption systems expected (based on the incidence for the random sample) given the number of galaxies in the real sample at a given \dtran. The square root of this quantity  represents the Poisson uncertainty in the number of absorbers that would be detected; the division by $n_{\rm gal}$ results in an expression for the error in \eabs\ \textit{per galaxy.} This value is akin to the shaded error bars surrounding the median value of the random sample in Figure \ref{whisker}. The signal-to-noise ratio is then taken to be:
\begin{equation}
\textrm{S/N} = \frac{\eta_{\rm abs}}{\sigma_{\rm sig} }
\end{equation}

The maps are generated using three different bin sizes. We consider the intermediate bin size (200 x 200 pkpc) for two cuts in \NHI ~(Figures \ref{asymMap14.5_17_200} and \ref{asymMap13_14.5_200}) which cover impact parameters out to 3 pMpc.  
For high-\NHI\ absorbers which exhibit a peak at small scales, we also consider maps with 100 x 100 pkpc bins (Figure \ref{asymMap14.5_17_zoom}). For low-\NHI\ absorbers, we provide a map with 400 x 400 pkpc bins to emphasize the large scale distribution (Figure \ref{asymMap13_14.5_400}). After the number of absorbers per bin is measured, the values are placed onto a 10 times finer grid and smoothed by a gaussian kernel with a full width half maximum equal to the bin size in impact parameter (x-axis) and a velocity scale of 140 \kms\ or 600 pkpc in Hubble distance (y-axis). Binning and smoothing is required to make maps of this type as the values of impact parameter and velocity are discrete. The scale of the smoothing, however, is motivated by the data. The smoothing in \dtran\ is selected to obtain a large enough sampling of the inner and outer bins such that the noise in the random distribution is reduced. The velocity scale smoothing corresponds roughly to the amplitude of the redshift uncertainties (see \S \ref{redshift}).

In Figures \ref{asymMap14.5_17_200} -- \ref{asymMap13_14.5_400}, the normalization of the color bar is re-scaled from the raw \eabs ~per 100 x 100, 200 x 200, or 400 x 400 pkpc bin by the multiplicative factor $\chi_v$, where:
\begin{equation}
\chi_v = \frac{D_{\rm Hubble}\left(300~{\rm km~s}^{-1}\right)}{D_{\rm Hubble}\left({\rm bin}\right)} \approx \frac{1.26~{\rm pMpc}}{D_{\rm Hubble}\left({\rm bin}\right)}
\end{equation}
where \dhub(bin) is either 100, 200, or 400 pkpc. For the 100 pkpc bins, $\chi_v = 12.6$. This renormalization brings the scale of \eabs\ into agreement with that shown in Figures \ref{vpack} and \ref{cf145}.

Recently, \citet{rak11b} constructed similar maps to those presented here using pixel optical depth analysis (instead of Voigt profile fits) for a subset of the data presented here. 
Many of the conclusions discussed in the following section are corroborated by the independent analysis of \citet[][\S4]{rak11b}.

\subsection{Absorbers with \NHI ~$> 10^{14.5}$ cm$^{-2}$}

Shown in Figure \ref{asymMap14.5_17_200} is a map of \eabs\ for 10$^{14.5}$ $<$ \NHI$~<$  10$^{17}$ cm$^{-2}$. Here we compare a map of the absorber distribution around galaxies (left-most panel) to a map which represents the general IGM as measured using the random distribution (third panel from left). Clearly, the incidence of \NHI $> 10^{14.5}$ \cm2 absorbers is higher near galaxies than in the general IGM. \eabs\ remains significantly and consistently higher than that seen in the random map within 300 \kms\ along the line of sight and out to 2 pMpc in impact parameter.  

The maps presented in Figure \ref{asymMap14.5_17_zoom} which show the inner distance and velocity ranges at higher resolution, allow one to compare the scales of the absorption excess in Hubble distance and in \dtran\ in order to detect redshift anisotropies indicative of particular kinematic patterns. Galaxy redshift errors and peculiar velocities of gas with respect to galaxies will generally expand the line-of-sight velocity distribution.\footnote{One exception to this ``rule'' is the case of inflowing gas on large scales where the inflow velocities counter the relative Hubble flow causing a reduction in the line-of-sight velocity distribution.} Therefore, if the signal along the line of sight is elongated more than expected given the redshift errors and the scale of the transverse distribution, the remaining velocity structure may be attributed to peculiar velocities of the gas with respect to galaxies.

Based on Figure  \ref{asymMap14.5_17_zoom}, adopting the green contours as representative of the most extreme portion of the overdensity of gas (conservatively) corresponds to a velocity scale of $\sim$300~\kms\ and a physical impact parameter $\sim$125 pkpc\footnote{Maps produced using a symmetric smoothing kernel of 200 pkpc in both the line-of-sight and transverse distance (not shown) also yield excess absorption to 300 \kms\ along the line of sight. The extent of the line-of-sight distribution is robust to reductions in the line-of-sight smoothing kernel; however, the S/N associated with the excess is improved by the adopted smoothing kernel.}. Subtracting in quadrature the 140 \kms\ smoothing (which corresponds roughly to our redshift errors) leaves a residual velocity scale of \absdv $\simeq 265$ \kms. This velocity, were it due to pure Hubble flow, would indicate a distance of $\sim 1.1$ pMpc. In order to consider the portion of the line-of-sight elongation likely due to peculiar velocities, we can subtract in quadrature the extent of the transverse distribution -- in effect subtracting a possible Hubble flow broadening. Subtracting the \dhub\ = 125 pkpc or $\sim 35$ \kms\ leaves $\simeq 260$ \kms; in other words, peculiar velocities of $\pm$260 \kms\ are indicated by the data.\footnote{Or, we have seriously underestimated the magnitude of our galaxy redshift errors.}

\label{high_N_map} 

\subsection{Absorbers with \NHI ~$< 10^{14.5}$ cm$^{-2}$}

\begin{figure}
\center
\includegraphics[width=0.5\textwidth]{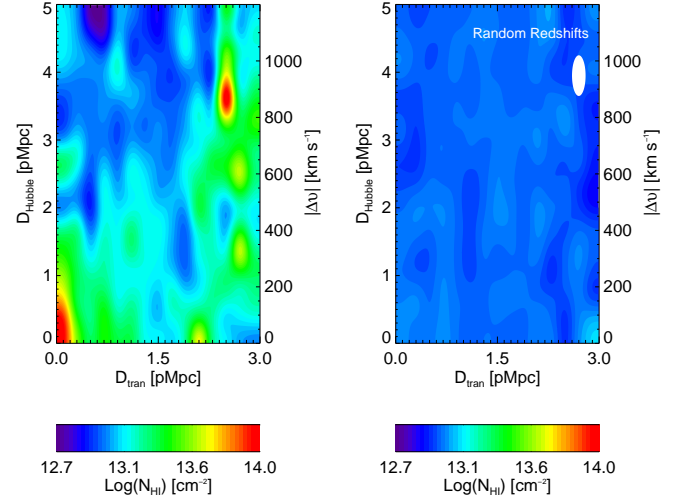}
\caption{Maps of the median \NHI\ with respect to the positions of galaxies. The ``bin'' size for this map is 200 x 200 kpc. The smoothing scale is 200 kpc in impact parameter and 140 \kms~(600 pkpc) along the line of sight. \textit{Left:} The map of the true distribution of column densities. \textit{Right:} The map of the column density distribution in the random sample. The significance with which we detect structures appearing in this map is quantified with the S/N maps presented in the preceding figures. }
\label{N_map}
\end{figure}

Figure \ref{asymMap13_14.5_200} displays a map of the incidence of lower-column density absorbers (10$^{13.0} <$ \NHI $< 10^{14.5}$ cm$^{-2}$). Again comparing the left-most panel (galaxy positions) to the third panel from the left (random IGM), one sees that the incidence of these absorbers is higher in the regions near galaxies than in the general IGM; however, the low-\NHI\ map exhibits no strong peak at small galactocentric distances.

Since there is little structure on small scales in Figure \ref{asymMap13_14.5_200}, the map has been more heavily smoothed to emphasize larger scales. Drawing attention to the edge of the red contours in Figure ~\ref{asymMap13_14.5_400}, it appears that the scale of the excess absorption is marginally \textit{smaller} along the \dhub\ (line-of-sight) axis (1.5 pMpc) than along the transverse distance (2 pMpc). This suggests there may be coherent infall motion on $\simgt$ Mpc scales compressing the distribution along the line of sight, the \citet{kai87} effect.  \citet{rak11b} analyze the pixel statistics in the KBSS survey and discuss the kinematics of the circumgalactic gas in detail. They find a significant detection of compression of the signal along the line of sight which they interpret as large-scale infall.

\subsection{Median Column Density Maps}

An alternative visualization of the distribution of \ion{H}{1} surrounding galaxies, avoiding the division of absorbers into low and high column density, is a map of the median \NHI\ in a given bin. Figure~ \ref{N_map} shows maps where the color bar represents the median \NHI\ in a given bin of \dtran\ and \dhub. On small scales, the behavior is similar to the high-\NHI\ absorbers, with significant elongation in the line of sight direction, due to a combination of peculiar motions of the gas and error in $z_{\rm gal}$.   

In this section we have shown that the kinematics of absorbers with log(\NHI)$>$14.5 are consistent with a peculiar velocity component of $\sim \pm 260$ \kms\ within $\sim 100$ pkpc of galaxies, and separately that absorbers with log(\NHI)$<$14.5 may exhibit the kinematic signature of retarded Hubble flow on scales $\simgt$ 1 pMpc.

\subsection{Explaining the Gas-Phase Kinematics}

The kinematics of the gas surrounding galaxies may provide important clues about the physical processes occurring near galaxies. The two most plausible scenarios for the origin of the observed gas is that it traces the large scale structure (in which case it is likely falling onto the halo) or that it is the result of galactic winds, known to operate in these galaxies. Most likely, it is some combination of the two phenomena. Unfortunately, the sign of the velocity offset (red-shifted or blue-shifted) cannot tell us what the relative motion of the gas is with respect to the galaxy since its position along the line of sight is not known. For example, gas which is redshifted with respect to the galaxy might be behind the galaxy, and thus be either outflowing or moving with the Hubble flow at larger distance. Conversely, redshifted absorbers could be foreground gas that is redshifted because it is falling onto the galaxy. There is no way to know \textit{a priori} which scenario holds in which cases.

We recall that in \S \ref{vel} it was shown that the full extent in $\Delta v$ spanned by absorbers comprising the net excess over random is \absdv $\simlt 700$ \kms, with the most significant portion of the excess within 
\absdv $\simlt 300$ \kms. Our map of high-\NHI\ systems confirms the 300 \kms\ excess and allows us to measure the extent of the peculiar velocities. After accounting for our estimated galaxy redshift errors (\S~\ref{high_N_map}), 
the range of significant excess becomes \absdv~$\simlt 260$ \kms.

In-falling cool gas is likely to be moving at relative velocities smaller than the galaxy 
circular velocity \citep{fau11a}:
\begin{equation}
v_{circ} = \sqrt{\frac{GM_{\rm DM}}{r_{\rm vir}}} = 220 ~\textrm{km s}^{-1},
\end{equation}
where we assume $M_{\rm DM} = 10^{12}$ \Msun\ and $r_{\rm vir} = 91$ pkpc (see \S \ref{gal_sample} and \S \ref{smallscale}).
Since we measure only the line-of-sight component of this velocity, $|v_{\rm los}| \sim 200$\kms\ might serve as an approximate upper limit on the line-of-sight velocity component due to graviationally induced peculiar velocities.   
Unless we have significantly underestimated the measurement uncertainties in $z_{\rm gal}$, accreting gas would
be expected to have a ``quieter'' velocity field than observed.   

On the other hand, outflows observed in the galaxy spectra regularly show blue-shifted velocities as high as 800 \kms, 
and large velocity widths appear to be required to explain the strength of strongly saturated absorption at modest
impact parameters (\dtran~$ \simlt 100$ pkpc), based on the galaxy-galaxy pairs analysis of \citet{ccs10}.  
In the context of their simple model, the envelope of $|\Delta v|$ is simply the component of the outflow speed $v_{\rm out}$ projected along an observers line of sight; this maximum velocity also dictates how strong the resulting absorption features will be, with $W_0$ increasing proportionally to $v_{\rm out}$. To model the behavior of
$W_0$ with impact parameter, $v_{\rm out}$ was the principle normalization factor and the inferred values 
were $650 \simlt v_{\rm out} \simlt 820$ \kms\ depending on the ion. 
In the present work, the first bin in \dtran\ extends from 
50-100 pkpc, with a median \dtran~$=80$ pkpc, with median log(\NHI)~$\simeq 16.5$. 
Assuming the \citet{ccs10} geometric/kinematic model (with $R_{\rm eff} = 90$ pkpc) would predict 
that a line of sight with \dtran$ = 80$ kpc would have \absdv$\simlt 0.46v_{\rm out}$ (equivalent to an observable velocity range between $\pm325$ and $\pm400$ \kms) 
and a threshold \NHI~$\sim 10^{17}$ \cm2 [analogous to our max(\NHI) statistic], both  
compatible with the observations from the QSO sightlines discussed above.



\section{The Doppler Width}

\label{dop}

\begin{figure}
\center
\includegraphics[width=0.45\textwidth]{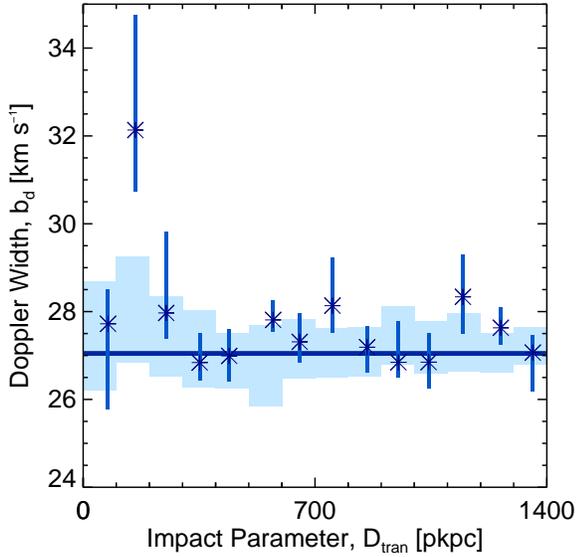}\caption{The Doppler width, \bd, of absorbers with log(\NHI) $> 13$ and \absdv $<$ 700 \kms\ as a function of transverse distance from a galaxy. The symbols have the same meaning as those in Figure \ref{whisker} except they refer to \bd\ instead of \NHI.  Note the peak in the second bin corresponding to absorbers with 100 $<$~\dtran~$<$ 200 pkpc.}
\label{b_all}
\end{figure}

\begin{figure*}
\center
\includegraphics[width=0.56\textwidth]{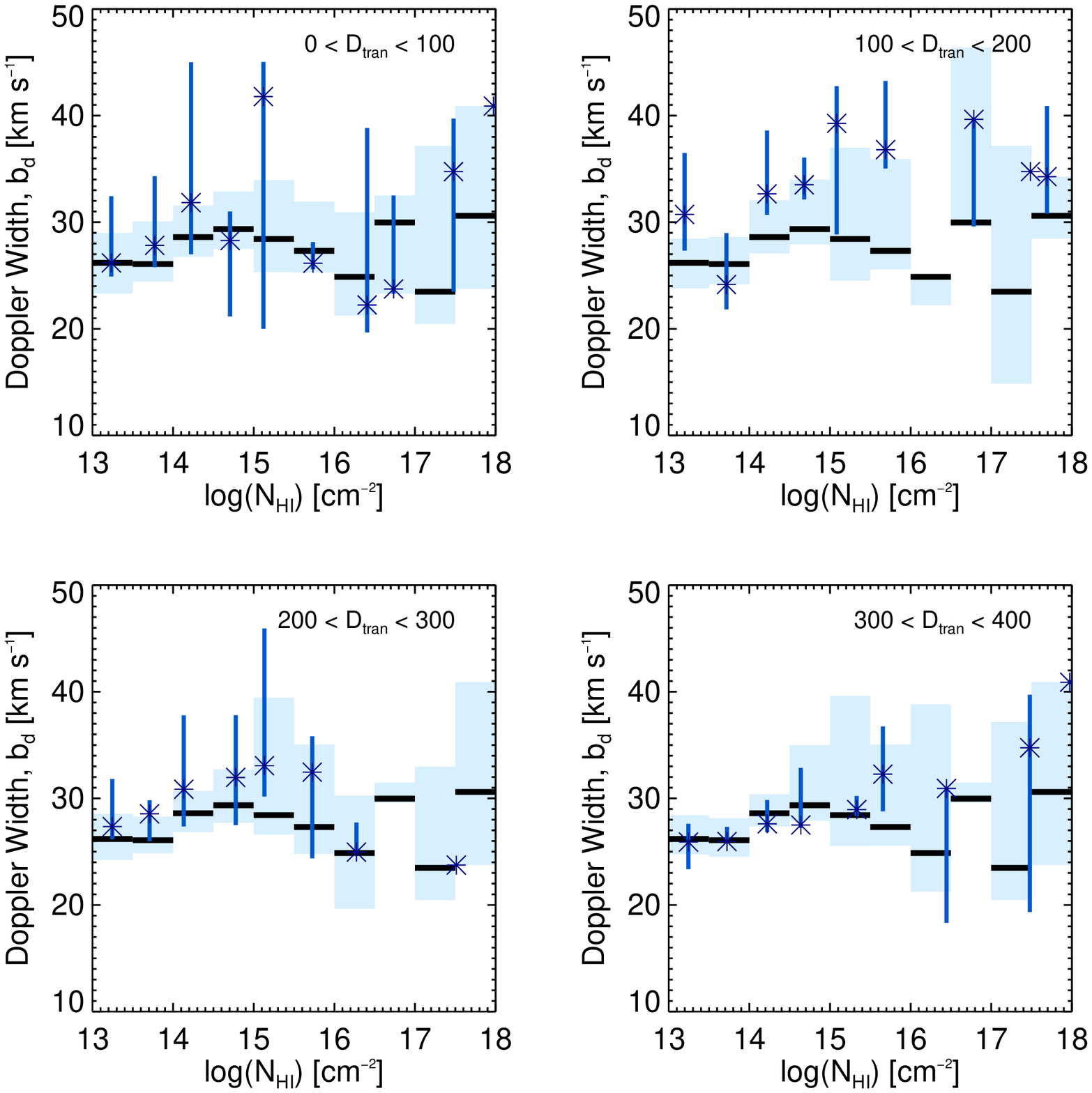}
\caption{Comparison at fixed \NHI\ of absorbers close to galaxies with absorbers in the full \ion{H}{1} catalog. The different panels show four bins of \dtran. Asterisks represent the median value of \bd\ in a bin of \NHI\ for absorbers close to galaxies. The horizontal black lines show the median \bd\ of absorbers from the full \ion{H}{1} catalog, and the light shaded boxes represent their dispersion calculated through bootstrap samples of the same size as those in the real sample. The velocity window is \absdv $<$ 700 \kms.  In the bins without asterisks, there were no absorbers which satisfied both the \dtran\ and \NHI\ criterion; similarly, asterisks without error bars represent bins containing only a single absorber satisflying both the \dtran\ and \NHI\ criterion. Recall from Figure \ref{b_all} that the strongest signal occurred at 100 $<$ \dtran\ $<$ 200 pkpc (shown in this Figure in the top right panel). Note that the median \bd\ of all absorbers close to galaxies with \NHI $> 10^{14}$ \cm2 are larger than those in the full absorber catalog.}
\label{b_whisker}
\end{figure*}

In addition to \NHI\ and $z_{\rm abs}$, the third component of a Voigt profile fit is the Doppler width, \bd.  In the case of an unsaturated line, the Voigt profile is well approximated by a Gaussian in optical depth, with a Doppler parameter:
\begin{equation}
\bd = \sqrt{2}\sigma = \frac{\textrm{FWHM}}{2\sqrt{\ln 2}}.
\end{equation}
In the case of purely thermal broadening, the gas temperature $T$ can be inferred directly from the Doppler parameter:
\begin{equation}
\bd (T)= \sqrt{\frac{2kT}{m}}
\end{equation}
where $m$ is the mass of the ion and $k$ is the Boltzmann constant. For \ion{H}{1},
\begin{equation}
T = \frac{m_{\rm H}}{2k}\bd^2 = 4 \times 10^4 ~\textrm{K} \left(\frac{\bd}{26~{\rm km~s}^{-1}}\right)^2;
\end{equation}
noting that 26 \kms\ is the median value of \bd\ in the full absorber catalog.

The physical state of the IGM is predicted to be governed by the balance of two principle processes: the adiabatic cooling caused by the expansion of the Universe and photoionization heating, generally from the UV background. In the regime where this holds and in the case that the baryonic overdensity roughly traces the dark matter overdensity, a natural consequence is a relationship between the temperature of the gas, $T$, and its density, $\rho$. Higher-density regions, having more gravitational resistance to the Hubble flow, cool less and thus have higher temperatures on average \citep[see e.g.,][]{hui97,sch99}. And indeed, observed distributions of thermally broadened absorption lines exhibit higher \bd\ for absorbers with higher \NHI\ \citep{pet90,sch00, bry00, ric00, mcd01}.

Other processes can also broaden individual absorption components. For the most diffuse and physically extended absorbers, generally with lower \NHI, the Hubble flow itself can contribute significantly to the line width. Turbulence (here meaning bulk motions of the gas) can also broaden absorption features such that:
\begin{equation}
\bd^2 = b_{\rm turb}^2 + \frac{2kT}{m}
\end{equation}
where $b_{\rm turb}$ is the turbulent component of the line width.

In this section we discuss the distribution of Doppler widths (\bd) observed in absorption systems as a function of their galactocentric distance.  The observed trends are much more subtle than those in \NHI, but their utility in discerning the physical state of the gas motivates a careful analysis.

\subsection{The Dependence of \bd\ on Proximity to Galaxies, \dtran}

\begin{figure}
\center
\includegraphics[width=0.45\textwidth]{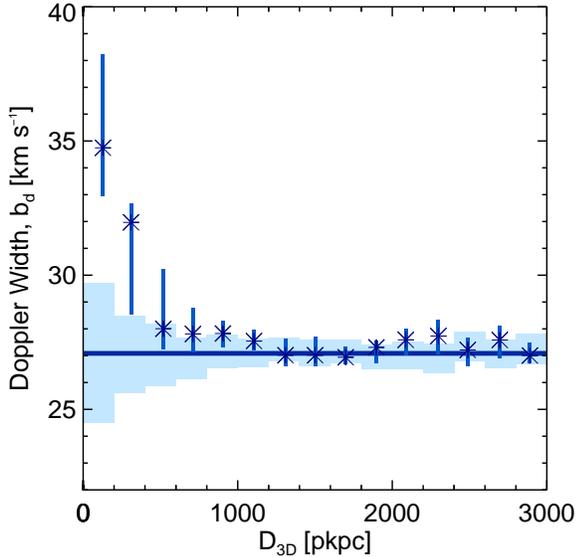}
\caption{Same as Figure \ref{b_all} but versus the 3D distance.}
\label{b_all_3D}
\end{figure}

\begin{figure*}
\center
\includegraphics[width=0.56\textwidth]{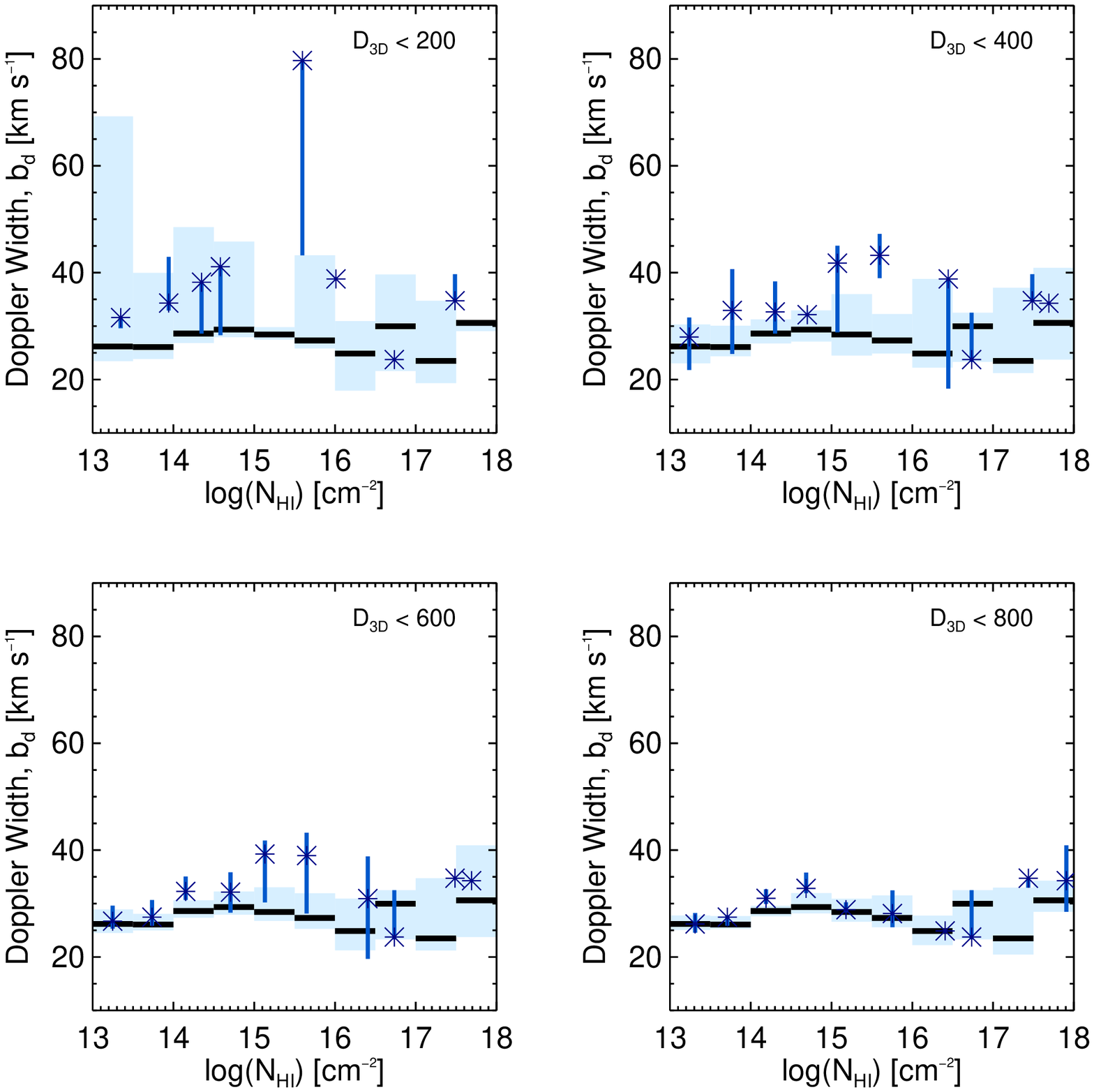}
\caption{Same as Figure \ref{b_whisker} but now divided into bins of 3D distance, \dtd, and with cumulative bins in \dtd. From Figure \ref{b_all_3D} one notes that the majority of the signal results from \dtd $<$ 400 pkpc (shown in this Figure in the upper two panels). Notably,  absorbers near galaxies have higher \bd\ at nearly all values of \NHI\ at these impact parameters. Because the signal from all bins is combined, higher \bd\ at fixed \NHI\ is evident for absorbers with \dtd $<$ 600 pkpc. }
\label{b_whisker_3D_cumm}
\end{figure*}

Figure \ref{b_all} shows the Doppler width (\bd) of absorbers with \NHI ~$> 10^{13}$ \cm2 versus \dtran.  While considerably more noisy than the column density trends in the previous section, it appears that within the second bin (corresponding to 100 $<$ \dtran $<$ 200 pkpc) the median value of \bd\ is considerably higher than that of the random sample.  

We consider the trend in \bd\ versus \NHI\ for both the sample of absorbers close to galaxies and that of the full absorber catalog.  In this way, we can examine the \bd\ of absorbers at \textit{fixed} \NHI, as shown in Figure \ref{b_whisker}, in order to evaluate the effect of the $T-\rho$ relationship on this result.  Here, each of the four closest bins in \dtran\ (Figure \ref{b_all}) is broken into individual panels. The asterisks represent the absorbers close to galaxies. The black horizontal lines and shading refer to the \textit{full absorber catalog}. \footnote{Note that within the full catalog for absorbers whose growth is expected to remain approximately linear (\NHI $< 10^{15}$ \cm2, $\rho/\overline\rho \lesssim 10$), the median values (black horizontal lines) exhibit increasing \bd\ with increasing \NHI, as one would expect for the $T-\rho$ relation.} Considering first the top right-hand panel with $100 <$ \dtran $< 200$ pkpc (the bin with the majority of the signal in Figure \ref{b_all}) for \NHI $> 10^{14}$ \cm2, the median \bd\ of absorbers close to galaxies is larger than the median of the full absorber sample at fixed \NHI. This suggests that the larger \bd\ cannot be attributed solely to a dependence on \NHI\ as in the $T-\rho$ relation. 

Figure \ref{b_whisker} shows that the median \bd\ for the sample of absorbers within \dtran\ $\sim 100-300$ pkpc of galaxies is systematically larger than that of absorbers from the full absorber catalogue across most bins in \NHI. However, the absorbers closest to galaxies (\dtran $<$ 100 pkpc; top left panel of Figure \ref{b_whisker}) do not appear to exhibit the same systematic ``excess'' in median \bd\ (cf. Figure \ref{b_all}).  There are too few galaxies in this inner bin to determine whether the result is statistically significant.

In order to quantify the statistical significance of the apparent excess \bd\ in the range \dtran\ $ < 300$ pkpc, we performed a Monte Carlo simulation. This was done by drawing random samples of absorbers from the full absorber catalogue with the same distribution in \NHI\  and of the same size as the absorber sample close to galaxies. 
The figure of merit was taken to be the average difference computed as follows: in each bin of \NHI\ the difference between the median b$_{\rm d, MC}$ from the Monte Carlo sample and the median \bd\ in the full absorber catalog was computed (asterisks minus black bars in Figure \ref{b_whisker}). The differences from the \NHI\ bins were then averaged. The fraction of Monte Carlo samples whose average \bd\ excess is greater than that of the \dtran\ $< 300$ pkpc sightline sample is 0.005, meaning the excess is
significant at approximately the $3\sigma$ level.

\subsection{Doppler Widths vs. 3D Distance}

In Figure \ref{b_all_3D} we consider the distribution of \bd\ as a function of \dtd, which allows one to consider trends as a function of distance without pre-selecting a velocity window. Here again, \bd\ is elevated for bins close to galaxies (\dtd~$\simlt$ 400~pkpc). In the first bin, the median \bd~$\approx$~35~\kms\ compared to \bd $=$27~\kms\ for the random sample. 

We again perform a comparison at fixed \NHI\ and since the median \bd\  monotonically decrease with increasing \dtd, we consider cumulative distance bins. For absorbers with \dtd $<$ 600 pkpc, the median \bd\ for absorbers with \NHI $< 10^{16}$ \cm2 is larger than the median measured in the full absorber catalog. Notably, for 15~$<$~log(\NHI)$<$~16, absorbers close to galaxies (\dtd $<$ 400 pkpc) have a median $\bd > 40$ \kms. If interpreted as the result of an increase in gas temperature, the change in \bd\ is equivalent to more than doubling $T$. We also note that these intermediate \NHI\ systems that are characteristic of the CGM tend also to be associated with the strongest high ionization metals (\ion{O}{6}, \ion{C}{4}), often exhibiting evidence that the ionization has been produced by shocks in addition to photoionization \citep{sim02}. 

As previously stated, the 3D distance is strongly influenced by the magnitude of \absdv. Notably, the first three bins in \dtd\ (those with the most significant deviation in \bd) correspond to absorbers with \absdv~$<$~150~\kms.

In summary, considering the trends illustrated in Figures \ref{b_all} -- \ref{b_whisker_3D_cumm}, it appears that absorbers of all \NHI\ with $100 < $~\dtran~$< 300$~pkpc and \absdv~$< 150$~\kms\ have significantly larger Doppler widths than absorbers of the same \NHI\ in the full absorber sample. 

\subsection{Doppler Parameter: Possible Physical Explanations} 
\label{dop_disc}

The trends in Doppler widths, \bd, could have a variety of physical origins. As previously discussed, elevated \bd\ could be related to the observed trend in \NHI\ through the $T-\rho$ relation in the IGM. However, we showed that at \textit{fixed} \NHI, the median \bd\ of absorbers close to galaxies is still significantly larger than in the full absorber catalog (Figures \ref{b_whisker} and \ref{b_whisker_3D_cumm}). This suggests that the $T-\rho$ relationship alone cannot account for the broader absorption lines close to galaxies. 

Another possible contributor to increasing \bd\ near galaxies is a change in the \textit{ionization} at a given \NHI. If the galaxy itself produces a significant fraction of the local ionizing radiation field (i.e., in excess of that of the metagalactic UV background), then \NHI/$N_\textrm{\footnotesize{H}}$ would be smaller near galaxies. The fraction of the ionizing photons which escape the ISM of the galaxy is small \citep[][Steidel et al. in prep]{nes11}, but their number, compared to the UV background is likely significant within $\sim$ 100 pkpc.  In this case, the $T-\rho$ relation could hold, but the mapping of \NHI\ to $\rho$ would be altered. If this were the case, we would expect the locations of the asterisks in Figures \ref{b_whisker} and \ref{b_whisker_3D_cumm} to be consistent with a leftward shift of the black bars. However, this scenario is at odds with the observation that the median \bd\ of absorbers with $10^{14} < $\NHI $< 10^{16}$ \cm2 and 100 $<$ \dtran $<$ 200 pkpc  is higher than the median \bd\ for any bin in \NHI\ observed in the full absorber catalog (Figure \ref{b_whisker}).  The same is true of absorbers with \dtd\ $<$ 400 pkpc and $10^{14} < $\NHI $< 10^{16}$ \cm2 (Figure \ref{b_whisker_3D_cumm}).

Assuming that overdensity is not the main cause of the increased \bd, we consider other plausible processes which could lead to an increase in the temperature or turbulence of the gas. In principle it is possible to disentangle the source of the broadening (turbulent vs. thermal) by comparing the widths of absorption lines arising from ions with different atomic weights. The thermal broadening component, \bd(\textit{T}),  depends on atomic mass, while the turbulent broadening component will remain constant so long as the ions reside in the same gas. While the HIRES QSO spectra in our sample include metal absorption lines which could be used for this purpose, the analysis of the metal lines is beyond the scope of this paper and will be presented elsewhere. At present, we consider the implications of both possible effects (temperature and turbulence).

Referring to Figure \ref{b_all_3D}, there is an increase in the median \bd\ from 27 \kms\ to 35 \kms. Assuming purely thermal broadening, these values would indicate $T = 4.4 \times 10^4$ K and $7.4 \times 10^4$ K, respectively, or an increase in temperature of 68\%. If we consider the absorbers with \bd $\simgt$ 40 \kms~as in Figure \ref{b_whisker_3D_cumm}, these suggest a temperature of $\sim 10^5$ K.  If instead the increase in line width is due to turbulence, an excess turbulent velocity of 20-30 \kms\ would account for the departure from the \bd = 27 \kms\ global median. In either case, it is curious that the elevated Doppler widths occur in gas with 100 $<$ \dtran $<$ 300 pkpc but are not detected in the bin from 0 - 100 pkpc. While the non-detection at the smallest impact parameters is likely not significant, clearly, whichever mechanism is responsible for this increase in width is acting significantly outside of the typical galaxy virial radius ($\sim$ 90 pkpc).

Unless AGN are involved (galaxies with detected AGN signatures were not used in our analysis), excess photo-heating for material closer to galaxies seems unlikely, since the UV radiation field from local stellar sources is likely to be (if anything) softer than the metagalactic background believed to dominate at an average IGM location. 

There are at least two plausible processes which could increase \bd\ in the CGM of galaxies: galactic winds (outflows) and baryonic accretion. Accreting gas can be heated by shocks forming as the gas falls into galaxy halos. The galaxies in our sample have halo masses comparable to the theoretically expected transition mass between ``cold mode'' and ``hot mode'' accretion \citep{van11a,fau11b,ker09b,ocv08}; most of these authors predict that both hot and cold accretion occur in such halos at $z > 2$. The virial temperature for galaxies with halos of mass $\sim 10^{12}$ \msun\ is expected to be $T_{\rm vir} \simeq 10^6$ K \citep{van11a}, and \citet{ker09b} suggest that this hot gas may fill the volume surrounding galaxies to a few \rvir. The \ion{H}{1} absorbers in our sample are unlikely to trace gas this hot, however gas cooling from this temperature would be detectable.

The interaction of cool dense gas with either a hot halo or a rarified wind fluid is also expected to produce turbulent boundary layers via both Rayleigh-Taylor\footnote{Rayleigh-Taylor instabilities would form in the case of a lighter, less-dense fluid lying deeper in the gravitational potential of the galaxy than a denser cold-stream.} and Kelvin Helmholtz\footnote{Kelvin Helmholtz instabilities would form in the condition where the dense stream was moving with respect to the hot medium. This is especially plausible in the case of a fast moving galactic wind or a filament moving at the free-fall velocity.} instabilities \citep{ker09, fau11a}. These boundary layers are also where metallic ions such as \ion{O}{6} are expected to be most abundant. 

Galactic super winds can naturally explain either thermal or turbulent broadening, potentially to large distances away from galaxies. The favored mechanism for large scale winds is shocks from supernovae which could easily increase the gas temperature of the ambient medium. Although the distance to which galactic winds propagate is not well constrained, there is evidence that they expand to at least \rvir\ \citep{ccs10}.
\citet{she11} recently considered high-resolution ``zoom-in'' simulations of a single Lyman Break Galaxy\footnote{Lyman Break Galaxies at $z=3$ have very similar properties to the galaxies studied in this paper.} at $z=3$. In their model, the LBG itself (as opposed to its satellite galaxies) enriches the IGM with metals from supernovae-driven winds to 3\rvir\ (see their Figure 9). \citet{kol06} considered the effect of winds on the CGM of star-forming galaxies at $z=3$ and found that winds could have a profound effect on the temperature of nearby gas: varying the energy released by supernovae in their simulations by a factor of 5 changed the temperature of gas close to galaxies by $\sim250$\% (see their Figure 10).  More recently, \citet{van11d} considered the physical properties of gas surrounding galaxies with halo masses M$_{\rm DM} = 10^{12}$ \Msun\ at $z=2$ as a function of the kinematics of the gas (i.e. inflowing vs. outflowing). They found a radial temperature profile of outflowing gas qualitatively consistent with the observed increase in \bd. Outflows are also predicted to drive gas turbulence outside of virial halos through the conversion of kinetic energy from supernova driven bubbles into random motions through gas instabilities \citep[see e.g.,][]{evo11}.

While the exact cause of the enlarged line widths is not apparent, it is clear that the physical state of the gas within $\sim$ 300 pkpc is markedly different from that in ``random'' locations in the IGM - suggesting that galaxies and their potential wells deeply affect their surrounding CGM. This fact may introduce significant complications to studies of the general IGM which will be discussed further in \S \ref{distinct_physics}.

\section{General Discussion}

In \S \ref{analysis} -- \ref{dop} we presented measurements and plausible interpretations of the properties of neutral hydrogen gas surrounding star-forming galaxies at high $z$. Here, we speculate on the possible implications of the observed trends.

\label{discussion}

\subsection{Gaseous ``Zones'' around Galaxies}

\label{def}

The distribution of strong absorbers (Figure \ref{whisker}) and of the incidence of absorbers (Figures \ref{vpack} and \ref{cf145})  suggest three distinct ``zones'' in the gaseous envelopes surrounding galaxies. \begin{enumerate}

\item{The first zone corresponds to the volume within \dtran $<$ 300 pkpc. For scale, the virial radius of a typical galaxy in our sample is $\sim90$ pkpc. Inside 300 pkpc, the values of the max(\NHI) statistic are significantly elevated and rise toward the position of the galaxy (Figure \ref{whisker}). Figure \ref{vpack} shows rising \eabs\ with decreasing \dtran\ for the three bins within 300 pkpc. Similarly, at 300 pkpc, \eabs\ for absorbers with \NHI\ $> 10^{14.5}$ drops sharply. In Figure \ref{cf145}, again at 300 pkpc one sees a rapid drop in the incidence of all absorbers with \NHI\ $ > 10^{14.5}$ cm$^{-2}$. }

\item{The second zone lies between 300 pkpc $<$ \dtran $\lesssim$ 2 pMpc. In Figure \ref{whisker}, we observed that max(\NHI), while high compared to the random distribution, plateaus at a roughly constant value within this zone. \eabs\ also plateaus over the same range as shown in Figures \ref{vpack} and \ref{cf145}.}

\item{The third zone corresponds to \dtran $\gtrsim$ 2 pMpc. The transition between the second and third zone is not a sharp feature; rather between $\sim 2 - 3$ pMpc, the distribution of absorbers as seen in the max(\NHI) and \eabs\ statistics drops to become consistent with (or below that of) the IGM median.}

\end{enumerate}

Three distinct zones are also evident in the velocity distribution of absorbers. \begin{enumerate}

\item{The first zone, corresponding to the strongest peak in the velocity distribution, is within \absdv\ $<$ 300 \kms\ (Figures \ref{vel_hist} and \ref{inner_dist_bin300}). }

\item{The second velocity zone includes gas to  \absdv\ $\approx$ 700 \kms, which encompasses the full extent of the excess \NHI\ (Figures  \ref{vel_hist_unweight} and \ref{vel_hist}). }

\item{The third zone is \absdv\ $>$ 700 \kms, where the distribution with respect to galaxies is consistent with that of random locations (Figures \ref{vel_hist_unweight} and \ref{vel_hist}).}

\end{enumerate}

Note that in Figure \ref{inner_dist_bin300}, the median \NHI\ of absorbers with \dtran\ $<$ 300 pkpc exhibits a sharp drop-off at \absdv\ $\ge$ 300 \kms\ suggesting that absorbers at \dtran\ $>$ 300 pkpc are likely responsible for the excess absorption at 300 $<$ \absdv\ $<$ 700 \kms. 

\subsection{Defining the CGM}

\label{def2}

\begin{figure}
\center
\includegraphics[width=0.45\textwidth]{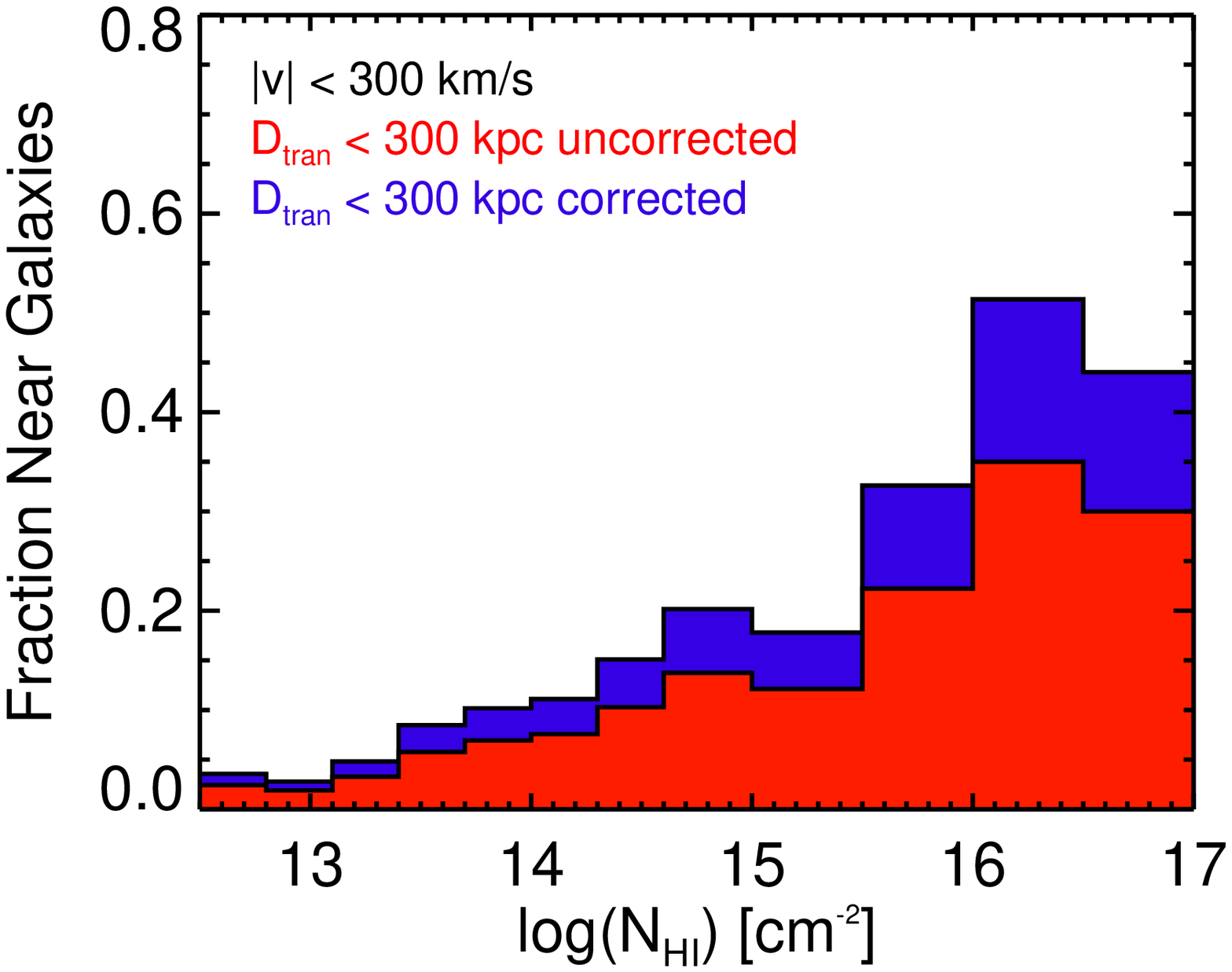}
\caption{The fraction of all \ion{H}{1} systems in our QSO sightlines that arise within \absdv\ $<$ 300 \kms\ and \dtran\ $<$ 300 pkpc of the position of a galaxy in our spectroscopic galaxy sample (lighter red histogram), as a function of \NHI. The darker (blue) histogram indicates the fraction after correcting for the fact that not all galaxies in the photometric sample have been observed spectroscopically. These corrected points thus reflect the fraction of absorbers arising within the CGM of galaxies that meet our {\it photometric} selection criteria.}
\label{N_frac}
\end{figure}

We adopt a working definition of the ``circumgalactic medium'' (CGM) as the region within \absdv\ $<$ 300 \kms\ and \dtran\ $<$ 300 pkpc of a galaxy. Figure \ref{N_frac} shows (in red) the fraction of absorbers from the full absorber catalog which fall within the CGM of a galaxy \textit{in our spectroscopic sample} as a function of \NHI. 

Our spectroscopic galaxy sample is 70\% complete with respect to the photometric parent sample within 300 pkpc of the line-of-sight to the QSOs.\footnote{Our absorption line catalogue is essentially 100\% complete for absorbers with log(\NHI) $\gtrsim$ 13.} The blue histogram in Figure \ref{N_frac} has been corrected for this incompleteness assuming that unobserved galaxies have the same overall redshift distribution and similar CGM to those that have been observed. Figure~\ref{N_frac} shows that $\simgt$ 20\% of all absorbers with log(\NHI) $>$ 14.5 and $\simgt 40$\% of those with log(\NHI) $>15.5$ arise in the CGM of galaxies that meet our selection criteria.  Noting that the comoving number density of identically selected galaxies (with the same limiting magnitude of ${\cal R}=25.5$) is $\Phi= 3.7 \times 10^{-3}$ cMpc$^{-3}$ \citep{red08}, the CGM of LBGs at $\langle z \rangle = 2.3$ contains only $\simeq 1.5$\% of the Universe's volume but accounts for nearly half of the total gas cross-section for \NHI\ $> 10^{15.5}$ \cm2. This is perhaps surprising given that our sample includes only galaxies with $L_{UV} \ge  0.25L_{UV}^*$ and the faint-end slope of the UV luminosity function is very steep at $z \sim 2.3$; however, it is consistent with previous results based on strong metal-line absorption around similar galaxies at comparable redshifts \citep{ade03,ade05,ccs10}.

The situation changes substantially for \NHI $\simlt 10^{14.5}$ \cm2; we have seen that these lower \NHI\ absorbers are more loosely associated with galaxies in our sample, though their incidence is enhanced by $\sim 10-20$\% relative to average locations in the IGM (e.g. see the top panels of Figure~\ref{vpack}). 

Stated another way, we have found that low and high- column density absorbers cluster differently with high-$z$, UV-bright galaxies. We showed (Figure \ref{vpack}) that absorbers with log(\NHI) $> 14.5$ exhibit a much stronger peak in incidence towards the positions of galaxies than do the absorbers with lower column densities. Similarly, in maps of the incidence (Figures \ref{asymMap14.5_17_200} -- \ref{asymMap13_14.5_400})  low-column density absorbers show only large scale correlation with the positions of galaxies while high-column density absorbers again showed a strong peak near galaxies both along the line of sight and in \dtran. 

Collectively, these lines of evidence point to a circumgalactic zone defined by the boundaries of \absdv $<$ 300 \kms\ and \dtran $<$ 300 pkpc and dominated by absorbers with log(\NHI) $> 14.5$. They also suggest that only systems with log(\NHI) $\simlt 14$ \cm2\ are confidently ``IGM''-- as \NHI\ increases beyond this limit, the likelihood that the gas lies within the CGM of a relatively bright galaxy increases very rapidly.

\subsection{Small Scale Distribution of \NHI}

\label{smallscale}

In an effort to understand the physics responsible for the distribution of \NHI\ as a function of transverse distance from galaxies, we compare the data to two theoretical models. First we consider the possibility that the \ion{H}{1} column density traces the dark matter halo density profile. 

The typical dark matter halo hosting a galaxy in our sample is theoretically expected to follow a NFW \citep{NFW} density profile:
\begin{equation}
\rho(r) = \frac{\rho_c(z)  ~\delta_\textrm{\footnotesize{NFW}}}{c\frac{r}{r_{\textrm{\footnotesize{vir}}}}\left(1+c \frac{r}{r_{\textrm{\footnotesize{vir}}}}\right)^2}
\end{equation}
where $\delta_\textrm{\footnotesize{NFW}}$ is a normalization defined as
\begin{equation}
\delta_\textrm{\footnotesize{NFW}} = \frac{200}{3} \frac{c^3}{\ln(1+c) -\frac{c}{1+c}}
\end{equation}
and $\rho_c$ is the critical density, $c$ is the concentration parameter of the dark matter halo which we take to be $c=4$ following \citet{duf08}, and \rvir\ is the virial radius for halos of average mass $ 10^{12}$ \Msun\ \citep[91 pkpc; Trainor \& Steidel, in prep;][]{con08, ade05b}.

\begin{figure}
\center
\includegraphics[width=0.5\textwidth]{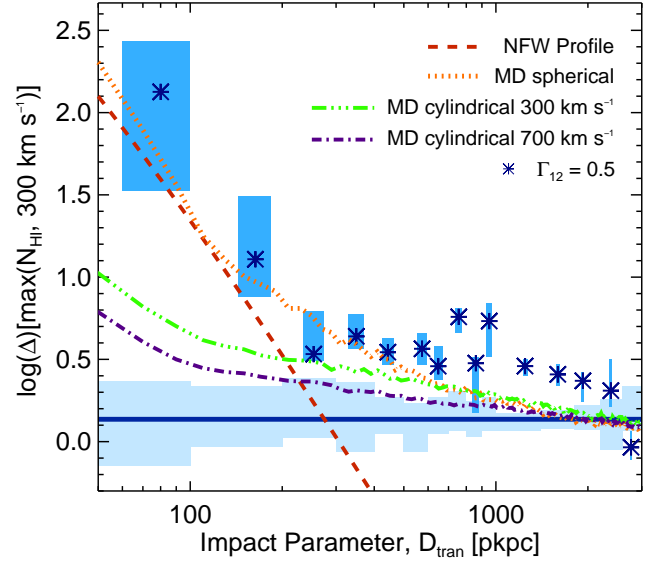}
\caption{The max(\NHI, 300 \kms) statistic, converted into a universal overdensity using \citet{sch01} plotted against transverse distance to a galaxy on a logarithmic scale. The same data shown in Figure \ref{whisker300} with two models over-plotted for comparison. Here we use 100 pkpc bins within 1 pMpc and 400 pkpc bins at larger \dtran\ to simplify the plot. Shown in the (red) dashed line is the NFW radial density profile plotted as a universal overdensity. See text for more details. Shown in the light (orange) dotted , light (green) dash-dotted, and dark (purple) dash-dotted lines are the average dark matter profiles from the MultiDark simulation for halos with M$_{\rm DM} > 10^{11.8}$ \Msun. 
The light (orange) dotted curve is the median value (across the halos considered) of the average density in concentric spherical shells of radius \dtran. The light (green) dash-dotted curve is the median (across the halos considered) of the average dark matter density in concentric cylindrical shells of radius \dtran\ and length $\pm$300 \kms $= \pm$ 1200 pkpc. The dark (purple) dash-dotted curve is analogous to the green one, but the velocity window is $\pm 700$ \kms.}
\label{maxN_curves_xlog}
\end{figure}

We define $\Delta$, the matter overdensity, as:
\begin{equation}
\Delta = \frac{\rho}{\overline \rho} = \frac{\rho}{\Omega_{\rm M} (1+z)^3 ~\rho_c(z=0)}
\end{equation}
Noting the definition for the critical density:
\begin{equation}
\rho_c(z) = \frac{3H^2(z)}{8\pi G}
\end{equation}
we can then express the NFW profile as a function of overdensity, rather than density:
\begin{equation}
\Delta_\textrm{\footnotesize{NFW}}(r) = \frac{1}{\Omega_{\rm M} (1+z)^3} \frac{\delta_\textrm{\footnotesize{NFW}}}{c\frac{r}{r_{\textrm{\footnotesize{vir}}}}\left(1+c \frac{r}{r_{\textrm{\footnotesize{vir}}}}\right)^2}\frac{H^2(z)}{H_0^2}
\label{NFW}
\end{equation}

Using Jeans scale arguments, \citet{sch01} defined a scaling function which relates the gas overdensity ($\rho_b/\overline \rho_b$) to an expected value of \NHI. While on the smallest scales where the growth is highly non-linear it is certainly not the case that $\rho_b/\overline \rho_b$ is equal to $\Delta$, the matter overdensity (which is dominated by dark matter), the expectation is that they should mirror each other well at low overdensity ($\Delta \approx 1 - 10$) where the growth is yet to become significantly non-linear. We therefore assume $\rho_b/\overline \rho_b \approx \Delta$, in order to compare our measured max(\NHI) to theoretical density profiles. For the cosmology used in this paper, at the redshifts of our sample, the scaling function from \citet{sch01} is: 
\begin{equation}
\label{jeans}
\Delta \approx \left(\frac{N_{\textrm{\footnotesize{HI}}}}  {10^{13.4}}\right)^{2/3}T_4^{0.17} \left( \frac{\Gamma_{12}}{0.5}   \right)^{2/3} \left( \frac{1+z}{3.3} \right)^{-3},
\end{equation}
where $T_4$ is the gas temperature in units of $10^4$ K and $\Gamma_{12}$ is the hydrogen photoionization rate in units of $10^{-12}$ s$^{-1}$ where the normalization is taken from \citet{fau08}.
Using this relation, we can convert our max(\NHI) statistic into $\Delta$ which can then be compared to the rescaled NFW profile. 

The values after converting max(\NHI) to $\Delta$ are shown in Figure \ref{maxN_curves_xlog}. The (red) dashed curve is the NFW density profile (Equation \ref{NFW}). The NFW halo has a radial distribution similar to that of the $\Delta$ inferred from our max(\NHI) statistic for small \dtran, but falls well below the Jeans-scale-inferred overdensity for \dtran~$\simgt 300$ pkpc. This is expected since the NFW profile is measured in simulations for dark matter (DM) within the virial radius. In the following section, we compare to the DM density on larger scales using DM profiles drawn directly from simulations.

\subsection{Large Scale Distribution of \NHI}

\begin{figure}
\center
\includegraphics[width=0.5\textwidth]{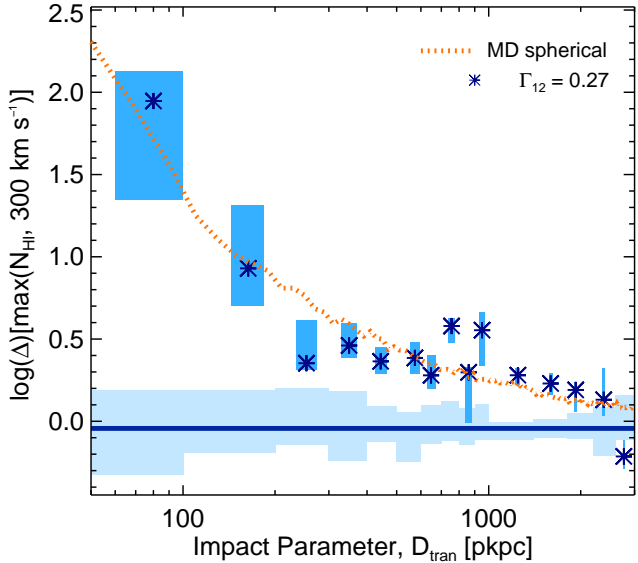}
\caption{Same as Figure \ref{maxN_curves_xlog} but using a different normalization of the photoionization rate. In Figure \ref{maxN_curves_xlog} the data values were converted into $\Delta$ assuming $\Gamma_{12} = 0.5$ as suggested by \citet{fau08}. Here, the data values are converted into $\Delta$ assuming $\Gamma_{12} = 0.27$ which bring them into the closest agreement with the MultiDark spherical density profile measurements.}
\label{maxN_curves}
\end{figure}

Considering the distribution of gas around galaxies on Mpc scales, it is interesting to note the similarity between the scale over which circumgalactic gas has higher density than an average location in the IGM (\dtran\ $\lesssim 2$ pMpc) and the galaxy-galaxy autocorrelation scale length recently measured {\it from the same galaxies},  $r_0 = (6.5\pm 0.5)h^{-1}$ cMpc ($\simeq 2.8$ pMpc at $\langle z \rangle =2.3$; Trainor \& Steidel, in prep). It would not be surprising if they were closely related. 

Most of the recent work on the dark matter density profiles at several \rvir\ and beyond have focused on the local universe. Since we are interested in comparing the gas-inferred density profiles with an average dark matter profile at $z \sim 2-3$, we created median DM density profiles using the MultiDark simulation \citep{kly11}.  The radial density profile within 3 pMpc was computed from 100 halos with M$_{\rm DM} > 10^{11.8}$ \Msun\footnote{The halo mass was inferred from the number of particles found to ``belong'' to a halo using the Friends-of-Friends algorithm. The halo-mass cut of $10^{11.8}$ \Msun\ is the same mass threshold which reproduces the clustering of galaxies in our spectroscopic sample, as described in Trainor \& Steidel (in prep). The mean and median halo masses of galaxies with this halo-mass threshold are $10^{12.2}$ \Msun\ and $10^{12}$ \Msun, respectively.}. Here we consider the dark matter density profile measured in two ways presented in terms of universal overdensity as shown in Figure \ref{maxN_curves_xlog}. (1) First, we consider the average density in concentric spherical shells of radius \dtran\ centered on the position of each halo.  We take the median of this profile across the 100 halos considered (light orange dotted curve) and compare it to the max(\NHI, 300\kms) statistic converted into overdensity using equation \ref{jeans}. (2) We also consider the average dark matter density profile computed within concentric cylindrical shells of radius \dtran\ and length \dv$=\pm300$ \kms\ ($\pm 1200$ pkpc) as shown in the light (green) dash-dotted curve. We consider the effect of the velocity window used as well - the dark (purple) dash-dotted curve is analogous to the green curve, but with a cylinder length of $\pm$700 \kms.

The spherically-averaged radial density profile is well-matched by the NFW profile on small scales as expected (Figure \ref{maxN_curves_xlog}). After 200--300 pkpc (the same scale as the first ``zone'' of the CGM, \S \ref{def}), the spherically averaged radial density profile flattens compared with NFW. From $\sim$200--300 pkpc onwards, the inferred density declines smoothly, falling to the mean density of the universe [log($\Delta$)=0] at \dtran $>$ 3 pMpc. 

The cylindrically averaged density profile significantly dilutes the signal in the inner bins of \dtran\ suggesting that the max(\NHI) statistic traces more of a radial distribution than a ``line-of-sight'' distribution on small scales. On larger scales, the cylindrical and spherically averaged density profiles look quite similar. Between \dtran\ $\approx$ 1.5 - 2 pMpc, the two cylindrically averaged curves become consistent with each other. This suggests that on scales of $\sim$ 2 pMpc the density along a sightline between $\pm300$ \kms\ and $\pm300$ to $\pm700$ \kms\ is similar -- as one would expect far from galaxies as the density approaches the universal mean.

The density inferred from the \ion{H}{1} measurements is markedly higher on $\sim$ pMpc scales than is predicted by either MultiDark DM density profile. There are three possible causes for this discrepancy: (1) It may be the case that the max(\NHI) statistic traces high-density substructure while the radial density profiles drawn from the simulations are (by construction) smoothed - averaging over these structures. The assumption of Jeans smoothing intrinsic to the arguments of \citet{sch01} should minimize the discrepancy; however, it is likely still that substructure may affect our max statistics. (2) One or more of the assumptions used in the \citet{sch01} conversion between  \NHI\ and $\Delta$ could be invalid. For example one of the more uncertain quantities required in the conversion is the \ion{H}{1} photoionization rate, $\Gamma$.\footnote{$\Gamma_{12}$ refers to the photoionization rate in units of $10^{-12}$ s$^{-1}$.} In Figure \ref{maxN_curves} we consider the change in the value of $\Gamma$ needed to bring the measured $\Delta$[max(\NHI, 300\kms)] into agreement with the spherically averaged dark matter density profile. This plot suggests a normalization $\sim$40\% lower than the estimates from \citet{fau08}.\footnote{The temperature of the gas is another quantity that could be adjusted. The temperature assumed currently is $10^4$ K. To match our observations we would need to significantly lower the temperature assumed to $T < 10^3$ K which is unlikely to be the case. If anything, 10$^4$ K is likely a \textit{lower} limit on the temperature of the IGM at $z\sim2$ \citep[see e.g.,][]{bec11a}.} (3) Another possibility is that Jeans scale arguments do not hold in detail on these scales.\footnote{We tested the agreement between the aforementioned models and our data converted using the Fluctuating Gunn-Peterson Approximation \citep[FGPA;][]{rau97}, finding that the disagreement is larger. \citet{rak11b} also compare several methods for converting between \ion{H}{1} optical depth (akin to \NHI) and overdensity (see their \S 4.5).}

\subsection{The Implications of the Distinct Physics of the CGM}

\label{distinct_physics}

The final point is a note of caution regarding the effect the CGM of luminous galaxies may have on measurements which rely on ``average'' locations in the IGM. In addition to the rising \NHI\ and \eabs\ observed in the CGM of galaxies, we have shown that the physical state of the absorbing gas in this region is also distinct from the general IGM. The Doppler widths, \bd, of individual absorbers are significantly larger in the CGM zone, even when compared at \textit{fixed} \NHI. Further, absorbers with \NHI $> 10^{14.5}$ cluster tightly with galaxies (Figures \ref{vpack}, \ref{cf145}, and \ref{asymMap14.5_17_zoom}), and these absorbers are found in the CGM zone with very high frequency (Figure \ref{N_frac}). 

Taken together, these observations suggest that the physics of the ``IGM'' which has long been leveraged for measurements varying from the metagalactic background to the matter power spectrum, may be significantly affected by the baryonic physics of galaxy formation; particularly, low-resolution IGM studies dominated by \ion{H}{1} absorbers $\sim$saturated in Ly$\alpha$ (\NHI $> 10^{14.5}$ \cm2) are likely to be affected by the presence of a nearby galaxy. Further observational and theoretical examination of this subject is crucial to a full understanding of the measurement power of the IGM. 

 In addition, studies of the distribution of metal absorption lines in the ``IGM'' often consider metallic species associated with HI absorbers of log(\NHI) $\gtrsim$ 14.5 \citep{cow95, son96, ell00, car02, sim11}. To a large degree, this is a practical choice as the associated metal absorbers occurring in sub-solar metallicity gas must be found and measured in spectra of finite S/N. However, the properties of the CGM discussed above suggest that, as many authors have speculated, these measurements more directly probe the metallicity of the CGM than that of the IGM. Observations at $z \gtrsim5$ generally cannot measure the \NHI\ associated with metal systems due to the line density of the forest, but they are likely to suffer from the same effect, adding credence to the idea that the recent discovery of the high-$z$ downturn of \ion{C}{4} \citep{rya06b, rya09,bec09, bec11b,sim11b} likely also traces CGM gas. 

The observations of circumgalactic and intergalactic gas and its association with forming galaxies have now progressed to the point that more stringent comparison to theoretical expectations is possible.  Even using the comparatively ``blunt'' tool of gas covering fraction, we have shown (\S\ref{sim}) that the distribution of HI gas around galaxies exceeds the expectations of the models to which they have been compared, with the discrepancy increasing as \NHI\ decreases below $10^{17}$ cm$^{-2}$. It is clear that a combination of high-resolution (to capture the small-scale baryon physics) and cosmological  (to provide sufficient statistical leverage) simulations will be required to make use of the more detailed physical measurements emerging from the observations: e.g., gas-phase kinematics, multi-phase velocity/density structure, redshift-space distortions, gas temperatures, turbulence, the distribution and mixing of metals, all as a function of galaxy properties. In order to match the real universe, simulations will need to realistically model the physics of feedback from star formation, supernovae, and AGN activity and its effect on gas-phase accretion from the CGM.

It will be important for future simulations to have sufficient dynamic range to make predictions about the distribution of neutral hydrogen with $13 < $ log(\NHI) $ < 20$. The physical picture advocated by most modern theoretical treatments of the ``IGM'' is that the Ly$\alpha$ forest traces gas of modest overdensity within the filaments of the ``cosmic web''.  The high covering fractions found in this study (e.g., the relatively large \fc\ of saturated HI extending $\gtrsim$2 pMpc from $\sim 10^{12}$ M$_{\odot}$ dark matter halos; \S \ref{cf}) suggest that if this picture holds, the size of these filaments may remain quite large even as they approach the positions of high-bias halos. These initial comparisons suggest that the topology of the neutral hydrogen gas may be less ``filamentary'' (at all column densities) than current simulations of the ``cosmic web'' and ``cold accretion'' predict. If such disagreement between simulations and observations are common and persist they may become key to a deeper understanding of the baryon physics that ultimately controls the process of galaxy formation.

\section{Summary}

\label{conclusions}

In this paper we have presented a detailed study of neutral hydrogen in the IGM surrounding 886 high-$z$ star-forming galaxies using spectra of background QSOs from the Keck Baryonic Structure Survey.   We draw conclusions from the analysis of 15 sightlines to hyper-luminous high-$z$ QSOs using high-resolution (7 \kms), high-signal-to-noise ratio (50-200) spectra. This study constitutes the largest absorption line catalog ever created at these redshifts through a complete Voigt profile decomposition of the Ly$\alpha$ forest for all 15 sightlines. The high S/N and excellent UV/blue-wavelength coverage of the HIRES QSO spectra allow precise determination of the column densities of saturated \ion{H}{1} absorption systems whose positions we have found to correlate strongly with those of the galaxies in our sample. This sample is also considered by \citet{rak11b} who analyze the pixel statistics of the \ion{H}{1} optical depth.

In this work, we study the absorption patterns of \ion{H}{1} gas in the velocity and spatial locations surrounding each galaxy redshift. We have examined the correlations among column density (\NHI), covering fraction (\fc), incidence (\eabs), absorption line width (\bd), galactocentric transverse distance (\dtran), velocity offset (\dv),  and 3D distance (\dtd). Our principal results are as follows:

\begin{enumerate}

\item{The redshifts of \ion{H}{1} absorption systems in the QSO spectra correlate strongly with the redshifts of galaxies (\S \ref{vel}). The distribution of relative velocities (\dv) is peaked at the systemic redshift of galaxies in our sample (Figure \ref{vel_hist_unweight}). The full excess absorption near galaxies falls within $\pm700$ \kms\ with the majority of the excess \NHI\ (i.e. the higher column density absorbers) falling within $\pm 300$ \kms\ (Figures \ref{vel_hist} and \ref{inner_dist_bin300}).}

\item{The column density (\NHI) of absorbers is strongly correlated with the impact parameter to a galaxy (\dtran, \S \ref{tran}). We consider the highest-\NHI\ absorber within $\pm 300$ \kms\  [max(\NHI)] of each of the 10 galaxies within 100 physical kpc  and find that the median value is 3 orders of magnitude higher than at random places in the IGM (Figure \ref{whisker300}). A trend of rising column densities as a function of decreasing \dtran\ is observed within the inner 300 physical kpc (1 cMpc) surrounding galaxies (Figures \ref{whisker}, \ref{whisker300}, and \ref{maxN_scatter}).}

\item{The column densities (\NHI) of absorbers as measured through the max(\NHI) statistic plateau in the range 300 pkpc $<$ \dtran $\lesssim$ 2 pMpc (Figures  \ref{whisker} and \ref{whisker300}, \S \ref{largescale}). Interestingly, the plateau is elevated in \NHI\ with respect to random places in the IGM by $0.3 - 0.6$ dex. For \dtran $\gtrsim$ 2 pMpc, the values of max(\NHI) decline and become consistent with random places in the IGM.}

\item{The median \NHI\ as a function of 3D distance\footnote{The 3D distance, \dtd, is computed using the quadratic sum of the physical impact parameter between the galaxy and the line of sight to the QSO (\dtran) and the line-of-sight distance one would calculate assuming the \dv\ of each absorber was due only to the Hubble flow, \dhub.} smoothly declines with increasing \dtd\ and becomes consistent with random places in the IGM at \dtd\ $\approx ~3$ pMpc (Figure \ref{3Dwhisker}, \S \ref{text3D}).}

\item{The incidence\footnote{The number of absorbers \textit{per galaxy} with \absdv $<$ 300 \kms\ and in a given bin of  \dtran.} (\eabs) of absorbers of all \NHI\ is higher within \dtran $\simlt$ 2 physical Mpc of the position of a galaxy than at random places in the IGM (Figure \ref{vpack}, \S \ref{cf}). For \dtran $\gtrsim$ 2 pMpc, \eabs\ is consistent with that of the random IGM.}

\item{For absorbers with log(\NHI) $> 14.5$ \cm2, there is a strong peak in the incidence, \eabs, close to galaxies at \dtran\ $<$ 300 pkpc (Figure \ref{vpack}, \S \ref{cf}). No such peak is seen in the distribution of lower-column density absorbers. This suggests that absorbers with log(\NHI) $> 14.5$ \cm2 are more directly related to galaxies than absorbers of lower \NHI. }

\item{The incidence, \eabs, of absorbers with \NHI\ $> 10^{14.5}$ \cm2 and \absdv $<$ 300 \kms\ shows rising values within \dtran\ $< 300$pkpc. From 300 pkpc $<$ \dtran $\lesssim$ 2 pMpc, the \eabs\ roughly plateaus  with \eabs\ $\gtrsim$ 0.5. After  $\sim$2 pMpc, the distribution becomes consistent with random places in the IGM (Figure \ref{cf145}, \S \ref{clustered}). }

\item{Within 300 pkpc and $\pm$350 \kms\ of a galaxy, the probability of intersecting an absorber with log(\NHI) $>$ 14.5 is $> 4$ times higher than at a random place in the IGM. (Figure \ref{exprob_noword} and Table \ref{excessProb}, \S \ref{clustered}).}

\item{Maps of the incidence, \eabs, of low and high-\NHI\ absorbers as a function of \dtran\ and \dhub\ allow for the measurement of redshift anisotropies (\S \ref{mapText}). In low-\NHI\ gas [log(\NHI) $<$ 14.5], on large scales (pMpc) the distribution along the line of sight is compressed compared to the distribution in the transverse direction. This is likely the signature of gas infall e.g. the \citet{kai87} effect  (Figure \ref{asymMap13_14.5_400}). In the high-\NHI\ gas [log(\NHI) $>$ 14.5], we find the ``finger of God'' effect, the elongation of the distribution along the line of sight compared to the transverse distribution (Figure \ref{asymMap14.5_17_zoom}). Taking into account the effect of our redshift errors, the data are consistent with peculiar velocities of $\sim250$ \kms. These conclusions are in agreement with those presented in \citet{rak11b} who used pixel optical depth analysis to demonstrate that the compression along the line of sight on large scales is highly significant ($> 3\sigma$).}

\item{The Doppler widths (\bd) of individual absorption lines in the HIRES spectra rise as the galactocentric distance is decreased (\S \ref{dop}). The majority of the increased width is found within \absdv $<150$ \kms\ and 100 $<$ \dtran $<$ 300 pkpc (Figures \ref{b_all} and \ref{b_all_3D}). The larger line widths could be caused by an increase in the gas temperature, the turbulence of the gas, or both. The median \bd\ close to galaxies are consistent with a gas temperature $\sim 10^5$ K, or a turbulent contribution of $\sim20-30$ \kms.  }

\item{Based on these measurements, we suggest a working definition of the circumgalactic medium (CGM) to be all locations within 300 pkpc and $\pm$300 \kms\ of a galaxy. We find that the CGM of our completeness corrected galaxy sample can account for nearly half of all absorbers with log(NHI) $>$ 15.5 (Figure \ref{N_frac}, \S \ref{def2}). Notably, the CGM of these galaxies comprises only 1.5\% of the universe's volume at these redshifts.}

\end{enumerate}

These findings not only define relevant scales of the CGM of luminous galaxies at redshift $z\sim2-3$, but also demonstrate that much of the moderately high-\NHI\ portion of the IGM likely originates in the regions surrounding luminous galaxies where the baryonic physics of galaxy formation appear to affect the physical state of the gas. This point may have important consequences for past and future studies of the ``IGM'' and should be considered in greater detail.

Our understanding of the physical properties of gas surrounding star-forming galaxies will greatly benefit from analysis of metallic absorption features in the HIRES spectra. These absorption lines probe the ionization state and metallicity of the gas and will provide direct constraints on the total mass in hydrogen and in metals that surround star-forming systems. This work is underway and will be published in the second paper in this series. 

With this further analysis, we will have for the first time a quantitative snapshot of the baryonic interplay between the intergalactic medium and forming galaxies which is evidently central to the processes of galaxy formation and evolution.

\acknowledgements

The authors would like to thank Ryan Cooke who contributed the fits to the Damped Profiles in our QSO spectra. We are grateful to Bob Carswell for his assistance with the code VPFIT.  Many thanks to Joop Schaye for his careful reading of the draft and insightful comments. We would also like to thank George Becker, Brian Siana, and Jean-Ren\'{e} Gauthier for many helpful and interesting discussion. Thanks to Michele Fumagalli for providing the values listed in Table \ref{fum_table}. We wish to acknowledge the staff of the the W.M. Keck Observatory whose efforts insure the telescopes and instruments perform reliably. Further, we extend our gratitude to those of Hawaiian ancestry on whose sacred mountain we are privileged to be guests. 

This work has been supported by the US National Science Foundation through grants AST-0606912 and AST- 0908805. CCS acknowledges additional support from the John D. and Catherine T. MacArthur Foundation and the Peter and Patricia Gruber Foundation. Support for N.A.R. was provided by NASA through Hubble Fellowship grant HST-HF-01223.01 awarded by the Space Telescope Science Institute, which is operated by the Association of Universities for Research in Astronomy, Inc., for NASA, under contract NAS 5-26555. This research has made use of the Keck Observatory Archive (KOA), which is operated by the W. M. Keck Observatory and the NASA Exoplanet Science Institute (NExScI), under contract with the National Aeronautics and Space Administration. 

\textit{Facilities:} \facility{Keck:I (LRIS)}, \facility{Keck:I (HIRES)},  \facility{Keck:II (NIRSPEC)}

\bibliographystyle{apj}
\bibliography{HI_cgm}



\end{document}